\documentclass[a4paper, 10pt, pdftex, fleqn]{book}

\usepackage{thesis}

\begin{document}


\pagenumbering{roman}



\hypersetup{pageanchor=false}

\includepdf{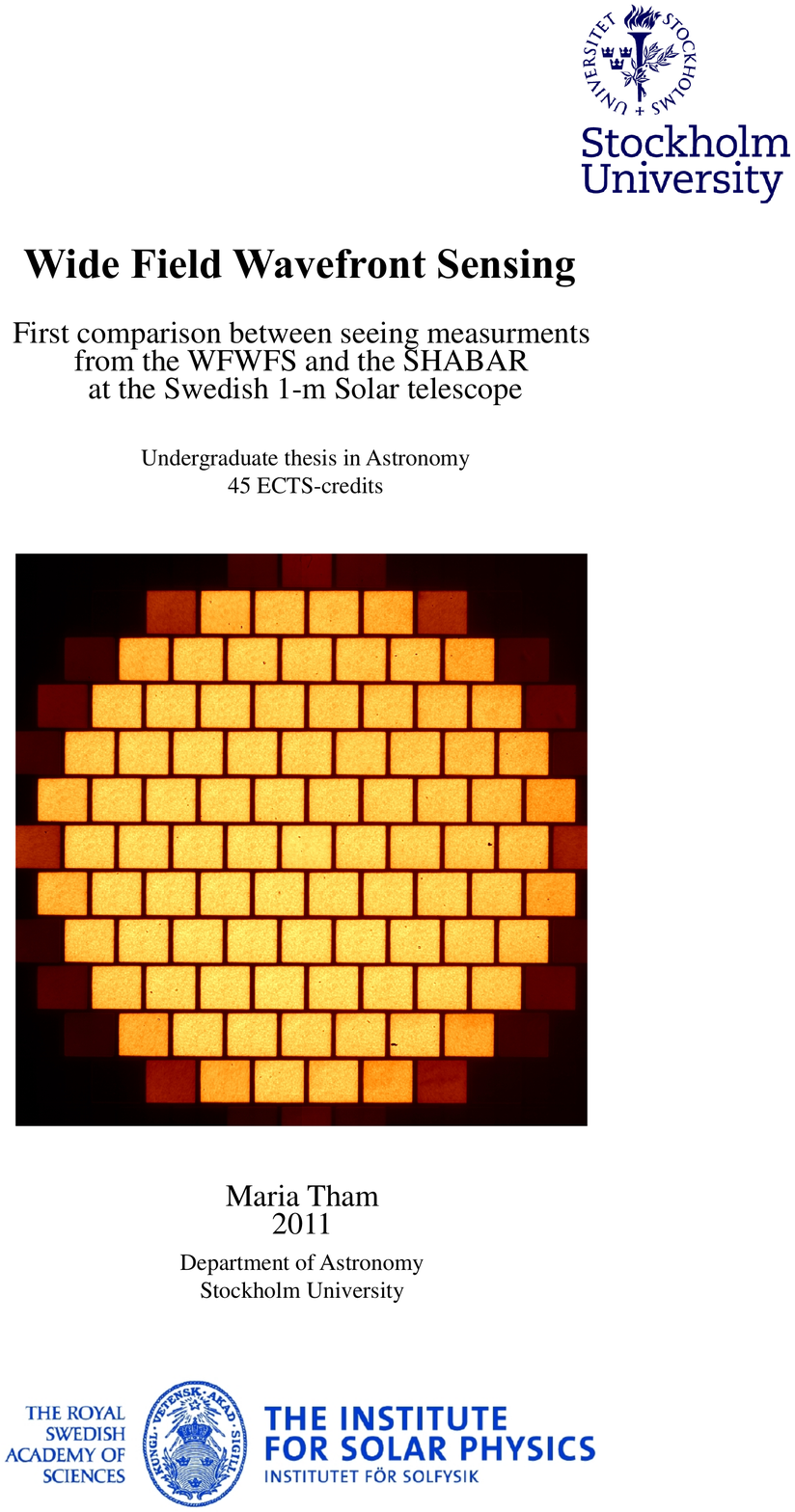}

\thispagestyle{empty}
\null
\vfill
\firstcoverinfo

\hypersetup{pageanchor=true}



\cleardoublepage
\null
\vfil
\begin{center}\textbf{\Large Abstract}\end{center}
      When observing from the ground one has to deal with the turbulence from the atmosphere and the turbulence caused by the telescope's surroundings. In order to find the best sites for future solar telescopes and develop multi-conjugate adaptive optics (MCAO) for them, the contributions to seeing have to be characterized for heights up to at least 12 km above the telescope.

      This characterization is done with a method that measures differential image displacements using several subapertures. By extending the ordinary DIMM- and S-DIMM-methods to measure displacements at different field angles, the height distribution of seeing can be measured. This extended method is called S-DIMM+, where the letter $S$ denotes the use of the Sun as source and the $+$ signifies that it is an extension of the former methods.

      Observations are made with the Wide Field Wavefront Sensor (WFWFS) mounted at the Swedish 1-m Solar Telescope (SST) at La Palma. The results from these measurements are compared with the Shadow Band Ranger (SHABAR), also mounted on the SST.

      The first results of these comparisons show good correlation between the two instruments for heights up to 500 m. Results from the WFWFS also show that the dominant seeing at this site comes from layers close to the ground and at high altitude and that Fried's parameter, $r_0$, is larger than 40 cm for the intermediate layers.
\vfil
\newpage

\null
\vspace{15cm}
\begin{center}\textbf{\Large Acknowledgment}\end{center}
      I would like to thank my supervisor Göran Scharmer who has guided me through my work and Tim van Werkhoven for introducing me to all this and for always answering questions about his previous work. I would also like to thank the rest of the Solar Physics Group for being around and making the atmosphere much nicer to work in. Especially I would like to thank Roald Schnerr for always helping out with everything and for giving good advises, Michiel van Noort for helping me with all sorts of computer trouble during the installation of the programs and Peter Sütterlin for his help with scripts and local updates on La Palma. I would of course also like to thank my boyfriend, Jakob Regberg, for putting up with me and always taking care of me. Thank you all!
\vfil
\newpage
\setcounter{tocdepth}{1}
\tableofcontents

\newpage


\pagenumbering{arabic}



\chap{Introduction}
\label{chap:intro}
Image quality is very important for all types of observations and it has been known for quite long that the quality of an image observed with a telescope does not only depend on which type of telescope that is used, the site where the telescope is located matters. The atmosphere, through which observations are made (with ground based telescopes), perturbs the wavefronts causing image degradation. 

Different methods exist to compensate for the wavefront distortion caused by the atmosphere. The nature of the distortion must be known in order to compensate for it. Different ways of quantifying these distortions exists as well. 

The distortion is different for different sites, making the location of the telescope very important. The distortions are also changing rapidly with time. Methods for quantifying this distortion and for compensating it in real time are therefore needed. 

A brief introduction to the origin of these distortions and how to compensate for them are given below together with a short description of the next European solar telescope. 

This chapter ends with a short outline of this report.

\sec{Seeing}
\label{sec:seeing}
The atmosphere is not, as one might imagine, homogeneous, it is turbulent and consists of turbulent eddies with different temperatures. The index of refraction is primarily a function of temperature and it will therefore be different for the different layers in the atmosphere. Light is refracted when traveling through the turbulent atmosphere and the optical path length for one line-of-sight will differ from another. Plane wavefronts will be disturbed after passing through the atmosphere making the observed objects blurred, destroying the visible details. 

\begin{figure}[ht!]
  \centering
    \includegraphics[width=0.47\textwidth]{\imgpath 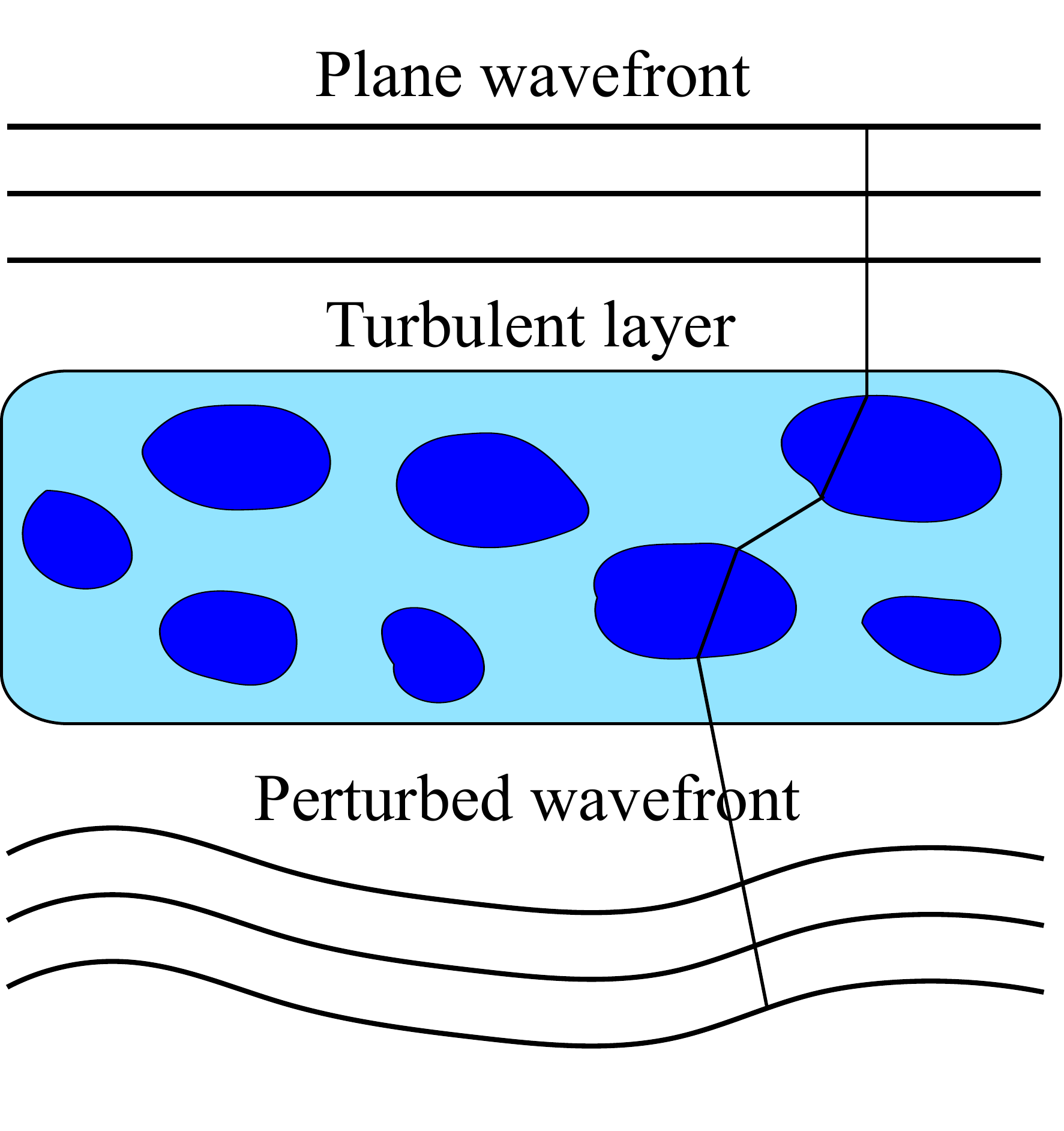}
  \caption{Plane wavefronts enter the atmosphere and are distorted by the turbulent layers. The different objects in the turbulent layer represent regions with different refraction indexes $n$. A sample light ray shows how refraction in these objects leads to the perturbed wavefront.}
  \label{fig:seeing}
\end{figure}

One example of this is stars that seem to twinkle when you observe them a clear night with your naked eye. The twinkling is not due to the stars themselves, it's caused by high-altitude turbulence in the atmosphere and the effect is called seeing. \refpicl{fig:seeing} schematically describes this.

These rippled wavefronts result in different forms of image degradation. Light that follows different paths through the atmosphere can form separate images of the same object. Labeyrie \citep{1970A&A.....6...85L} referred to this as speckle patterns. 

\begin{figure}[ht!]
\centering
\subfloat[]{\includegraphics[bb=0 0 220 189, width=0.33\textwidth]{\imgpath 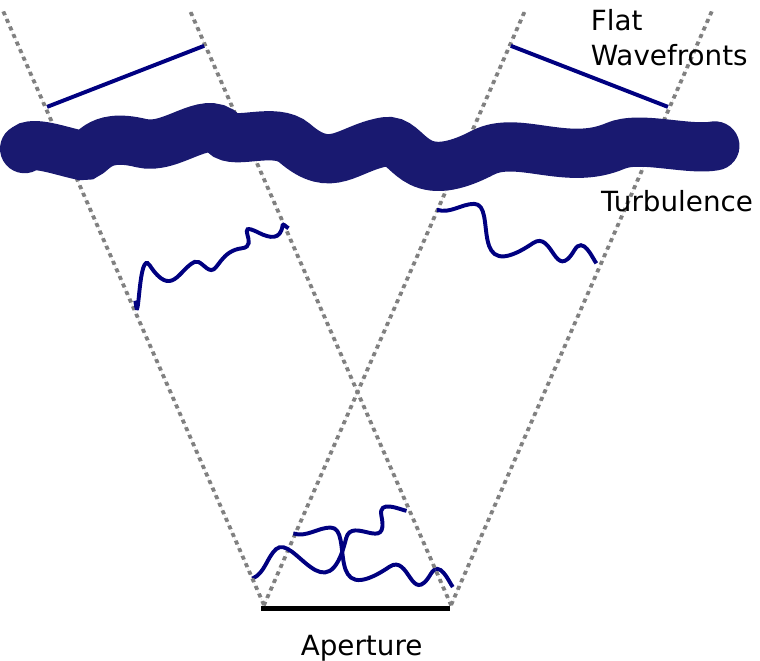}
\label{fig:high_seeing}}
\centering
\subfloat[]{\includegraphics[bb=0 0 220 189, width=0.33\textwidth]{\imgpath 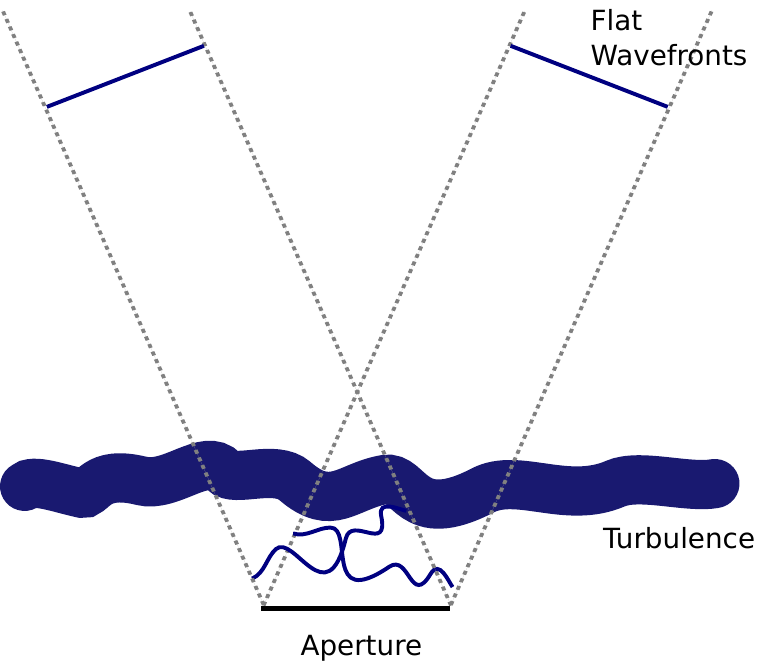}
\label{fig:low_seeing}}
\centering
\subfloat[]{\includegraphics[bb=0 0 220 189, width=0.33\textwidth]{\imgpath 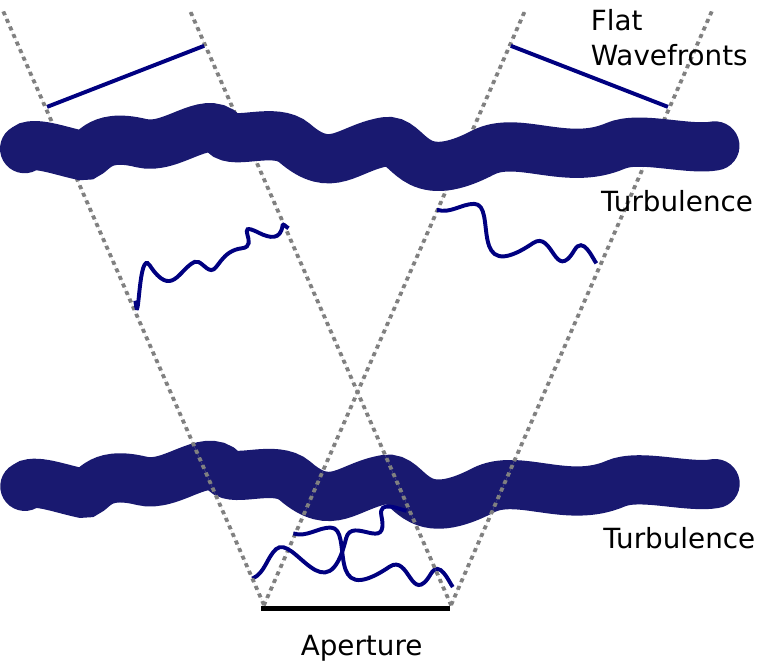}
\label{fig:multi_seeing}}
\caption{(a) High-layer seeing. Different parts of the atmosphere affect different wavefronts so that the distortion varies across the field of view. (b) Ground-layer seeing. The field of view is affected equally since wavefronts from all directions traverse through the same turbulence. (c) Multi-layer seeing. It is most common that seeing occurs both in higher layers and closer to the ground. High-layer seeing is normally less intense than ground-layer seeing. Pictures are adapted from T.I.M. van Werkhoven.} \label{fig:seeing_layers}.
\end{figure}

The turbulence that causes bad seeing can be anywhere, from the air inside the telescope up to the high atmosphere. High-layer seeing, \refpic{fig:high_seeing}, is significantly less intense than ground-layer seeing, \refpic{fig:low_seeing}, for daytime observations.
The different layers will cause different image distortions. High-layer seeing causes differential distortions because different parts of the field of view (FOV) will cross different regions of air at high altitude. 

\newpage
\sec{Seeing correction}
\label{sec:corr}
Short exposure images need to be taken in order to observe speckles, otherwise they will be averaged out. Long exposures will cause overall image degradation since the speckle patterns will overlap. The resolution of the image will therefore be degraded. In 1970, Labeyrie \citep{1970A&A.....6...85L} presented techniques to partially restore the original image in post processing from a burst of short images. A disadvantage with this method is that it is limited to good seeing and to small telescope apertures.

\begin{figure}[ht!]
  \centering
    \includegraphics[width=0.5\textwidth]{\imgpath 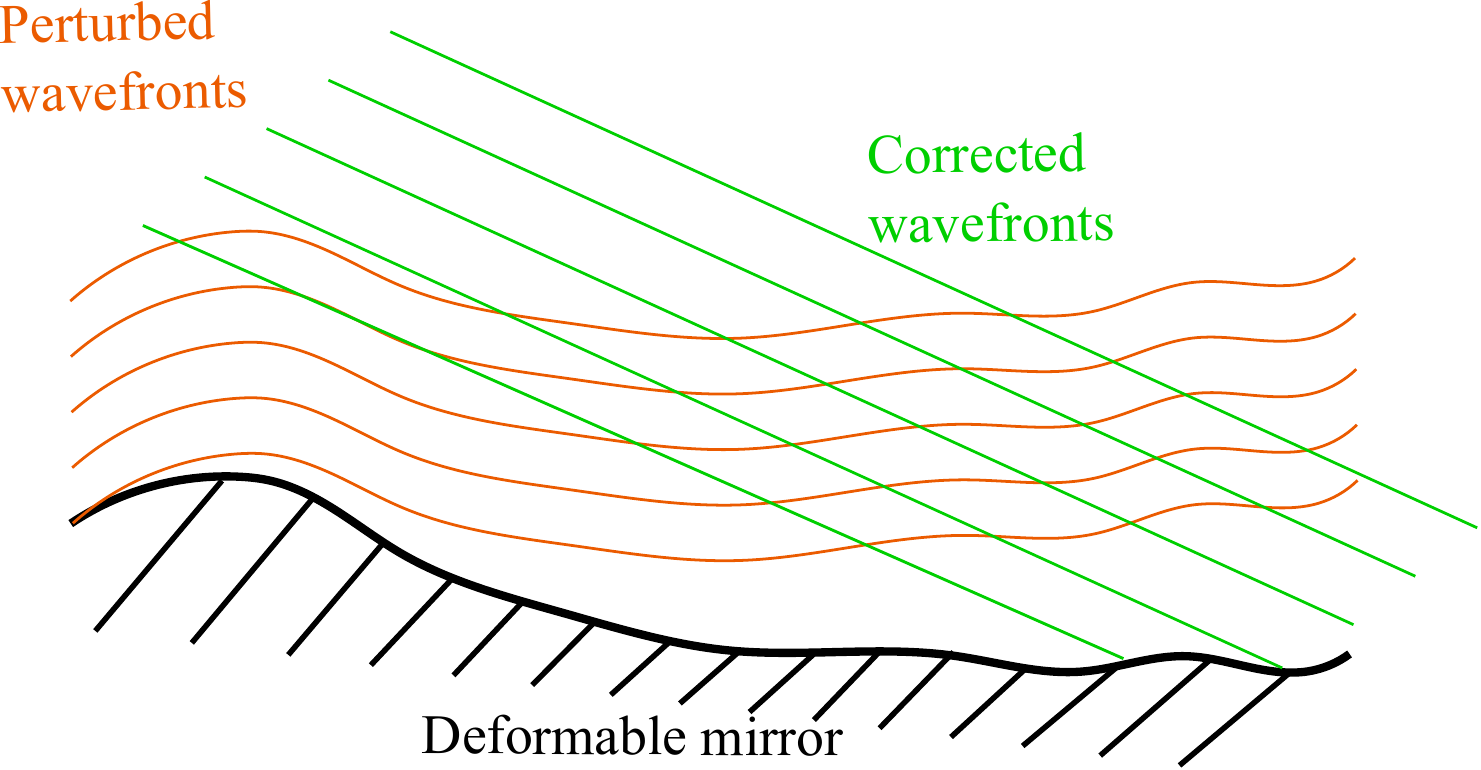}
  \caption{The correcting device of an adaptive optics system, a deformable mirror. The deformable mirror is deformed so that the wavefront distortions caused by the atmosphere are compensated for.}
  \label{fig:dm}
\end{figure}

Another method is adaptive optics, which is the most powerful technique for all astronomical applications \citep{1993ARA&A..31...13B}. This technique uses a wavefront sensor that controls an optical component, a mirror whose surface can be deformed, introducing a controllable counter-distortion to the wavefront. The sensor will measure the distortion of the incoming wavefront and feed this information to the mirror. The mirror will then correct the distortion in real-time and the large scale structure of the original wavefront will be restored. \refpicl{fig:dm} schematically describes the corrector in this technique. 

By the use of several deformable mirrors, a Multi-conjugate Adaptive Optics (MCAO) system can be built. The deformable mirrors conjugate different altitudes and the turbulence can be compensated for in a three-dimensional way.

A more detailed description of adaptive optics is given in \refchap{chap:ao}.

\sec{European Solar Telescope}
\label{sec:EST}
The European Solar Telescope (EST) is a future 4-meter solar telescope that will be located in the Canary Islands. The project involves institutions and partners from 15 European countries. The project is divided into different workpackages and one of them (WP08000) is responsible for the site characterization \citep{EST}. The turbulence above the observatories of the Canary Islands are studied in order to achieve turbulence profiles for the MCAO system planned for the EST \citep{2010A&A...513A..25S}.

The site will be characterized with two different instruments, a long-base SHABAR (see \refsec{sec:shabar}) and a wide-field wavefront sensor (see \refsec{sub:wfwfs}), operating at the two sites, the Observatorio del Teide on Tenerife and the Observatorio del Roque de los Muchachos on La Palma \citep{EST}. One SHABAR and one WFWFS are placed on the Swedish 1-m Solar Telescope on La Palma and the other WFWFS and SHABAR will be placed on VTT (the Vacuum Tower Telescope) on Tenerife. The advantage of placing both a WFWFS and a SHABAR on the same building is that the seeing measured by the both instruments can be compared \citep{WP08100}. This comparison is done in section \refsec{sec:shab_comp}.

The aim of the observations and the data analysis is to understand the isoplanatic patch that is attainable with the MCAO and to compare La Palma's and Tenerife's high-altitude seeing \citep{EST}.

\vspace{1cm}
This report will mainly cover the data reduction and analysis from the WFWFS mounted on the Swedish 1-meter Solar Telescope (SST) and comparison with the SHABAR, also mounted on the SST. \refchap{chap:seeing} gives a theoretical background to seeing and atmospheric turbulence. \refchap{chap:ao} describes the different components of an adaptive optics system. \refchap{chap:dimm} describes different methods to quantifying seeing and ends with the optical setup at the SST. \refchap{chap:process} deals with the data processing made after data were taken. \refchap{chap:results} provides some initial results and comparison between the WFWFS and the SHABAR mounted on the SST. In \refchap{chap:conclusions} some conclusions from these results are drawn and some future aspects are discussed.



\chap{Quantifying seeing}
\label{chap:seeing}
The characteristics of the wavefront aberrations that need to be compensated for must be well known in order to design a good AO system. These aberrations are random and can therefore only be described statistically. The statistics describe the seeing conditions, which evolve with time. In order to design MCAO (and other instruments) the statistics of their evolution, their mean value and their standard deviation, need to be known. 

This chapter will explain the nature of the wavefront distortions, starting with a short analysis of atmospheric turbulence. A more detailed analysis can be found in Roddier's \textit{Adaptive Optics in Astronomy} \citep{2004aoa..book.....R} and Beckers' \textit{Adaptive Optics for Astronomy} \citep{1993ARA&A..31...13B}. Methods for quantifying seeing are described in \refchap{chap:dimm}.

\sec{Atmospheric turbulence}
\label{sec:tubrulence}
Changes in the refractive index of air are essentially proportional to changes in the air temperature. Temperature inhomogeneities are produced when layers of different temperatures are mixed due to wind shears. The statistics of refractive index inhomogeneities follow the inhomogeneities of temperature which are described by Kolmogorov \citep{1941DoSSR..30..301K}. The model assumes that energy is inserted at low frequencies on large scales which are characterized by the \textit{outer scale of turbulence}. There is also an \textit{inner scale} set by molecular friction. Energy is transported from the outer to the inner scale in a cascading way and is finally converted into heat \citep{1993ARA&A..31...13B}. 

The value of the refractive index $n$ is of little interest when studying wavefront perturbations caused by the variation of $n$. The Kolmogorov model states that the variance of the difference between the refractive index in two different locations, separated by the three-dimensional separation vector $\vec{\rho}$, is given by 
\begin{equation}
D_n(\vec{\rho})=\average{|n(\vec{r}) - n(\vec{r} + \vec{\rho})|^2} = C_N^2(z)\rho^{2/3},
\label{eq:isf} 
\end{equation}
where \vec{r} is a three-dimensional positions vector and $\rho = |\vec{\rho}|$. $D_n$ is called the index structure function and the approximation is only valid when $\rho$ is smaller than the \textit{outer scale of turbulence}. The value for this outer scale is highly debated and quoted values span from a few tens of centimeters up to kilometers \citep{1987ApOpt..26.4106C}.
%

The power law of \refeqn{eq:isf} has experimentally been found to be quite accurate over distances less than 1 meter. That is, \refeqn{eq:isf} is certainly valid for small telescopes but it is likely to be more inaccurate for large telescopes \citep{2004aoa..book.....R}.

$C_N^2(z)$ in \refeqn{eq:isf} is the index structure coefficient and varies over distances at a much larger scale than over the inhomogeneities. The integral over $C_N^2(z)$ along the line of sight gives a measure of the total wavefront distortion at ground level.

The temporal variance, how fast the index fluctuates with time, is as important as the spatial variance for adaptive optics systems. This temporal variance can be described in the same way as the spatial, by a temporal structure function describing the variance of the difference between $n$ at time $t$ and a later time $t + \tau$. 
Assuming that variations in $n$ live longer than the time it takes for wind driven inhomogeneities to cross over the telescope, Taylor expansion gives
\begin{equation}
n(\vec{r}, t+\tau) = n(\vec{r} - \vec{v}{\tau}, t),
\label{eq:wind}
\end{equation}
where $\vec{v}$ is the wind velocity. \refeqnl{eq:isf} can with this assumption be rewritten as a temporal structure function
\begin{eqnarray}
D_n(\vec{v}\tau)&=&\average{|n(\vec{r},t) - n(\vec{r}, t+\tau)|^2} = \nonumber \\
	 &=&\average{|n(\vec{r},t) - n(\vec{r} - \vec{v}{\tau}, t)|^2} = C_N^2(z)|\vec{v}\tau|^{2/3},
\label{eq:tsf} 
\end{eqnarray}
and it can be concluded that the temporal structure function can be obtained by simply substituting $|\vec{v}\tau|$ for $\rho$ in the spatial structure function \refeqn{eq:isf}.

\sec{Wavefront distortions}
\label{sec:distortions}
The effect of the temporal and spatial variation of the index of refraction in a turbulent atmosphere on the wavefront perturbation are of great interest for adaptive optics. As described in \refsec{sec:seeing} the distortion is caused by varying optical path lengths through the atmosphere, due to the inhomogeneity in $n$. The optical path length is given by
\begin{equation}
\delta=\int n(z) \, \mathrm{d} z,
\label{eq:opl}
\end{equation}
where $n$ is integrated over the line of sight. The air refractive index is fairly wavelength independent in the range from visible to near infrared and hence also the optical path length. However, the wavefront phase fluctuation is wavelength dependent and given by 
\begin{equation}
\varphi=k \int n(z) \, \mathrm{d} z \equiv k\delta,
\label{eq:wfp}
\end{equation}
where $k$ is the wave number which varies as the inverse of the wavelength, $k=2 \pi / \lambda$. The distortions that are corrected for are however the difference in path length and they can therefore be compensated for at all wavelengths (since the path length is wavelength independent).

The difference between two points at the telescope entrance aperture is (again) of more interest than the absolute wave-front phase. The variance of the difference in the structure function between the phase in two different locations, separated by the two-dimensional separation vector $\vec{\xi}$ is given by
\begin{equation}
D_{\varphi}(\vec{\xi})=\average{|\varphi(\vec{x}) - \varphi(\vec{x} + \vec{\xi})|^2},
\label{eq:sfp}
\end{equation}
where $\vec{x}$ is the two-dimensional position vector and $\xi=|\vec{\xi}|$ is the distance between the two points at the telescope entrance aperture. Using \refeqn{eq:wfp} in \refeqn{eq:sfp} makes it possible to express the phase structure function in terms of the index structure function integrated along the line of sight. Using \refeqn{eq:isf} and integrating gives an equation for $D_{\varphi}$ in terms of $C_N^2(z)$
\begin{eqnarray}
D_{\varphi}(\vec{\xi})=k^2 \int C_N^2(z) \, \mathrm{d}z \, \underbrace{\int\limits_{-\infty}^{\infty}(\xi^2+z^2)^{1/3}-z^{2/3} \, \mathrm{d}z}_I,
\label{eq:sfp_int}
\end{eqnarray}
where the integral $I$ is evaluated as
\begin{equation}
I=\frac{2}{5}\frac{\Gamma(\frac{1}{2})\Gamma(\frac{1}{6})}{\Gamma(\frac{2}{3})}\xi^{5/3} \approx 2.91 \, \xi^{5/3}, 
\end{equation}
giving
\begin{equation}
D_{\varphi}(\vec{\xi})=2.91 k^2 \int C_N^2(z) \, \mathrm{d}z \, \xi^{5/3}.
\label{eq:sfp2}
\end{equation}
$C_N^2(z)$ is only dependent on height since the atmosphere generally is considered to be stratified in plane parallel layers. \refeqnl{eq:sfp2} can therefore be rewritten as
\begin{equation}
D_{\varphi}(\vec{\xi})=2.91 \frac{k^2}{\cos(\gamma)} \int C_N^2(h) \, \mathrm{d}h \, \xi^{5/3}.
\label{eq:sfp3}
\end{equation}
where $\frac{1}{\cos(\gamma)}$ is the air mass and $\gamma$ is the angular distance from zenith. \refeqnl{eq:sfp3} is usually rewritten as 
\begin{equation}
D_{\varphi}(\vec{\xi})=6.88(\xi/r_0)^{5/3},
\label{eq:sfp4}
\end{equation}
where the Fried parameter \citep{1965JOSA...55.1427F},
\begin{equation}
r_0=\left[0.423 \frac{k^2}{\cos(\gamma)} \int C_N^2(h) \, \mathrm{d}h \right]^{-3/5},
\label{eq:r0}
\end{equation}
is a useful quantity when characterizing seeing quality.

\refeqnl{eq:sfp4} describes the spatial distribution of the wave-front distortions. 

Using \refeqn{eq:tsf}, a temporal structure function for the phase can be found by substituting $\bar{v}\tau$ for $\xi$ in \refeqn{eq:sfp4},
\begin{equation}
D_{\varphi}(\bar{v}\tau)=6.88(\bar{v}\tau/r_0)^{5/3},
\label{eq:tsfp}
\end{equation}
where $\bar{v}$ is the velocity with which the wavefront-phase propagates. This velocity is a weighted average of the layers' velocities. The effect of the finite response time of an AO system can be calculated with \refeqn{eq:tsfp}. If a correction to the wavefront is applied at $t + \tau$ when the phase is measured at $t$, the mean square phase error is given by
\begin{equation}
\sigma_{\mathrm{time}}^2(\tau)=6.88(\bar{v}\tau/r_0)^{5/3}.
\label{eq:mspe}
\end{equation}

\refeqnl{eq:mspe} can be used to calculate a maximum allowed time delay before applying correction
\begin{equation}
\tau_0=6.88^{-\frac{3}{5}} \frac{r_0}{\bar{v}}=0.314\frac{r_0}{\bar{v}}.
\label{eq:mtd}
\end{equation}
This delay is called the Greenwood time delay and the corresponding frequency, $f_G$ is called the Greenwood frequency \citep{1977JOSA...67..390G}.

The isoplanatic angle is also a frequently used quantity to describe atmospheric turbulence. It defines an angular distance $\theta$ from the source that is used to sense the wavefront to objects around this source for which the compensation still will be good. The image quality decreases as the angular distance increases. The mean square error in the wavefront for a layer at a distance $h/\cos(\gamma)$ is obtained by replacing $\xi$ with $\theta h / \cos(\gamma)$ in \refeqn{eq:sfp4}. For this to be valid not only for one layer, $h$ should be replaced by the weighted average $\bar{h}$ of the layer altitudes (similar to what was done with $\bar{v}$ in \refeqn{eq:mspe}) giving
\begin{equation}
\sigma_{\mathrm{aniso}}^2(\theta)=6.88 \left(\frac{\theta \bar{h}}{r_0 \cos(\gamma)}\right)^{5/3}.
\label{eq:aniso}
\end{equation}
To have an anisoplanicity rms error less than 1 radian the isoplanatic angle must be less than
\begin{equation}
\theta_0=6.88^{-\frac{3}{5}} \frac{r_0 \cos(\gamma)}{\bar{h}}=0.314\frac{r_0 \cos(\gamma)}{\bar{h}}.
\label{eq:isoang}
\end{equation}

\sec{Wavefront representation}
\label{sec:representation}
It is often useful to describe the wavefront for a circular aperture in terms of the orthogonal set of Zernike polynomials, $Z_j(\rho, \theta)$. They can be expressed either in Cartesian $(x,y)$ or Polar $(r, \theta)$ coordinates. The polar form is most common and will be used here.
\begin{equation}
\begin{array}{lr}
  \left.\begin{array}{l}
   Z_{even\,j}= \sqrt{n+1}R_n^m(r)\sqrt{2}\cos(m\theta)\\
   Z_{odd\,j}= \sqrt{n+1}R_n^m(r)\sqrt{2}\sin(m\theta)\\
  \end{array} \right\} & m\neq0\\
  \begin{array}{l}
  Z_j=\sqrt{n+1}R_n^0(r)
  \end{array} & m=0
\end{array}
\label{eq:zernike}
\end{equation}
where
\begin{equation}
R_n^m(r)=\sum_{s=0}^{(n-m)/2}\frac{(-1)^s(n-s)!}{s![(n+m)/2-s]![(n-m)/2-s]!}r^{n-2s}
\label{eq:zerniker}
\end{equation}
A wavefront phase-distortion over a circular aperture of radius $R$ can be expressed as a Zernike polynomial expansion,
\begin{equation}
\varphi(R\rho,\theta)=\sum_j a_j Z_j(\rho,\theta),
\label{eq:zernikesum}
\end{equation}
where $\rho=r/R$ and $a_j$ some coefficients (for more details see \citep{1976JOSA...66..207N}).

\begin{figure}[ht!]
  \centering
    \includegraphics[width=0.7\textwidth]{\imgpath 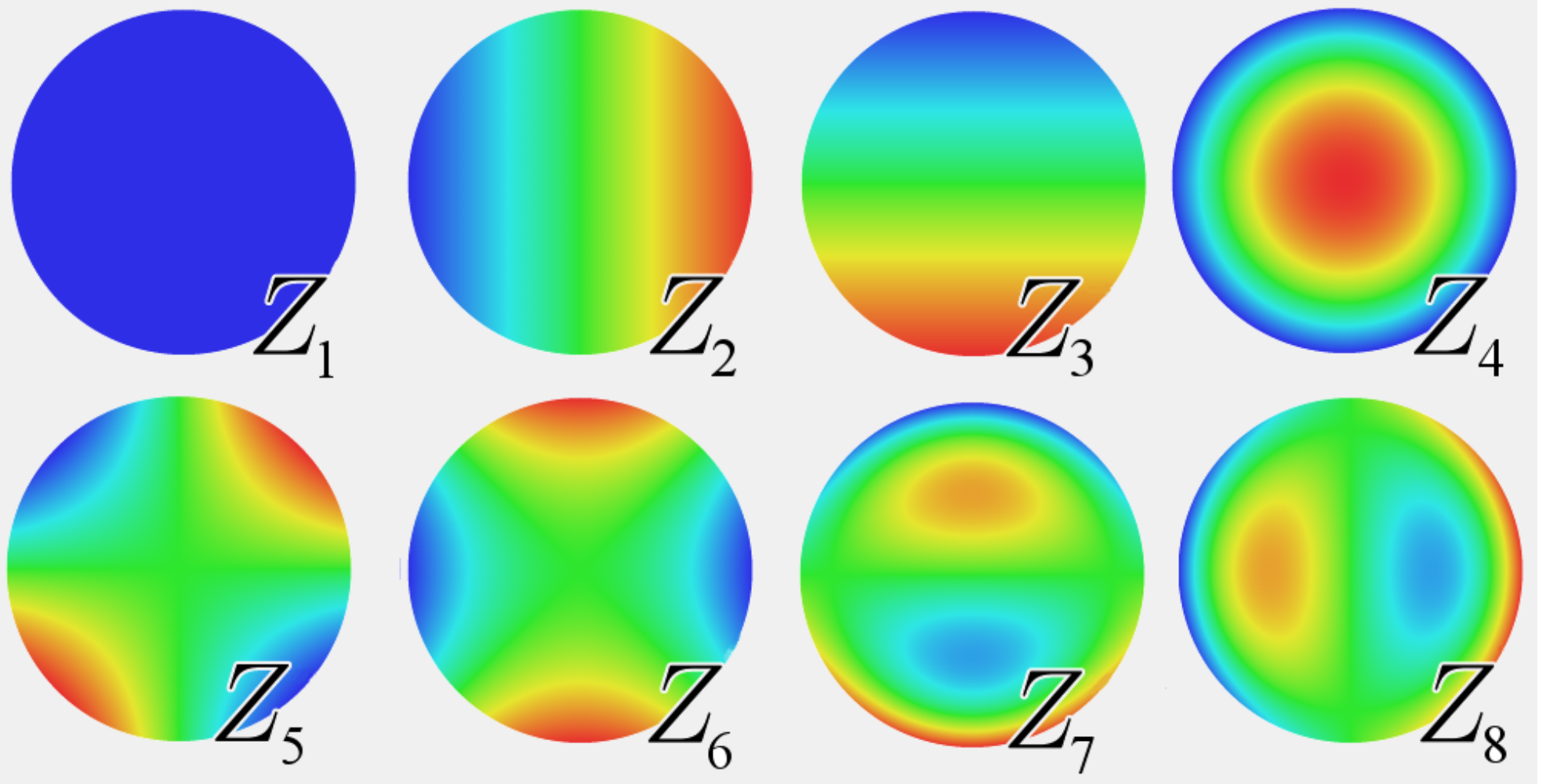}
  \caption{Visualisation of the first eight Zernike modes $Z_j$. From left to right: Piston ($Z_1$), Tip/tilt ($Z_2$ and $Z_3$), Defocus ($Z_4$), Astigmatism ($Z_5$ and $Z_6$), Coma ($Z_7$ and $Z_8$). Picture adapted from \protect\url{http://en.academic.ru/dic.nsf/enwiki/973636}}
  \label{fig:zernike}
\end{figure}

The polynomials are set up so that the first few directly relates to optical phenomena such as tip, tilt, defocus, astigmatism etc. The first eight Zernike polynomials are described in \reftab{tab:zernike} together with their optical error. \refpicl{fig:zernike} shows visualisations of the first eight modes.

\begin{table}[ht!]
  \begin{center}
    \begin{tabular}{llll}\hline
      \textbf{$j$}&\textbf{Zernike mode $Z_j$}&\textbf{$A_j$ (rad$^2$)}&\textbf{Optical aberration} \\\hline
      1&	$1$&				1.0299&	Piston\\
      2&	$2r\cos(\theta)$&		0.582&	Tip/tilt (X-axis)\\
      3&	$2r\sin(\theta)$&		0.134&	Tip/tilt (Y-axis)\\
      4&	$\sqrt{3}(2r^2-1)$&		0.111&	Defocus\\
      5&	$\sqrt{6}r^2\sin(2\theta)$&	0.0880&Astigmatism ($\pm 45^{\circ}$)\\
      6&	$\sqrt{6}r^2\cos(2\theta)$&	0.0648&	Astigmatism ($ 0^{\circ}/90^{\circ}$)\\
      7&	$\sqrt{8}(3r^2-2r)\sin(\theta)$&0.0587&	Coma (Y-axis)\\
      8&	$\sqrt{8}(3r^2-2r)\cos(\theta)$&0.0525&	Coma (X-axis)\\\hline
    \end{tabular}
  \end{center}
  \caption{The eight first Zernike polynomials. The constant $A_j$ is used to determine the mean square residual error after correcting $j$ Zernike modes, see \refeqn{eq:msre}. The last column describes the optical aberration that is associated with mode $j$. Table adapted from \citep{1976JOSA...66..207N}.
}\label{tab:zernike}
\end{table}

The correction for the first $N$ modes can be written as
\begin{equation}
\varphi_c(R\rho,\theta)=\sum_j^N a_j Z_j(\rho,\theta).
\label{eq:zernikesumcorr}
\end{equation}
Fried \citep{1965JOSA...55.1427F} and Noll \citep{1976JOSA...66..207N} found that if this correction is applied to a distorted wavefront, and assuming that \refeqn{eq:sfp4} holds, the mean square residual error can be defined as
\begin{equation}
\sigma^2_j=\average{\varphi^2}-\sum_j^N \average{|a_j|^2}= A_j \left(\frac{D}{r_0}\right)^{5/3}
\label{eq:msre}
\end{equation}
where $D$ is the aperture diameter and $r_0$ the Fried parameter from \refeqn{eq:r0}. The first eight values of $A_j$ are found in \reftab{tab:zernike}. Noll \citep{1976JOSA...66..207N} found that for $N>10$ the the mean square residual error is given by
\begin{equation}
\sigma^2_j \approx 0.2944 N^{-\sqrt{3}/2} \left(\frac{D}{r_0}\right)^{5/3}.
\label{eq:msre10}
\end{equation}

The residual error without correction ($\sigma^2_1$) is given by
\begin{equation}
\sigma^2_1 = 1.0299 \left(\frac{D}{r_0}\right)^{5/3}, 
\label{eq:msre1}
\end{equation}
so the rms phase distortions will be about 1 radian if a telescope aperture of size $r_0$ is used.

Another method for decomposing wavefront errors is the Karhunen-Loève expansion. Like Zernike polynomials, Karhunen-Loève (K-L) modes are orthogonal. In addition, they are statistically independent modes which are numerical instead of analytical (Zernike modes) and which can be approximated as sums of Zernike modes by diagonalization of the Zernike covariance matrix. 

The difference between a Zernike mode and its related K-L mode becomes larger with increasing order. The mean square residual error for an AO system compensating for the first $N$ K-L modes decreases faster than for a system compensating for Zernike modes. This means that it is more efficient to compensate K-L modes than Zernike modes.



\chap{Adaptive Optics}
\label{chap:ao}
Adaptive Optics (AO) is real time compensation of the image degradation, that is a technique to correct distorted wavefronts in real time. The technique was first proposed in 1953 by Babcock and later in 1957 by Linnick (independently). An adaptive optics system uses a deformable mirror to instantaneously correct the wavefront distortions \citep{2004aoa..book.....R}.

\begin{figure}[ht!]
  \centering
    \includegraphics[width=0.7\textwidth]{\imgpath 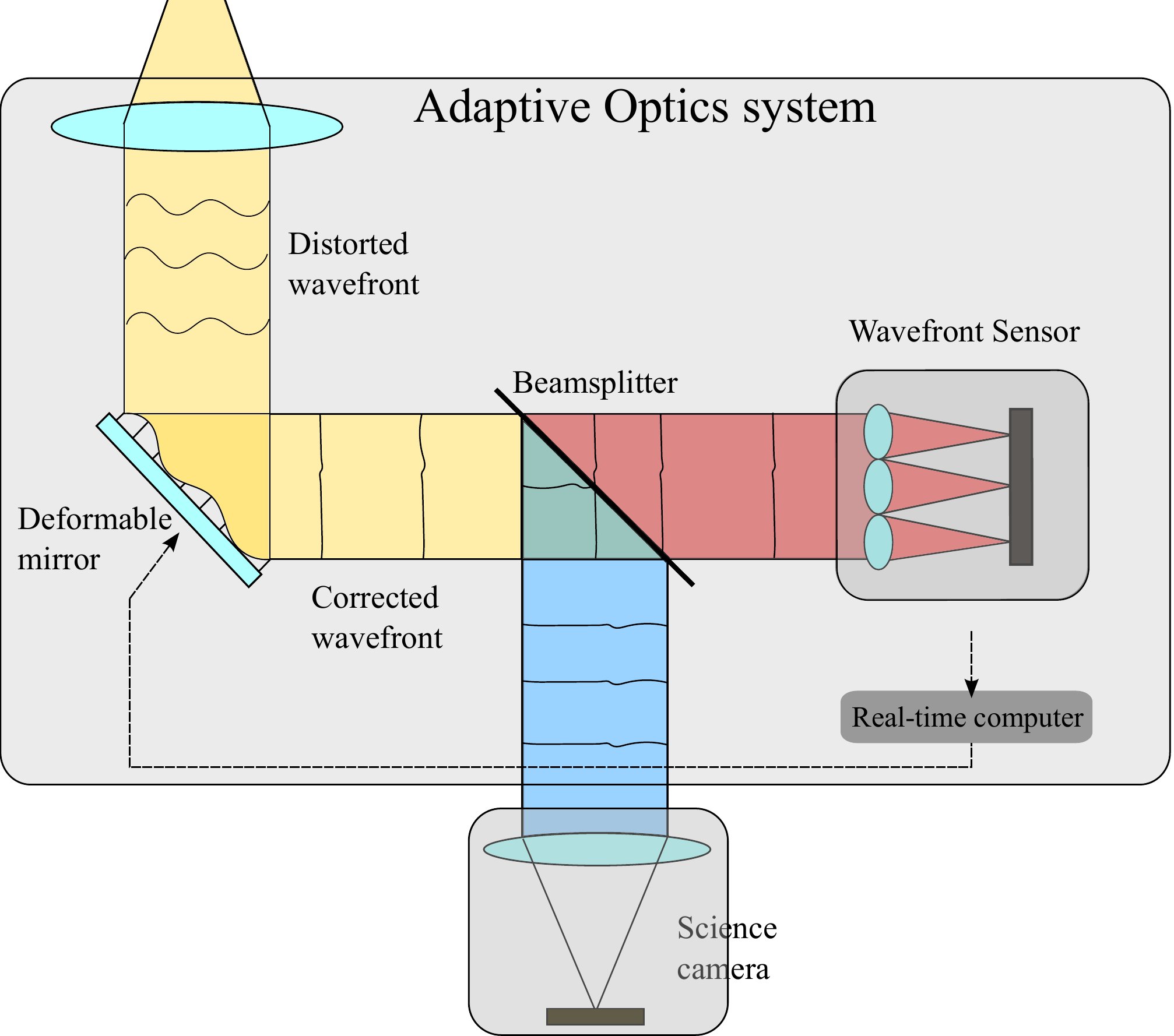}
  \caption{A sketch describing a typical adaptive optics system using a deformable mirror as wavefront corrector and a Shack-Hartmann wavefront sensor to detect the distortions. The wavefront sensor sends signals to the deformable mirror forming a closed-feedback loop. Most light is transmitted by a beam splitter into the science camera while some is transmitted into the wavefront sensor.}
  \label{fig:aosys}
\end{figure}

\refpicl{fig:aosys} schematically describes an adaptive optics system.

This chapter will describe the main parts of an adaptive optics a system but the first section will clear out the confusion between adaptive and active optics.

\sec{Active Optics}
\label{sec:active}
The difference between adaptive and active optics can be a bit confusing. Active Optics is commonly used to describe ways of controlling wavefront distortions that are caused by optical, thermal or mechanical effects in the telescope itself. Active Optics is slow compared to Adaptive Optics. The effects that the active optics system is built to compensate for vary on a long time scale compared to seeing which changes rapidly.

Active Optics works at bandwidths less than 1 Hz while Adaptive Optics uses bandwidths from 10 to 1000 Hz. Active Optics can use large primary mirrors for wavefront correction whereas Adaptive Optics need to use small mirrors optically conjugated to a lenslet array. \citep{1993ARA&A..31...13B}

\sec{Wavefront sensor}
\label{sec:wfs}
The wavefront sensor (WFS) is one of the basic elements in an adaptive optics system. The wavefront phase aberration to be corrected must first be measured. This is done by a wavefront sensor. The wavefront has to be sensed with enough spatial resolution and enough speed for real time compensation of the atmospheric seeing.

No wavefront-phase sensors for visible light exist today \citep{2004aoa..book.....R}. The deformation is measured indirectly. This can be done in various ways.

\begin{figure}[ht!]
  \centering
    \includegraphics[width=0.5\textwidth]{\imgpath 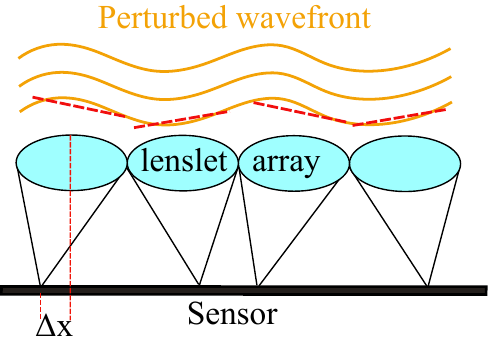}
  \caption{Principle of a Shack-Hartmann wavefront sensor. The perturbed wavefronts arrives at top. The picture shows a 1-dimensional cut through the subaperture array with 4 individual subapertures (SA). Each of these SA samples a different part of the incoming wavefront, sensing only the local slope. The slopes are drawn with dashed lines. The image displacement is shown in the leftmost subaperture. The image would have been centered if the wavefront had been plane, it is now offset by $\Delta x$.
}
  \label{fig:SH}
\end{figure}

\sub{Shack-Hartmann wavefront sensor}
\label{sub:sh}
The Shack-Hartmann\footnote{The design of the sensor was developed in 1900 by Johannes Franz Hartmann and was modified in late 1960s by Roland Shack and Ben C. Platt who replaced the opaque screen with a lenslet array. The wavefront sensor is therefore also known as Hartmann-Shack wavefront sensor.} (SH) wavefront sensors are the most commonly used in astronomy \citep{1993ARA&A..31...13B}. A SH WFS consists of an array of small lenses (subapertures) in the pupil plane. The subapertures (SA) image only a part of the wavefront so that they only sense the local wavefront tilt. Each SA forms an image of the source at its focus, if the wavefront is plane. If the wavefront is disturbed, each SA receives a tilted wavefront and forms an off-axis image in its focal plane. The measurement of this displacement is a direct measure of the local wavefront tilt. By using an array of subapertures the wavefront slope can be obtained at a fixed number of points and the shape of the wavefront can be reconstructed. \refpicl{fig:SH} shows the principle of a Shack-Hartmann wavefront sensor.

There are several methods for measuring the positions of the images formed by the lenslet array. 
Calculating the center of gravity for each subaperture, by tracking the peak of the image with a artificial (or natural) point source, is one method that works for point-like objects. \citep{2004aoa..book.....R}.

\newpage
For extended objects, like the Sun, one has to measure relative image motion by comparing the same features in different subpupil images. Subaperture images are cross-correlated with a chosen reference image for various displacements. The displacement that yields the highest cross-correlation is used as the subaperture offset. 

Löfdahl \citep{2010A&A...524A..90L} has evaluated the inherent accuracy of the most commonly used methods.

\sub{Wide Field Wavefront Sensor}
\label{sub:wfwfs}
The Wide Field Wavefront Sensor (WFWFS) is a SH with a large field of view. Because of this large FOV the seeing distribution from different layers (see \refpic{fig:seeing_layers}) in the atmosphere can be tomographically mapped using the angular dependence between the different wavefront slopes. A wide FOV lowers the height for the first seeing layer but a narrow FOV gives better height resolution.

The height variation is explained further in \refsec{sub:height}.

\sec{Wavefront correctors}
\label{sec:correctors}
When the wavefront distortion is known it is up to the optical components of the AO system to correct it. The optical components are usually deformable  mirrors. They are designed so that their shape can be adjusted in order to match the incoming wavefront distortion. The function of the optical components of an AO system is often divided among two components, a tip-tilt mirror 
and an adaptive mirror that corrects the higher order wavefront distortions \citep{1993ARA&A..31...13B}.

The tip-tilt mirror is usually the first corrector. It is a flat mirror that can rotate on two perpendicular axes. The tip-tilt mirror only corrects the tip and tilt modes while the deformable mirror compensates the high-order aberrations. Most of the atmospheric wavefront aberrations are located in the tip and tilt modes (see \reftab{tab:zernike}) and the tip-tilt mirror will therefore need a larger stroke than the deformable mirror.

The deformable mirror is a high-order corrector and typically has less stroke than the tip-tilt mirror. The efficiency of the AO system will be reduced if this mirror has to correct for low order modes as well as high order modes since there will be less stroke left for the high-order modes.

\sec{Multi-conjugate adaptive optics}
\label{sec:mcao}
As described above, a conventional adaptive optics system consists of a deformable mirror and a tip-tilt mirror integrated in the telescope's optical path, a wavefront sensor and a control computer. 

This can be extended to a Multi-conjugate Adaptive Optics system by inserting several deformable mirrors as well as wavefront sensors. The deformable mirrors in the MCAO system are optically conjugated at different altitudes.

The Sun is well suited for MCAO since it is an extended object and it displays small scale structure everywhere on its surface. A ''natural guide star'' can be found wherever the telescope points. \citep{2006SPIE.6272E...4B}



\chap{Methods for quantifying seeing}
\label{chap:dimm}
Several methods for quantifying seeing exist. Differential image motion monitors and shadow band ranging will be described in this chapter.

The chapter ends with a description of the optical setup at the SST.

\sec{Differential image motion monitor}
\label{sec:dimm}
The differential image motion monitor is the most widely used method for quantifying seeing. Wavefront slope differences are measured over two small pupils some distance apart. The differential method makes it insensitive to tracking errors. It has been shown that the differential motion exceeds the absolute motion when the distance between the two apertures equals a few times their diameter \citep{1990A&A...227..294S}.

\begin{figure}[ht!]
  \centering
    \includegraphics[bb=0 0 272 210, width=0.55\textwidth]{\imgpath 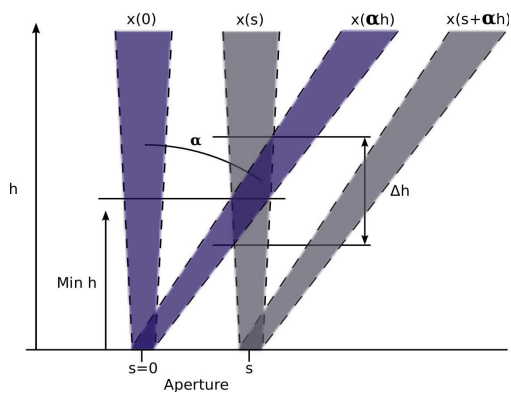}
  \caption{The relation between the wavefront sensor geometry and the contributions to the two differential image displacement measurements from a height $h$. $s$ is the separation between the two subapertures and $\alpha$ is the measured relative field angle between the two subfields. The beams diverge with height due to the non-zero FOV. Picture adapted from \citep{2010A&A...513A..25S}.}
  \label{fig:wfwfs_geom}
\end{figure}

Sarazin and Roddier \citep{1990A&A...227..294S} presented the following approximate equations for the variance of differential image displacements measured with two subapertures with a diameter $D$ and separated, along the x-axis, by a distance $s$ (nomenclature is chosen to match \refsec{sub:s-dimm+} and \refpic{fig:wfwfs_geom})
\begin{equation}
\sigma^2_l=\average{(x(s)-x(0))^2}=0.358\lambda^2r_0^{-5/3}D^{-1/3}\left(1-0.541\left(\frac{s}{D}\right)^{-1/3}\right),
\label{eq:var_long}
\end{equation}
for longitudinal displacement (a long a subaperture row) and where $r_0$ is Fried's parameter. The transverse displacements (perpendicular to the subaperture row, a long a column) are given by
\begin{equation}
\sigma^2_t=\average{(y(s)-y(0))^2}=0.358\lambda^2r_0^{-5/3}D^{-1/3}\left(1-0.811\left(\frac{s}{D}\right)^{-1/3}\right).
\label{eq:var_trans}
\end{equation}
Using the notation of Fried \citep{1975RaSc...10...71F} \refeqnl{eq:var_long} and \refeqn{eq:var_trans} can be written
\begin{eqnarray}
\average{(x(s)-x(0))^2}&=&0.358\lambda^2r_0^{-5/3}D^{-1/3}I(s/D,0), \label{eq:var_long2}\\
\average{(y(s)-y(0))^2}&=&0.358\lambda^2r_0^{-5/3}D^{-1/3}I(s/D,\frac{\pi}{2}), \label{eq:var_trans2}
\end{eqnarray}
where $I(s/D,0)$ and $I(s/D,\frac{\pi}{2})$ are tabulated in Fried's paper \citep{1975RaSc...10...71F}. The function $I$ is normalized so that it approaches unity when $s$ approaches infinity. It is also symmetric so that $I(s/D,0)=I(-s/D,0)$ and $I(s/D,\frac{\pi}{2})=I(-s/D,\frac{\pi}{2})$

\sec{S-DIMM and S-DIMM+}
\label{sub:s-dimm+}
The Solar DIMM (S-DIMM) was first used in the Yunnan Observatory and uses differential motions of the solar limb measured with two small subapertures \citep{2001SoPh..198..197L}. The S-DIMM+, which can be seen as a natural extension of the DIMM and S-DIMM, includes measurements that are sensitive to height variation of seeing. The calculations made by Fried must be extended to account for the averaging effect of using a large field of view for wavefront sensing and to get more accurate estimates of $I$ when $s/D$ approaches zero \citep{2010A&A...513A..25S}.

\sub{Extension of Fried's calculations}
Scharmer and van Werkhoven \citep{2010A&A...513A..25S} extended the calculations of Fried \citep{1975RaSc...10...71F} to include $s/D<1$ and $s/D$ as large as $50$, the latter corresponding to a distance of 5 m for subapertures with a diameter of approximately 10 cm. Their calculations agree perfectly with those of Fried for $s/D=1$ but they deviate with increasing $s/D$ and differ by 10 \% for $s/D=10$. This suggest uncertainties in Fried's numerical integration and was suspected already by Sarazin and Roddier \citep{1990A&A...227..294S}. For $s/D>7.5 $, the calculations of Fried show a small decrease of longitudinal differential image displacements. Scharmer and van Werkhovens's calculations behaved similar when the step size of one of the integrating variables were to large and they therefore decreased that step size with approximately a factor of 80 compared to the calculations that essentially reproduced the results of Fried. After that improvement the calculations showed monotonous increase of $I(s/D,0)$ and $I(s/D,\frac{\pi}{2})$ with $s$ as long as $s/D<300$. That is a much larger range than needed for their analysis and it could also be concluded that their calculations must be more accurate than Fried's.


Scharmer and van Werkhoven conclude that the theory developed by Fried can be used for the S-DIMM+ by replacing the separation $s$ with $s+\alpha h$ and the pupil diameter $D$ with an effective diameter $\effdim$. An approximately round binary mask is applied when measuring image displacements from the granulation with cross-correlation techniques in order to make the averaging area roundish.

\refpicl{fig:wfwfs_geom} describes the relation between the wavefront sensor geometry and the two differential wavefront slope measurements at a certain height $h$. It is assumed that the first subaperture is located in the origin with a separation of $s$ to the second subaperture. The field angle of the first measurement is furthermore assumed to be zero while that of the second measurement is $\alpha$ and that the tilt only is along the x-component of the image displacement. The measured quantity $\delta x_1$ for the first measurement, corresponding to contributions added from the $N$ layers located at heights $h_n$, is
\begin{equation}
\delta x_1(s,0)=\sum_{n=1}^N(x_n(s)-x_n(0)),
\label{eq:m1}
\end{equation}
while the quantity for the second measurement will be
\begin{equation}
\delta x_2(s,\alpha)=\sum_{n=1}^N(x_n(s+\alpha h_n)-x_n(\alpha h_n)).
\label{eq:m2}
\end{equation}
The covariance between $\delta x_1$ and $\delta x_2$ is given by
\begin{equation}
\average{\delta x_1\delta x_2}=\sum_{n=1}^N\average{(x_n(s)-x_n(0))(x_n(s+\alpha h_n)-x_n(\alpha h_n))},
\label{eq:cov}
\end{equation}
because of the assumed independence of contributions from different layers. The average brackets, $\average{\ldots}$, in \refeqn{eq:cov} denote averages over many exposed frames. By evaluating the four terms in \refeqn{eq:cov} one by one and by taking advantage of the assumed homogeneity of the statistical averages at each height $h_n$, this equation can be rewritten as a combination of three variances of differential image displacements,
\begin{eqnarray}
\average{\delta x_1\delta x_2}=\sum_{n=1}^N&&\!\!\!\!\!\!\!\!\!\!\!\!\average{(x_n(\alpha h_n-s)-x_n(0))^2}/2 + \nonumber \\ 
&&\!\!\!\!\!\!\!\!\!\!\!\!\average{(x_n(\alpha h_n+s)-x_n(0))^2}/2-\average{(x_n(\alpha h_n)-x_n(0))^2}.
\label{eq:cov2}
\end{eqnarray}
When $\alpha=0$, the last term disappears and the first two terms are equal which means that the equation describes a conventional DIMM, except that the contributions from the high layers are reduced by the averaging of image displacements from the relatively large FOV used with the S-DIMM+ method.

By combining \refeqns{eq:var_long2}{eq:var_trans2} and (\ref{eq:cov2}), the following equations can be obtained:
\begin{eqnarray}
\average{\delta x_1\delta x_2}=\sum_{n=1}^N c_nF_x(s,\alpha,h_n), \label{eq:cov_long}\\
\average{\delta y_1\delta y_2}=\sum_{n=1}^N c_nF_y(s,\alpha,h_n), \label{eq:cov_trans} 
\end{eqnarray}
where 
\begin{eqnarray}
F_x(s,\alpha,h_n)&=&I((\alpha h_n-s)/\effdim, 0)/2+  \nonumber \\ 
&&I((\alpha h_n+s)/\effdim, 0)/2-I((\alpha h_n)/\effdim, 0), \label{eq:Fx} \\
F_y(s,\alpha,h_n)&=&I\left((\alpha h_n-s)/\effdim, \frac{\pi}{2}\right)/2+  \nonumber \\ 
&&I\left((\alpha h_n+s)/\effdim, \frac{\pi}{2}\right)/2-I\left((\alpha h_n)/\effdim, \frac{\pi}{2} +\right),
\label{eq:Fy}
\end{eqnarray}
and the coefficients $c_n$ come from the original DIMM equations \refeqns{eq:var_long2}{eq:var_trans2}, rewritten to match the S-DIMM+ method,
\begin{equation}
c_n=0.358\left(\frac{\lambda}{r_0(h_n)}\right)^{5/3}\left(\frac{\lambda}{\effdim(h_n)}\right)^{1/3}.
\label{eq:var_tot_coff}
\end{equation}

\newpage

\sub{Height variation}
\label{sub:height}
\refpicl{fig:wfwfs_geom} describes the relation between the WFWFS geometry and the contribution to the image displacement measurements from a specific height. Large separation between the subapertures gives better height resolution but smaller separation gives a higher maximum height. A large FOV lowers the minimum height but a small FOV, on the other hand, gives better height resolution.

The size of the subapertures is also important. Smaller subapertures give better height resolution, but it is limited by the FOV for large heights and bigger subapertures lowers the noise.

\refeqns{eq:cov_long}{eq:cov_trans} can also be expressed in terms of the turbulent strength of each layer, $C_N^2\, \mathrm{d}h$. The atmospheric structure constant $C_N^2$ can be found by rearranging \refeqn{eq:r0} into
\begin{equation}
C_N^2\, \mathrm{d}h=0.0599\cos(\gamma)\left(\frac{\lambda}{r_0}\right)^{5/3}\lambda^{1/3}, 
\label{eq:cn2}
\end{equation}
with which \refeqn{eq:var_tot_coff} can be rewritten as
\begin{equation}
c_n=\frac{5.98}{\cos(\gamma)}\left(\frac{1}{\effdim(h_n)}\right)^{1/3}C_N^2 \, \mathrm{d}h.
\label{eq:var_tot_coff2}
\end{equation}
Scharmer and van Werkhoven found that the S-DIMM+ (for the SST) should be able to distinguish seeing at the pupil where $h=0$ and at a height of $h=500$ m above the pupil and that the angular resolution is enough to allow measurements up to a height of approximately 20--30 km. 

\sec{Shadow band ranging}
\label{sec:shabar}
As mentioned in \refsec{sec:seeing} stars seem to twinkle when observed with your naked eye. The proper name for this twinkling is astronomical scintillation. The light from a star is focused as a point at the back of your eye. As shown in \refpic{fig:wfwfs_geom} the beams gradually widens closer to the star (due to a non zero FOV). The size of this beam is still very small at the height of the turbulent layers and the star seems to twinkle since the light is disturbed.

Seykora \citep{1993SoPh..145..389S} showed that the scintillation of solar light was closely correlated to seeing and measuring this scintillation using a photodiode is hence another method for quantifying seeing \citep{1993SoPh..145..399B, 1993SoPh..145..389S}. This is done by a scintillometer which detects and evaluates the intensity fluctuations of the transmitted signal. 

By assembling several detectors in a scintillometer array, information about the structure function $C_N^2$ can be obtained \citep{1997SoPh..176...23B}. This was further developed into an instrument called SHadow BAnd Ranger (SHABAR) which consists of a non-redundant linear array of scintillometers \citep{2001ExA....12....1B}.

A SHABAR measures the scintillation index, $\sigma_I^2$, which is defined as the variance of the relative irradiance fluctuations. The covariance between the SHABARs scintillometer units are then inverted to achieve the $C_N^2$ profile. 

Higher layer contribute less to he scintillation index and the SHABAR has therefore a limited range. A longer baseline for the scintillometer array is needed to achieve sensitivity on higher heights.

\begin{figure}[ht!]
\centering
\subfloat[]{\includegraphics[trim = 5mm 70mm 0mm 40mm, clip, width=0.4\textwidth]{\imgpath 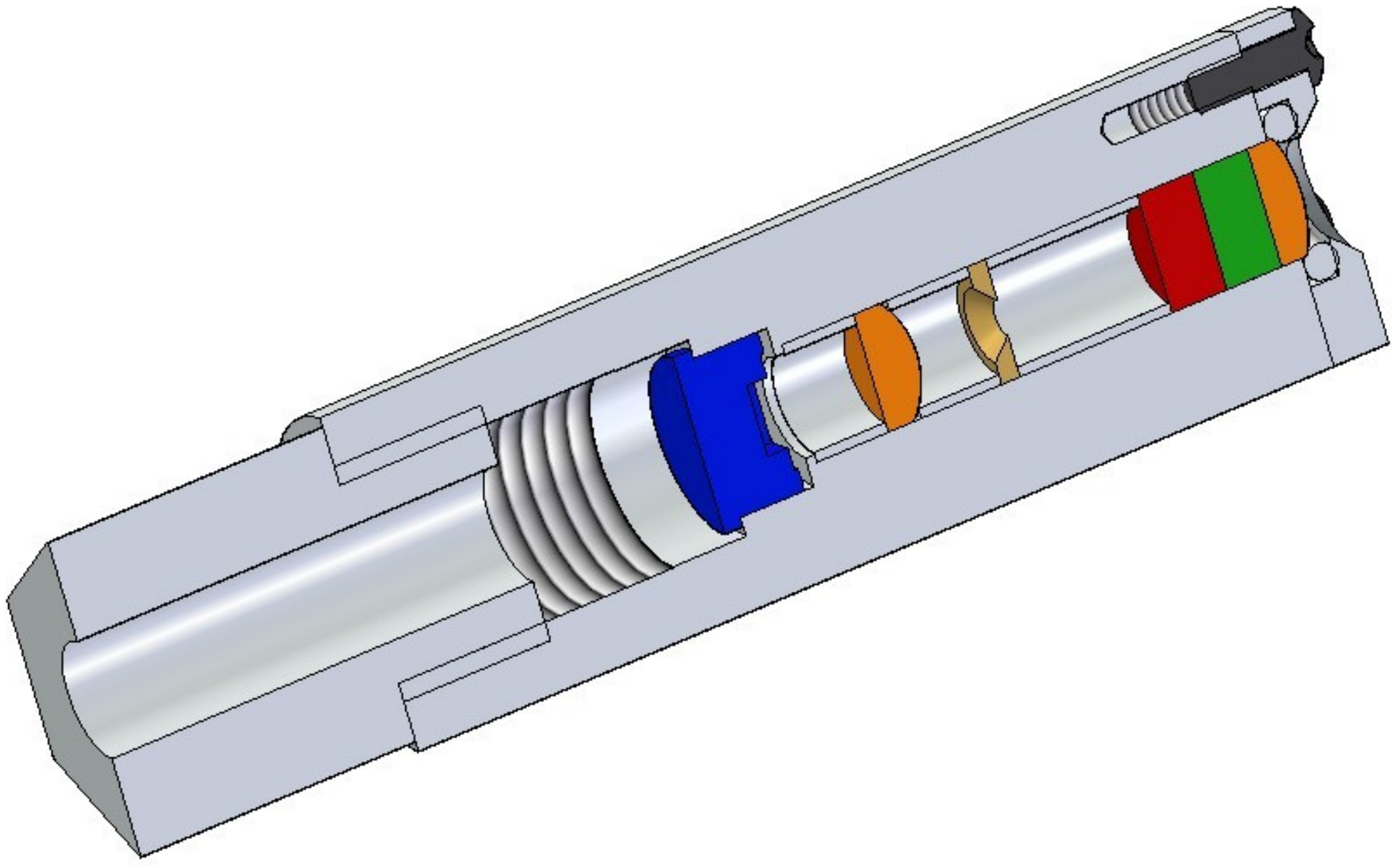}
\label{fig:shabar_scint}}
\centering
\subfloat[]{\includegraphics[trim = 5mm 50mm 0mm 40mm, clip,width=0.4\textwidth]{\imgpath 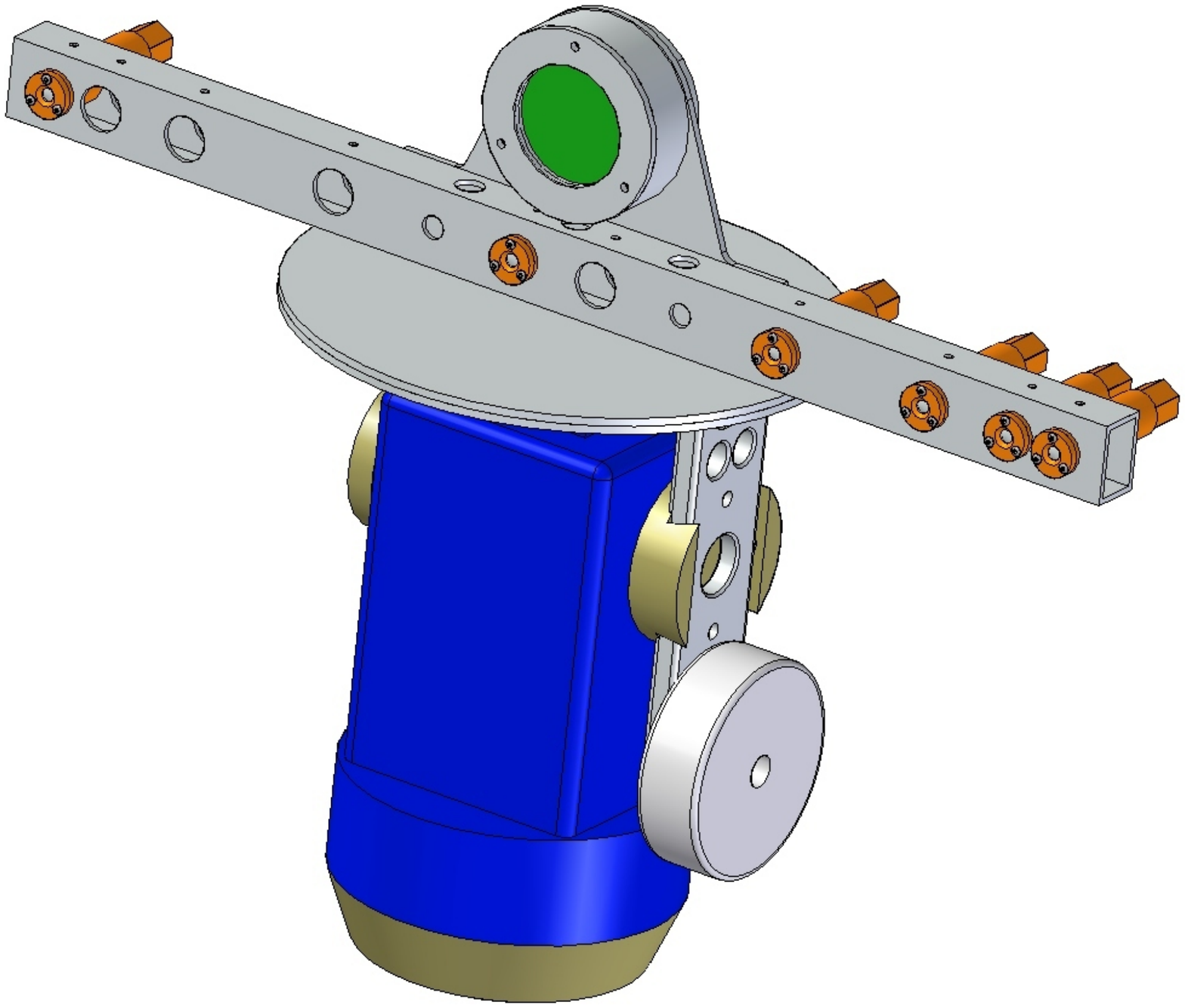}
\label{fig:shabar_mount}}
\caption{(a) Shabar scintillometer unit. (b) The 6 scintillometers mounted on the bar. Figures adapted from \citep{2010SPIE.7733E.144S}.}\label{fig:shabar}
\end{figure}

\begin{figure}[ht!]
  \centering
    \includegraphics[width=\textwidth]{\imgpath 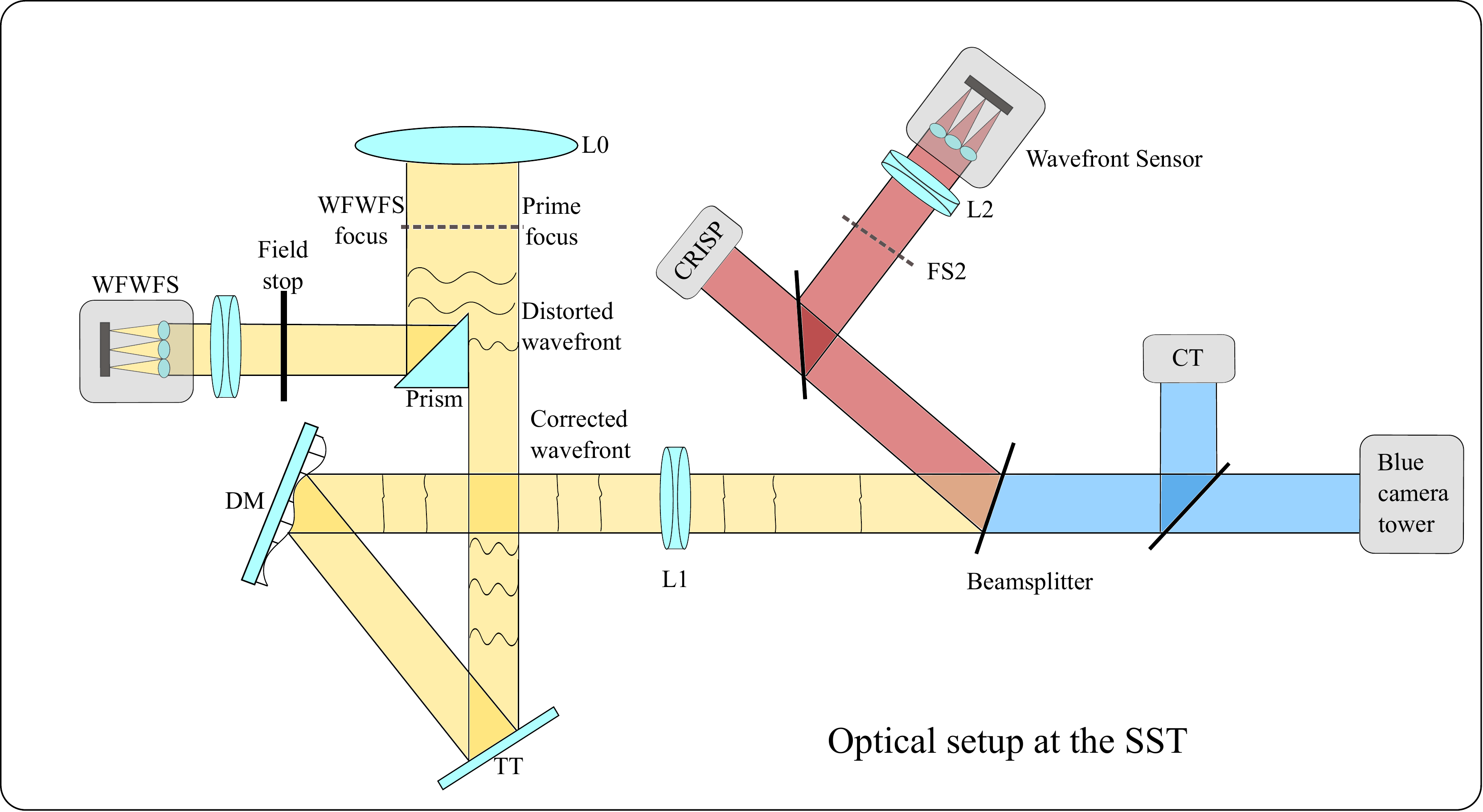}
  \caption{The present optical setup for the SST. The incoming light is folded from almost vertical to horizontal when passing the tip/tilt (TT) mirror and the deformable mirror (DM). The light is split at $\sim 500$ nm into two different science beams. A correlation tracker (CT) is located in the blue beam while CRISP and the WFS are located in the red. The WFWFS is mounted directly under the vacuum system so that the WFWFS beam doesn't pass the AO-system. Drawing adapted from \citep{SST}.}
  \label{fig:AO_SST}
\end{figure}

The short baseline SHABAR that is mounted on the SST has its 6 scintillometer cells inserted in a 50 cm long bar shown in \refpic{fig:shabar} \citep{WP08100}.
\newpage
\sec{Optical setup at the SST}
\label{sec:os_sst}
\refpicl{fig:AO_SST} describes the present optical setup used at the SST.

The telescope pupil is re-imaged onto a deformable mirror (DM) via a field lens (L0) and a tip-tilt mirror (TT). 
The tip-tilt mirror deflects the beam by 60$^{\circ}$ and the deformable mirror deflects it by 30$^{\circ}$, that is the beam is folded from nearly vertical to horizontal. The incident angle of 15$^{\circ}$ on the adaptive mirror makes the pupil image nearly circular (cos(15$^{\circ}$)=0.97). A re-imaging lens (L1) magnifies the image $\sim 2.2$ times, making an image scale of $\sim 0.041$ arc sec per pixel on the CCD in the blue beam and $\sim 0.059$ arc sec per pixel in the red beam. \citep{2003SPIE.4853..370S, 2003SPIE.4853..341S}

The electrode pattern 
of the 37-electrode bimorph deformable mirror is placed upon each of two thin plates of piezo electric material which are glued to the mirror. When voltage is applied to an electrode, one of the plates will expand while another will  contract, bending the bimorph. \citep{SST}

The horizontal part of the beam is split by a dichroic beam splitter into two different science beams. A correlation tracker is located in the blue beam and the SH WFS is located in the red beam (see \refpic{fig:AO_SST}). The correlation tracker controls the tip-tilt mirror and the WFS controls the differential mirror.

Another beam splitter (partially polarizing) splits the light so that 8 \% goes into the WFS. The wavelength range is restricted by a 100 nm wide interference filter that is centered at 550 nm. This filter will also improve the image quality by preventing chromatic errors. The field of view is defined by an adjustable field stop at the secondary focus (FS2) so that the subaperture images don't overlap. 
The collimator lens (L2) that is located after the secondary focus, but before the subaperture array, produces a parallel beam and re-images the DM onto the subaperture array. The subaperture array re-image the secondary focus onto the detector. \citep{SST}

\begin{figure}[ht!]
\centering
\subfloat[]{\includegraphics[width=0.5\textwidth]{\imgpath 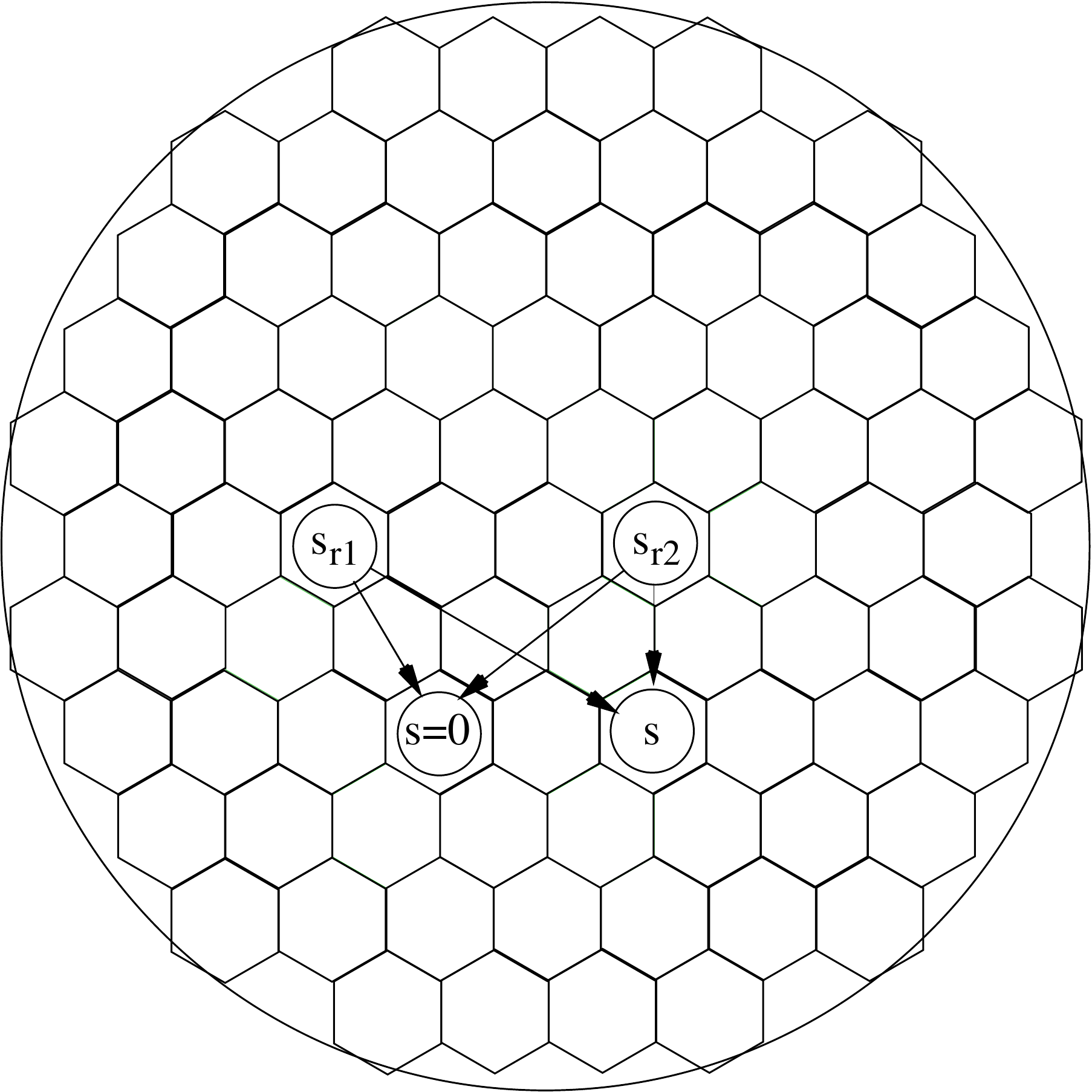}
\label{fig:wfwfs_layout}} \\
\subfloat[]{\includegraphics[width=0.7\textwidth]{\imgpath 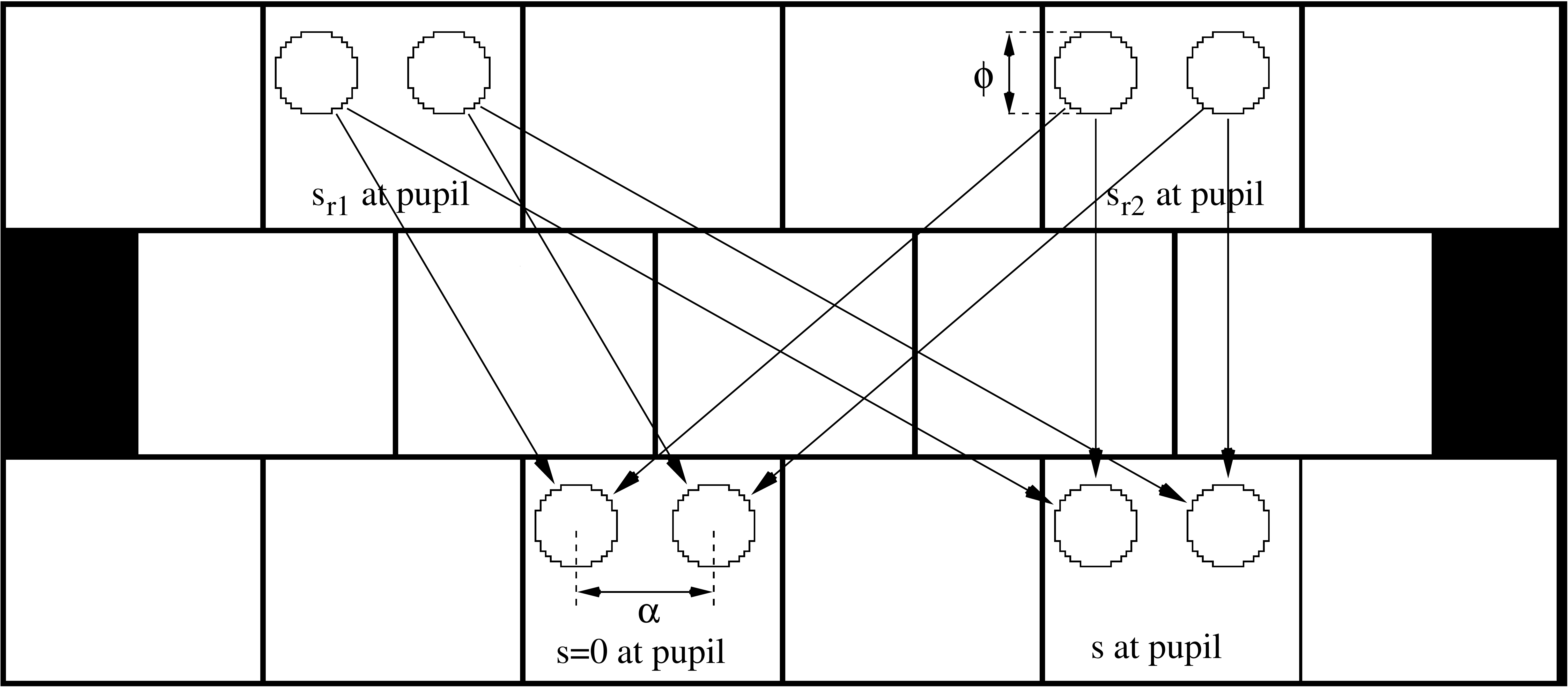}
\label{fig:si_layout}}
\caption{(a) Layout of the 85 fully illuminated hexagonal subapertures. The circle represents the 98 cm SST aperture. The two subapertures $s_{r1}$ and $s_{r2}$ corresponds to selected reference subimages with high RMS contrast. Arrows point to the subapertures $s=0$ and $s$ for which differential image shifts are measured. (b) Corresponding subimages with masks (not to scale), indicate the two subfields at field angles separated by $\alpha$. Figures adapted from \citep{2010A&A...513A..25S}.}\label{fig:wfwfs}
\end{figure}

\sub{WFWFS description}
\label{sub:wfwfs_desc}
The WFWFS is mounted immediately below the vacuum system. The field of view of the WFWFS is placed outside the science field of view and light is deflected horizontally so that the WFWFS beam does not pass the SST AO-system. 
The optics of the wavefront sensor consists of a field stop, a collimator lens and an array with 85 hexagonal subapertures within the 98 cm diameter pupil. This layout is shown in \refpic{fig:wfwfs_layout}. The diameter of the subapertures is equivalent to 9.8 cm. The WFWFS has a maximum field angle of 46.4 arc sec. The image scale is 0.344 arc sec per pixel.

\refpicl{fig:wfwfs_layout} shows the 85 hexagonal subapertures (SA) and \refpic{fig:si_layout} show the corresponding subimages. Subimages are located on the CCD while subapertures are located in the aperture plane (on the lenslet array). Every subimage is divided into several subfields. This is shown in \refpic{fig:wfwfs}.


\chap{Data processing}
\label{chap:process}
After observing with the telescope, the images taken with the WFWFS are inspected to see if they are good so that they can be used for further analysis.  If not, a warning or even an error is triggered. This inspection is done automatically directly after a data set is taken but can of course be started manually afterwards. The different data sets are then reduced using a program called \Astooki \, which, after several steps, returns a S-DIMM+ covariance map.

This chapter describes the inspection and some statistics made out of the inspection as well as the different reduction steps. The chapter includes a description of how data from the SHABAR (also mounted on the SST) were processed before they were compared with the WFWFS data and ends with a description of changes in the reduction.

\sec{Inspection}
\label{sec:autoproxy}
\verb!Autoproxy! is an automated proxy software, written by T.I.M. van Werkhoven, which acquires, checks and stores WFWFS data \citep{Autoproxy}. The checking or inspection starts after the data has been acquired. The quality of the data is checked and some statistics on the frames are calculated, for example min, max and average intensity, RMS, saturation etc. Some global statistics, such as the standard deviation of the frame statistics, are also computed.

The values of theses parameters are used to judge the quality of each frame and assign a status code. Warnings are indicated by positive status codes and errors give negative codes. Status code 0 implies a \textit{good} frame. The threshold values for each status code can be found in \reftab{tab:crit}.

A log file is written in the end of the inspection including the statistics of each frame as well as the assigned status code. The log file also states the first, last and best frame, together with information about the number of frames that where classified as \textit{good}, \textit{warning} or \textit{bad}. This can then be used to classify the whole data set.

Frames with errors, which during the inspection were marked as \textit{bad}, are removed by \verb!Autoproxy! to save storage space. This setting can of course be changed when, for some reason, also bad data need to be stored. 

\setlength{\tabcolsep}{2pt}
\begin{table}[ht!]
\renewcommand{\arraystretch}{1.3}
\begin{center}
\begin{tabular}{rp{4.3cm}p{1cm}p{0.9cm}p{1cm}p{0.9cm}p{1cm}p{0.9cm}} \hline
\multicolumn{1}{c}{\textbf{Status}}&\multirow{2}{*}{\textbf{Criteria}}		& \multicolumn{2}{c}{\textbf{Image}}		& \multicolumn{2}{c}{\textbf{Flats}}		& \multicolumn{2}{c}{\textbf{Darks}}\\
\multicolumn{1}{c}{\textbf{code}}&		 				& \textbf{Warn} 	& \textbf{Bad} 		&\textbf{Warn} 		& \textbf{Bad}		& \textbf{Warn} 		& \textbf{Bad}\\ \hline
$\pm$ 4 			& Too many saturated pixels			& 0.05 \% 		& 0.2 \%		& 0.1 \% 		& 0.2 \% 		& 0.01 \% 			& 0.1 \% \\
\rowcolor[gray]{.8} $\pm$ 8 	& Too low max. intensity 			& $<$2000 		& $<$800 		& $<$2000 		& $<$800 		& - 				& - \\ 
$\pm$ 16			& Intensity problems, average of frame 		& $>$3200 $<$2000	& $>$3600 $<$600	& $>$2800 $<$1000	& $>$3200 $<$400 	& \multirow{2}{*}{$>$150} 	& \multirow{2}{*}{$>$500} \\
\rowcolor[gray]{.8}$\pm$ 32 	& Intensity problems, average of SA		& $>$3200 $<$2000	& $>$3600 $<$600	& $>$3200 $<$1600	& $>$3600 $<$600	& \multirow{2}{*}{-} 		&\multirow{2}{*}{-}\\
$\pm$ 64 			& Frame avg. $>$ SA avg.			& 70 \%			& 64 \%			& 70 \%			& 64 \%			& -				&- \\
\rowcolor[gray]{.8}+128		& Frame after series of bad frames, probably also bad &		&			&\multirow{2}{*}{-} 	&\multirow{2}{*}{-} 	&\multirow{2}{*}{-} 		&\multirow{2}{*}{-}\\
$\pm$ 256 			& Too large SA avg. deviation 			& $>$80 		& $>$200 		&- 			&- 			&- 				&- \\
\rowcolor[gray]{.8}$\pm$ 512 	& Too high RMS 					& 6 \% 			& 10 \% 		& 2 \% 			& 4 \% 			&- 				&- \\ \hline
\end{tabular}
\caption{Criteria for file inspection. The plus and minus sign indicate that a frame has a warning status ($+$) or a bad status ($-$). If a frame (image, flat or dark) has saturated pixels it will be assigned the status $4$ (warning or bad depending on the amount of saturated pixels). Status $8$, $16$ and $32$ are applied to frames if they have too low or too high intensity (maximum or average). Status $64$ is assigned if the frame has an average intensity that is higher than 64 \% (\textit{warning}) or 70 \% (\textit{bad}) of the average intensity of the subaperture. Status code $128$ is only assigned as a warning to frames that follow long series of bad frames. Status $256$ is applied if the difference between the frames subaperture (SA) value and the average value for all SA in that set is too big. Status $512$ is assigned if the relative RMS value of the pixels within the subapertures are too large. A dash (-) indicates that the criterium is not available for that file type.}\label{tab:crit}
\end{center}
\end{table}
During this master thesis project, several scripts were written in order to make it easier to inspect the result of the automated inspection. A script that checks the number of bad frames for every status code is shown in \refapp{code:bad}. There is a corresponding script for warnings. 

\sub{Status codes}
\label{sub:codes}
The 8 available status codes are described in \reftab{tab:crit} with their corresponding thresholds. A frame can be assigned with one of these status codes or with a combination of them. Status codes are "combined" with the binary OR operator in order to get unique status codes. Status code $48$ for example is assigned to a file that has a waring due to to low or to high average intensity in the frame (status $16$) and in the subaperture (status $32$), that is $16 | 32 = 48$. 

The combination of status codes is unique as long as the frame only has warnings. When the frames start to get errors, some status codes can be a ''combination'' of different statuses. Status $-200$ for example can either be $16|32|-256|8$ or $16|32|256|-512|8$. A frame with one error status code will though always have a resulting code that is negative (since e.g. $512 | -8 = -8$). 

\sub{Change in inspection}
\label{sub:change}
A lot of frames from 2009 were browsed through with the \textit{ANA browser} and \textit{SST Java Browser} available on the local computers \citep{SST} to see if the thresholds for the inspection seemed to have the right values. During this browsing, it was found that some of the frames marked as \textit{good} were shifted or had overlapping subapertures.

Inserting more field stops in the optical setup and making sure that the WFWFS does not take data when something is blocking its field of view would decrease the number of such frames. However, the remaining bad frames still need to be found without manually inspect a million files.

A small change in the threshold for one of the already existing status codes ($64$) marks these frames as \textit{warning} or \textit{bad} frames depending on the amount of badness.

Status code $-64$ (indicating a bad frame) originally had following threshold \verb!frame average > subaperture average!. This was changed to the more strict thresholds \verb!frame average > 0.7 * subaperture average! resulting in status code $-64$ (\textit{bad}) and \verb!frame average > 0.64 * subaperture average! resulting in $+64$ (\textit{warning}). 

\sub{Intensity and RMS}
\label{sub:intensity}
Frames are assigned with a warning (or an error) if they have too low intensity. Since quite many frames were assigned with this type of warning, a script was made to check which intensities these frames actually had, that is if they were close to the given thresholds or not. Some frames were then examined by eye. 

Frames which have a too large relative RMS intensity for the pixels within a subaperture are assigned with a warning (or an error). A similar script was written to check if some frames where deleted because of high RMS and if they where close to the threshold.

\sec{Reduction}
\label{sec:astooki}
The reduction of WFWFS data is done in several steps using \textit{Astooki - the Astronomical Toolkit}, which is a Python program made to process astronomical data \citep{Astooki}. The steps are explained below.

During this master thesis project a reduction script was written in order to make all of the following steps in one go by calling the different tools of \Astooki. The reduction script loops over all data sets for all specified days so that data from a whole year can be reduced in one go if wanted to.

At first the script tries to find the flat field and dark frame that are closest in time to a particular data set. The flats and darks, together with the image frames, are needed as input for some of the \Astooki \, tools.

Code snippets from the reduction script can be found in \refapp{app:reduction}.

\sub{Initial setup}
\subsub*{Making a subimage mask}
The first step is to create a subimage mask. A set of coordinates for where the subimage is located are calculated. The units are arbitrary but they should be consistent, pixels (on the CCD) were used for this mask. 

The radius (1024 px, with a 2k by 2k pixel CCD) of the pattern  is specified as well as the pitch (194 px, 166 px) of subimage positions. The centroid coordinate, that is the coordinate in the middle of the subimage, and the origin coordinate (the coordinate in the lower left corner of the subimage) are calculated for every subimage. In order to match the subimage pattern with the corresponding hexagonal subaperture pattern, an offset is specified so that every odd row is shifted a half subimage in x-direction. 

It is checked if the obtained coordinate (with offset) lies within the aperture bounds. The telescope aperture is circular, and since the subimages on the CCD correspond to subapertures in the telescope aperture, the subimages also have to fit within the circular bound.

The coordinates (with applied offset) for the subimages that where located within the circular bound are saved in two arrays of subimages (one with origin positions and one with centroid positions) that describes the subimage pattern. Part of the resulting pattern is shown is \refpic{fig:si_layout}.

The subimage mask is made with equal parameters for all data sets and days. The mask will be optimized for a particular data set in next step.

\subsub{Optimizing the subimage mask}
After the ''standard'' subimage mask is generated, it needs to be optimized since it is possible that even a mask created with the right parameters won't match the subimages perfectly. In this step, the mask is matched to a flat field image (the one closest to that data set).

\begin{figure}[ht!]
  \centering
  \subfloat[]{\includegraphics[width=0.5\textwidth]{\imgpath 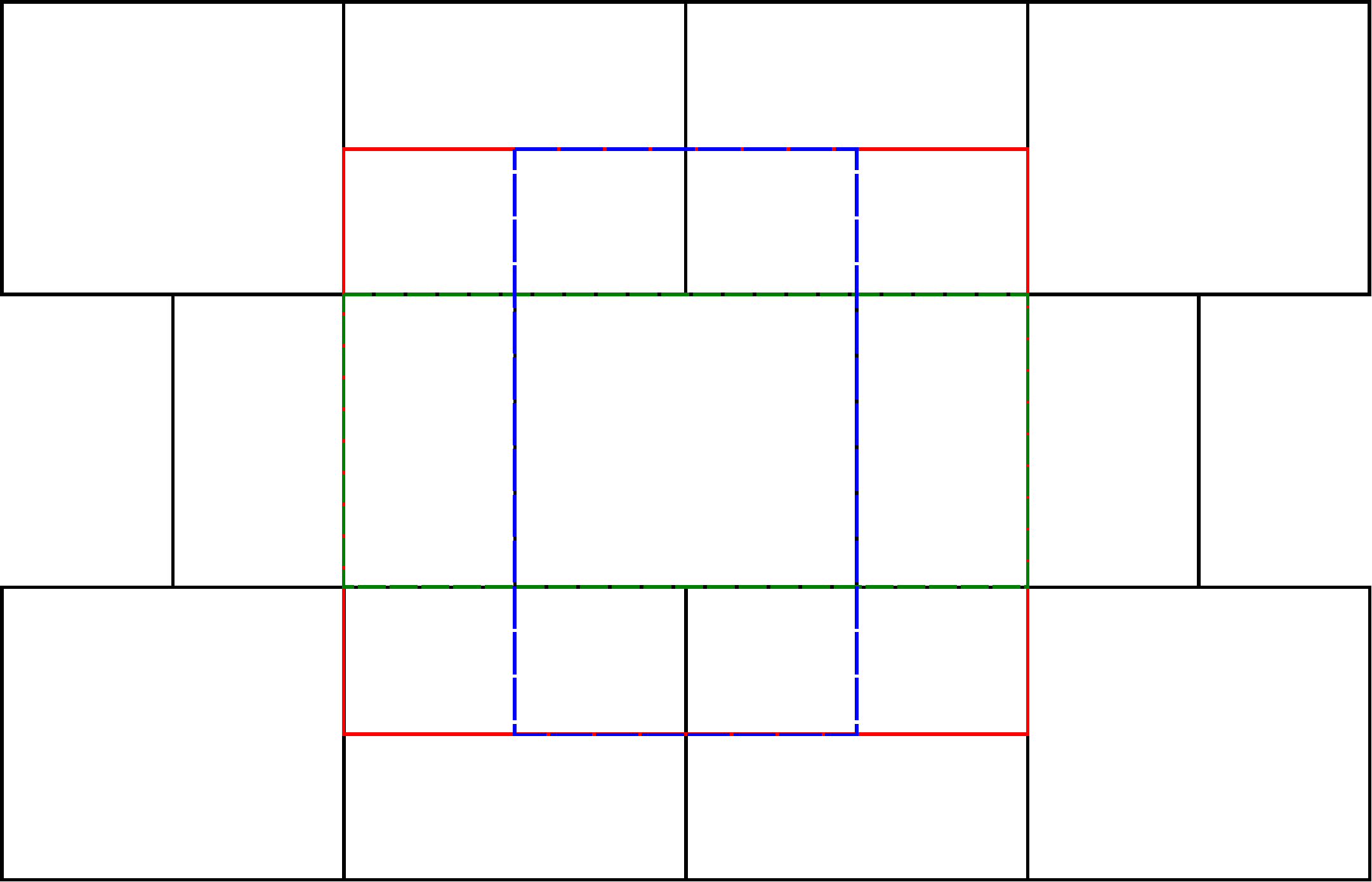}
  \label{fig:simask-fopt}}
  \subfloat[]{\includegraphics[width=0.5\textwidth]{\imgpath 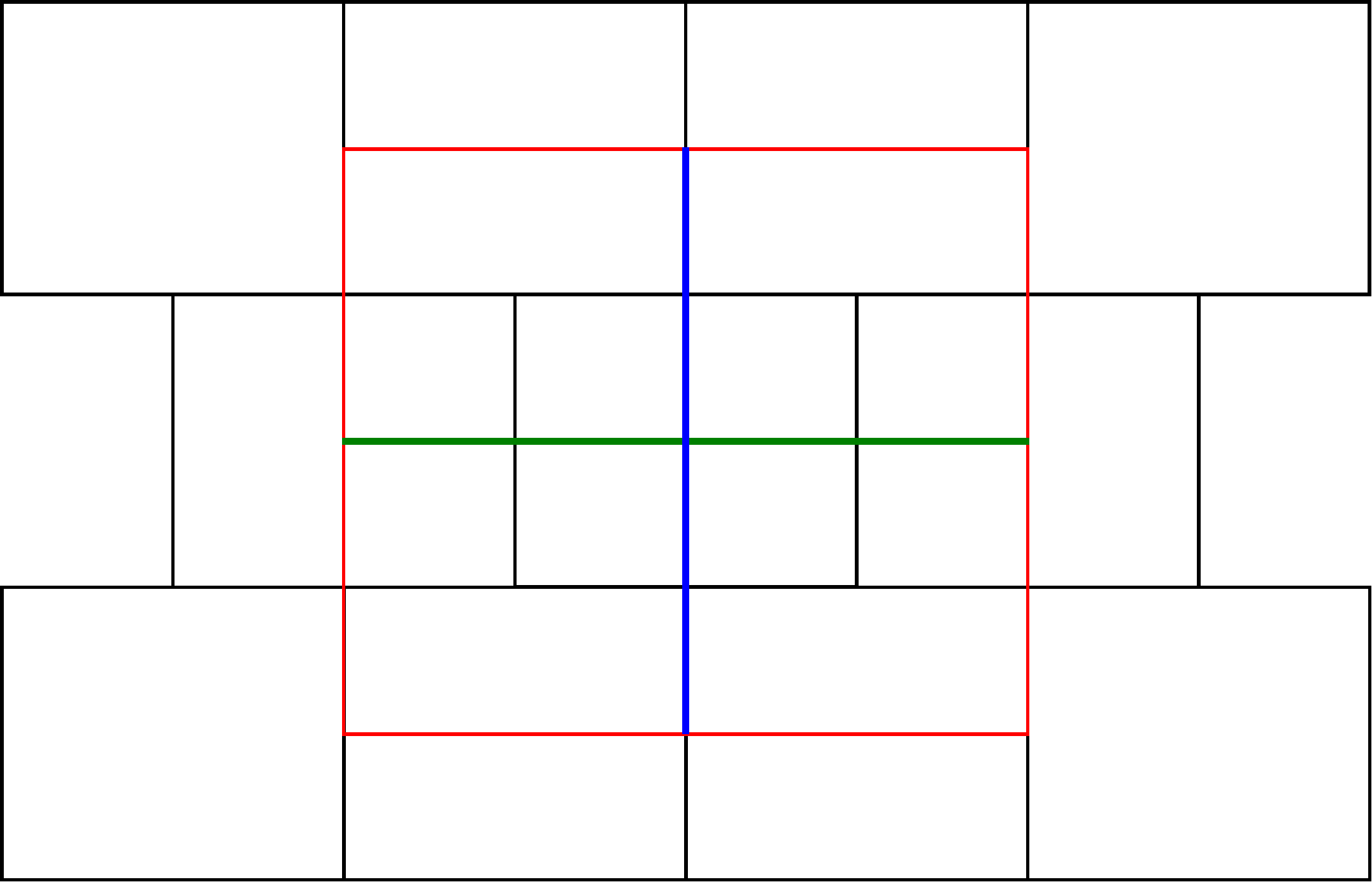}
  \label{fig:simask-fopt-1px}}
  \caption{(a) Slices in the optimized subimage mask. The red square indicates the range in which the two slices are taken out. The green rectangular area shows the horizontal slice and the blue rectangular area shows the vertical slice. (b) Slices are averaged down to one pixel to faster find the borders.}
  \label{fig:simask}
\end{figure}

\refpicl{fig:simask} explains how this fitting is done. The range, in which the vertical and horizontal slices are taken, is $1/2$ subfield larger, in every direction, than the compared subfield. This is to make sure that the range covers the whole flat fielded image. Two slices are taken out within this range, one in the x-direction and one in the y-direction (see \refpic{fig:simask-fopt}). 

In order to quickly find the borders between the subimages, these slices are averaged (the x-slice over the height and the y-slice over the width) to one pixel as shown in \refpic{fig:simask-fopt-1px}. The routine defines the border as the first pixel to left and right from the center of the x-slice with an intensity less than $80 \%$ of the maximum intensity. This procedure gives the width of the subimage. The height of the subimage is then given by the first pixel to the top and bottom from the center of the y-slice that is less than $80 \%$ of the maximum intensity.

Optimizing is done for all subimages. All sizes are stored and the mean subimage size is calculated as well as the standard deviation.

This routine is slightly sensitive to specks of dust on the flat field.

\newpage
\sub{Calibration}
The reference position of each subimage needs to be known in order to compare the corresponding subfields in different subimages. It can not be assumed that pixel (x,y) in subimage A corresponds to the same position in subimage B (due to static aberrations). The problem can also be addressed the other way around: given a granule on the Sun, it has to be known at what pixel that granule is located in each of the subimages. 

The ''static offsets'' are measured by taking a large field of view (almost the complete subimage) in one reference subimage and then cross-correlate that with all subimages. That results in N shift vectors, where N is the number of subapertures. It is possible to use multiple subapertures as references in order to get better results. This should provide the same data and will give an indication of the reability of the shift measurement. \citep{Astooki}

\subsub{Making a big subfield mask}
The subfield mask is a grid of coordinates relative to the subimage and subfields cannot lie outside the subimage. The size of the subimage, which was calculated in the step before, is therefore used as an input parameter.

A border of $30 \times 30$ pixels is also supplied in order to keep a guard range free at the edge of the subimage. This is needed in next step when the cross-correlation is done.

The effective size of the big subfield mask will be the size of the area inside the border ($30 \times 30$ px) that is applied to the optimized subimage size (from the step before).

\newpage
\subsub{Measure static offset shifts}
This part of \Astooki, that calculates the image shifts between different subfields and subimages, is implemented in C.

The big subimage mask and the subfield mask are, together with flat fielded and dark corrected data, used as input. The four subimages with the highest RMS value are selected as reference subimages. Each reference is compared with all subimages (even with itself) and a correlation map is calculated for every subimage. The cross-correlation is done with the absolute difference squared method, \refeqn{eq:adsq}, and the subpixel maximum of the correlation map is calculated with a 9 point interpolation method. 
\begin{equation}
\textrm{Correlation map}=-\left( \sum_{x,y} | g(x,y)-g_{\textrm{ref}}(x+i,y+j)| \right)^2,
\label{eq:adsq}
\end{equation}
where $i$ and $j$ is the shift (in $x$ and $y$ direction) and $g_{\textrm{ref}}$ is one of the reference images. The minus sign is to ensure a maximum for best match.

The image shift is calculated for all subimages in all frames and stored in a quite big 5 dimensional matrix of data (\#frames, \#references, \#subimages, \#subfields, \#directions). 

The differential image shifts that are calculated with cross-correlation techniques suffer from a lack of absolute zero-point reference. It is in practice impossible to define such an absolute reference precisely, which is furthermore not needed since one can rely on the differential image shifts to average to zero over a sufficiently long time interval. \citep{2010A&A...513A..25S}

One data set consists of 1000 frames, which corresponds to approximately 110 seconds of WFWFS data. It is assumed that the seeing induced differential image shift averaged over these 110 seconds is zero. The average shift measured is therefore subtracted from the 1000 frames of the burst.

The variance and the mean of these 1000 frames are calculated. The mean is the shift offset used in the next step (x and y offset for all subimages, averaged over frames).

\subsub{Updating subimage mask with calculated offset}
The previously calculated static offsets are now used to correct the subimage mask (that was calculated earlier). The origin position for every subimage mask is subtracted by the offset (that was obtained in the previous step) for that subimage. 

The maximum offset for all subimages (in one frame) is calculated and added to every subimage mask position. All subimage mask sizes are cropped with twice the maximum offset and the value for the smallest subimage size is stored.

Each subimage mask will now point at the same location on the Sun. 

\sub{Measure actual subfield/subimage shifts}
The subimage mask is now corrected with the static offsets and can be subdivided in different subfields so that the actual shifts for the data set can be calculated. 

\subsub*{Generate a subfield mask with smaller subfields}
This mask is created with the same routine as the previous big subfield but the resulting subfield mask will consist of several small subfields (instead of one big). 

\begin{figure}[ht!]
  \centering
  \includegraphics[width=0.5\textwidth]{\imgpath 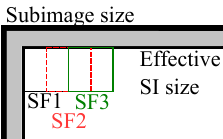}
  \caption{Upper left corner of a subimage mask with three subfields indicated. The border is 7 pixels wide and the subfields are located within the effective subimage size. The subfields are $16 \times 16$ pixels and overlap with 50 \%.}
  \label{fig:sfmask}
\end{figure}

Subfields can not lie outside the subimage and the smallest subimage size (found in the previous step) is used as input so that the subimage mask can be used for all frames. A 7 pixel wide border is applied to the subimage to keep a guard range at the edges. The effective size of the subimage mask (inside the border) is divided into subfields with a size of $16 \times 16$ pixels. The subfields overlap with 50 \%. \refpic{fig:sfmask} shows how the subfields are related to the subimage.

\subsub*{Measure subfield/subimage shifts}
The optimized and updated subimage mask and the $16 \times 16$-subfield mask are used to calculate the image shift for each subfield in each subimage. The calculations are done in the same way as for the static offset shifts described above, but with a circular mask applied to every subfield.

Four subimages, the ones with highest RMS, are chosen as references. Every reference subimage, with all subfields, is compared with all subimages (and their subfields) creating a cross-correlation map with image shifts for all subfield-subimage pairs. 

The cross-correlation is done with the absolute difference squared method, \refeqn{eq:adsq-sf} and with the circular mask applied. The subpixel maximum of the correlation map is calculated with a 9 point interpolation method. 
\begin{equation}
\textrm{Correlation map}=-\left( \sum_{x,y} \textrm{mask}\cdot| g(x,y)-g_{\textrm{ref}}(x+i,y+j)| \right)^2,
\label{eq:adsq-sf}
\end{equation}
where $i$ and $j$ is the shift (in $x$ and $y$ direction) and $g_{\textrm{ref}}$ is one of the reference images. The minus sign is to ensure a maximum for best match.

The image shift is calculated for all subimages in all frames and stored in a quite big 5 dimensional matrix of data (\#frames, \#references, \#subimages, \#subfields, \#directions). 

\sub{S-DIMM+ analysis}
\label{sub:sdimm+_analys}
By using the calculated image shifts, the data can be fitted to a model in order to produce information about the actual seeing. This is done by the statistical S-DIMM+ method described in \refsec{sub:s-dimm+}.

\subsub{Making a lenslet mask}
A subaperture mask of the lenslet coordinates is made with the same routine as the first generated subimage mask. Input parameters this time are the diameter of the microlenses and a radius that is slightly larger than the real aperture radius. Every subaperture has to fit 100 \% within this radius and the radius chosen is therefore a little bit bigger. Some apertures in the real optical setup are a bit cropped.

The centroid coordinate and the origin coordinate are calculated for every subaperture. In order to achieve a hexagonal subaperture pattern, an offset is specified so that every odd row is shifted a half subaperture in the x-direction. It is checked if the obtained coordinates lie within the circular aperture bound. 

The coordinates (with applied offset) for the subapertures that where located within the circular bound are saved in two arrays of subaperture positions (one with origin positions and one with centroid positions) that describe the subaperture pattern. This pattern is shown in \refpic{fig:wfwfs_layout}.

The subaperture mask is made with equal parameters for all data sets and days.

\subsub{S-DIMM+ analysis}
This tool computes a covariance map for all subfield-subimage/subaperture pairs. Part of this \Astooki \, tool is implemented in C.
The previously calculated image shifts are, together with the lenslet and subfield masks, used as input. 

The tool loops over all rows of subapertures (having the same y position). Within each subaperture, it loops over all subfield rows (subfields with the same y coordinate) and chooses a reference subfield. All other subapertures in that row, and their subfields are then compared to to the reference subaperture chosen. The comparison between the the subaperture-subfield pair found are compared as described in \ref{sub:s-dimm+}.

The same procedure is then repeated for all columns. After this a S-DIMM+ covariance map is produced. 

\sub{Inversion in ANA}
\label{sub:ana}
The actual inversion is done in ANA and written by Göran Scharmer \citep{2010A&A...513A..25S}. This step is not a part of \Astooki. The unknown coefficients $c_n$ in \refeqns{eq:cov_long}{eq:cov_trans} are obtained by solving a linear least-square fit problem by minimizing the badness parameter given by
\begin{eqnarray}
L&=&\sum_{s,\alpha}\left([C_x(s,\alpha)-\sum_{n=1}^N c_nF_x(s,\alpha,h_n)]^2 \right.\nonumber \\ 
&&\left. + [C_y(s,\alpha)-\sum_{n=1}^N c_nF_y(s,\alpha,h_n)]^2 \right)W^2(s,\alpha),
\label{eq:badness}
\end{eqnarray}
where $C_x(s,\alpha)$ is the covariance matrix in absence of noise and the weight $W(s,\alpha)$ equals the number of independent measurements for each $(s,\alpha)$ \citep{2010A&A...513A..25S}.

As can be seen in \refpic{fig:wfwfs_layout}, a large number of independent samples (subapertures) exist for a small separation $s$ but only a small number exist for a large separation. This is of course similar for the small and large field angles $\alpha$. To balance this varying number of measurements for different separation and angle, a weight ($W(s,\alpha)$) is inserted in \refeqn{eq:badness}. This weighting is important especially for high-altitude seeing when the signatures are at small $\alpha$ and $W$ is large \citep{2010A&A...513A..25S}.

Minimization of \refeqn{eq:badness} leads to a linear matrix equation $\mat{A}\mat{c}=\mat{b}$ for the coefficients $c_n$. But this allows solutions with negative values of $c_n$. A variable substitution 
\begin{equation}
c_n=\mathrm{e}^{y_n}
\label{eq:subst}
\end{equation}
restricts the solutions to only positive values of $c_n$. The problem is now a non-linear least-square problem, which is solved for the parameter $y_n$. \citep{2010A&A...513A..25S}

The height grids used for the inversion are found by calculating the inverse of matrix $\mat{A}$, corresponding to a linear solution for $c_n$. The grid is then chosen so that the noise sensitivity is minimized. 

The highest height possible with a FOV of $5.5 \times 5.5$ arc sec is 30 km and the node below that must be placed at 16 km in order to avoid high noise amplification. Better height resolution at large distances can only be improved with a smaller FOV. 

The heights used in this comparison are from 7 different layers (6 above the pupil plane). The correlation coefficient (\refeqn{eq:corr_r}) was calculated to see how much these layers correlate with each other. The height grid used was $h=[0,500,1300,3200,5800,10800,20000]$ m.

\newpage
\sec{SHABAR data}
\label{sec:shab_data}
Data from the short-baseline SHABAR that is installed on the SST has been processed in order to make some comparisons with the WFWFS.

The $C_n^2$ profiles measured by the SHABAR were integrated in order to obtain $r_0$ for a number of layers above $h=0$. The boundaries between layers first used were $h_{shabar}=[1,30,60,90,120,200,500,1000,3000,8000,20000]$, giving $r_0$ for 10 different SHABAR layers after integration.

The correlation between these different layers was calculated using the linear correlation coefficient ($r$) also known as \textit{Pearson's r} \citep{2002nrc..book.....P}. The correlation coefficient is given by
\begin{equation}
r=\frac{\sum\limits_i (x_i-\bar{x})(y_i-\bar{y})}{\sqrt{\sum\limits_i (x_i-\bar{x})^2}\sqrt{\sum\limits_i (y_i-\bar{y})^2}}.
\label{eq:corr_r}
\end{equation}
%
The value of $r$ lies between -1 (complete negative correlation) and 1 (complete positive correlation), and 0 indicates that there is no correlation at all between the variables (in this case no correlation between layers).

The correlation matrix produced with these correlation coefficients shows the correlation between the different layers. Little correlation between the layers will result in a diagonal matrix. The location and number of boundary layers used for the integration of $C_N^2$ were changed in order to get as little correlation as possible between the layers. 

In order to compare $r_0$ calculated from SHABAR measurements with $r_0$ calculated from WFWFS following quantity was minimized
\begin{equation}
r_0 (WFWFS) - \left(\sum_n c_n \cdot r_{0,n} (SHABAR)^{-5/3} \right)^{-3/5},
\label{eq:r0fit}
\end{equation}
where $r_0(WFWFS)$ is the Fried parameter calculated from WFWFS measurement at different heights and $r_{0,n} (SHABAR)$ the Fried parameter for different SHABAR layers. The coefficients $c_n$ are the unknowns and represent the best fit contributions from $C_N^2$ at different SHABAR layers used to fit WFWFS data at the pre-selected heights. 

\sec{Change of subfield size}
\label{sec:sfsize}
The size of the circular mask used together with the subfield mask when calculating the image shifts was changed in order to improve the sensitivity at large heights. Better height resolution can be achieved with a smaller subfield size, but a larger subfield size on the other hand, gives lower noise and fewer failures.

\begin{figure}[ht!]
  \centering
  \includegraphics[width=0.7\textwidth]{\imgpath 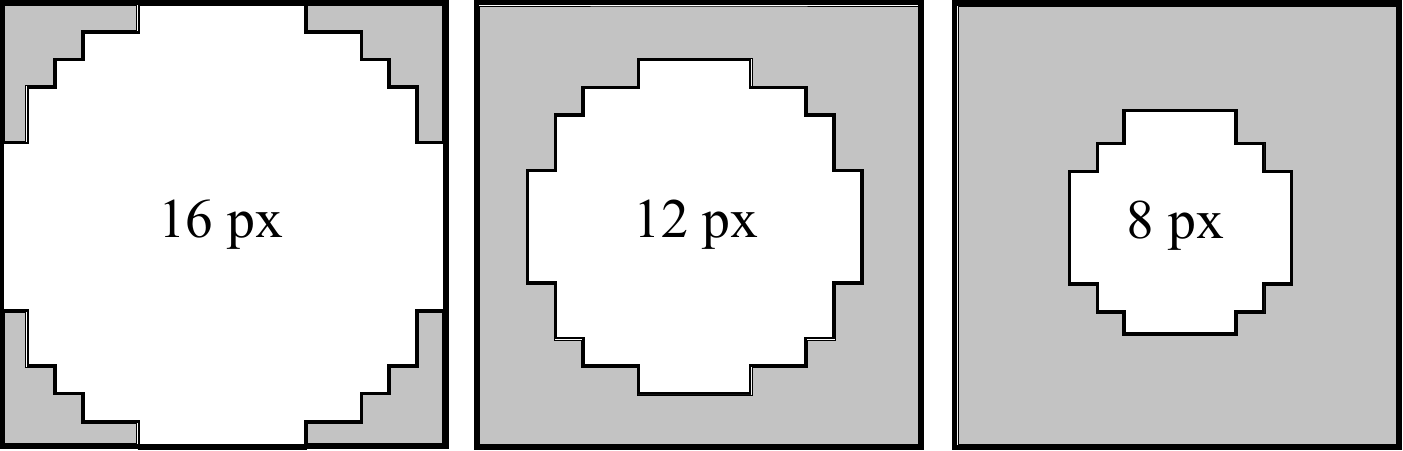}
  \caption{The different ''circular'' masks that where applied to the subfields when calculating the image shifts. The subfield size was not changed.}
  \label{fig:masks}
\end{figure}

A circular mask with a diameter of 16 px, the same size as the subfield, was used in the first reduction. A second reduction was made with a 12 px mask and a third with a 8 px mask. The subfield size ($16 \times 16$ pixels) was kept constant for easier comparison between the changes. \refpicl{fig:masks} shows how the different  masks are related to the subfield size.

\sec{Removal of outliers}
\label{sec:outliers}
Shift measurements that fail are considered as outliers. Sometimes when the shift calculations are made, the wrong set of granules is chosen for measuring the image offset. This means that the position in the reference subimage is compared to another position in the other subimage. These positions do not correspond to the same granules and the shift measured will be completely wrong. Including these measurements when calculating the covariances strongly reduces the resulting $r_0$.

A filter for outliers where therefore implemented in the code, before the covariances are calculated. The standard deviation for all image shifts (for a particular data set) is calculated. Subfields with a larger shift than 3 standard deviations are considered as outliers and ignored in the covariance calculations. 


\chap{Results}
\label{chap:results}
Data from all observing days in 2009 and for observing days to the 13th of September in 2010 have been run through the different inspection scripts created. 
Statistics presented in this chapter are made from these days.

WFWFS data from ten days in 2010 have been fully processed. Five of these days (9th, 21st, 22nd and 28th of June as well as 16th of July) have been compared with data from the SHABAR. Comparisons, results and discussion from the first day, the 9th of June, are presented in this chapter.

\begin{figure}[ht!]
\hspace*{-1.5cm}
\centering
\subfloat[]{\includegraphics[width=0.6\textwidth]{\plotpath 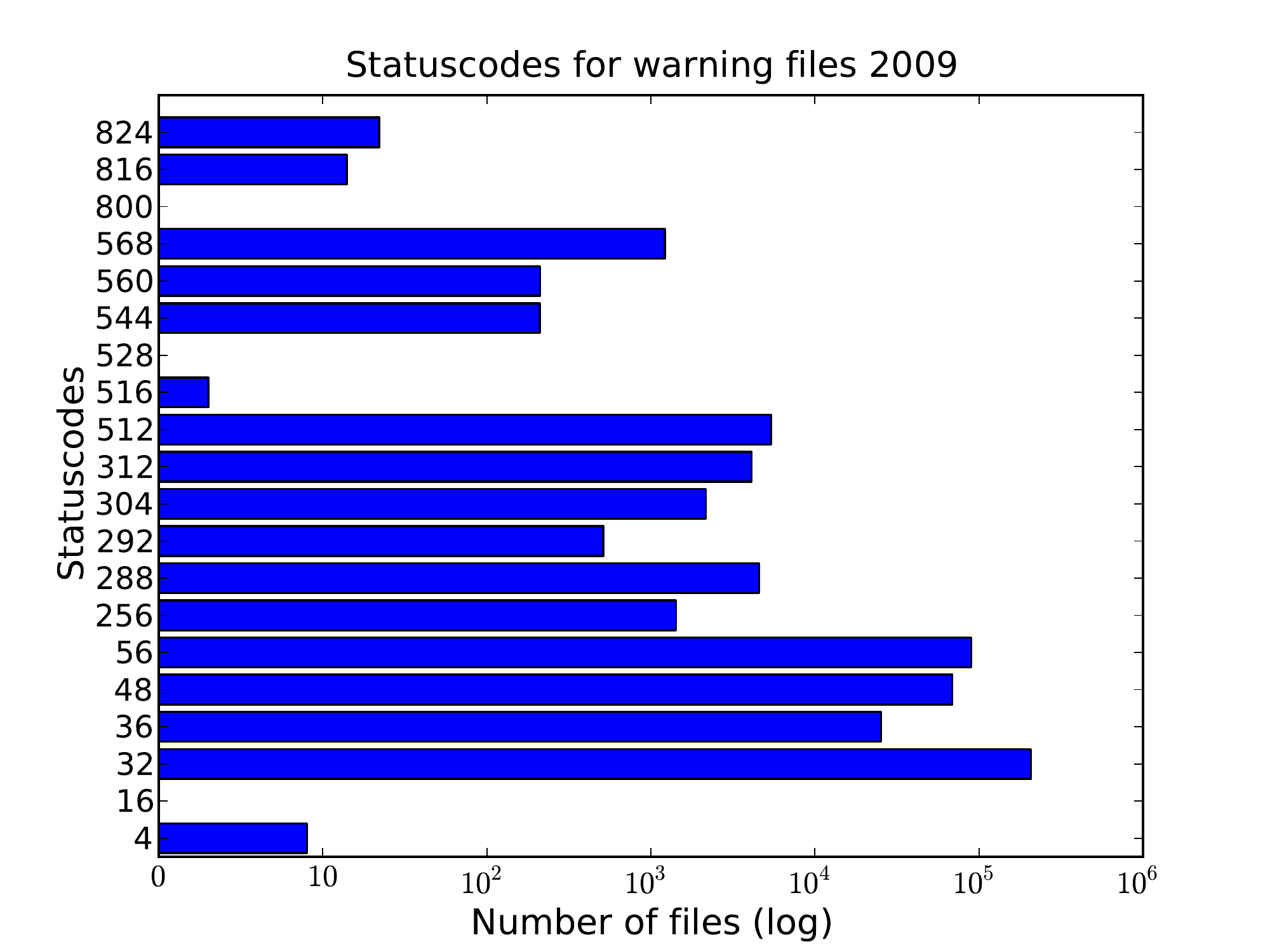}
\label{fig:warn_stat}}
\centering
\subfloat[]{\includegraphics[width=0.6\textwidth]{\plotpath 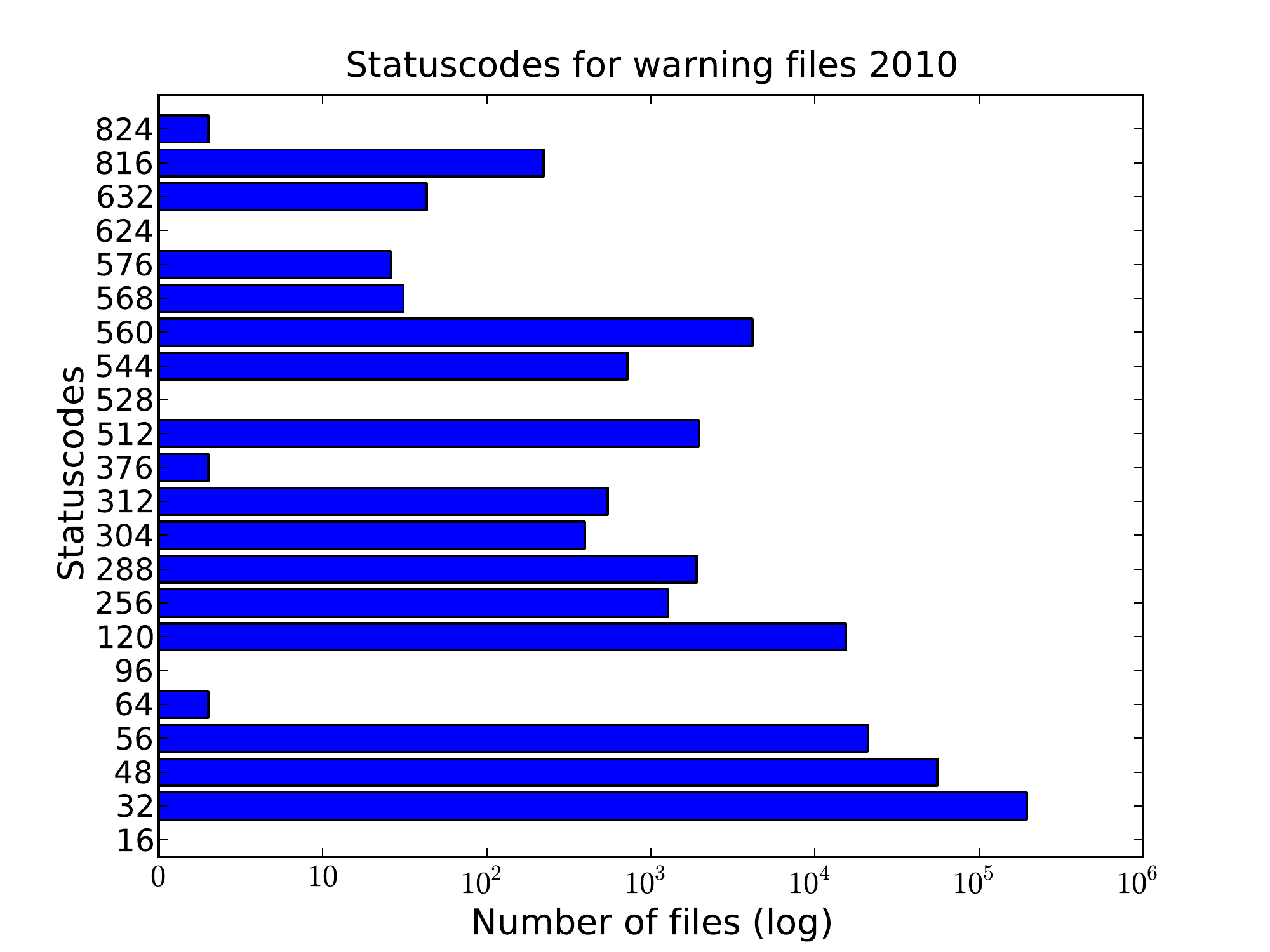}
\label{fig:warn_stat_log}}
\caption{(a) Number of frames for each warning status code for data taken in 2009. (b) Number of frames for each warning status code for data taken in 2010. Status codes that occur in 2010 but didn't in 2009 are combinations with status $64$ that didn't exist before due to the threshold change. Some statuses combined of statuses assigned to frames with low intensity didn't exist for frames in 2010 when the inspections where done. These might occur later in the observation season.}\label{fig:warn_status}
\end{figure}

\begin{figure}[ht!]
\hspace*{-1.5cm}
\centering
\subfloat[]{\includegraphics[width=0.6\textwidth]{\plotpath 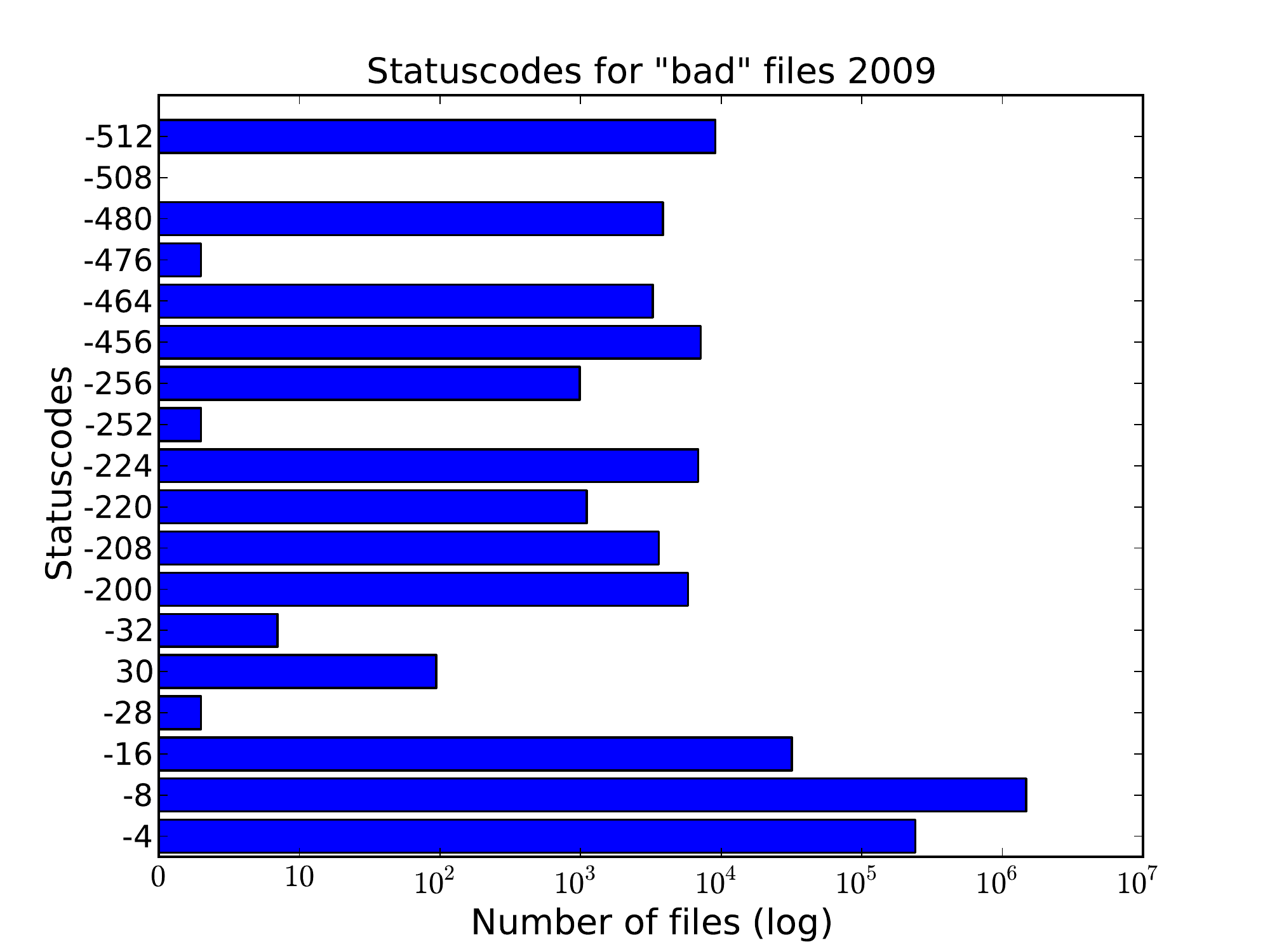}
\label{fig:bad_stat}}
\centering
\subfloat[]{\includegraphics[width=0.6\textwidth]{\plotpath 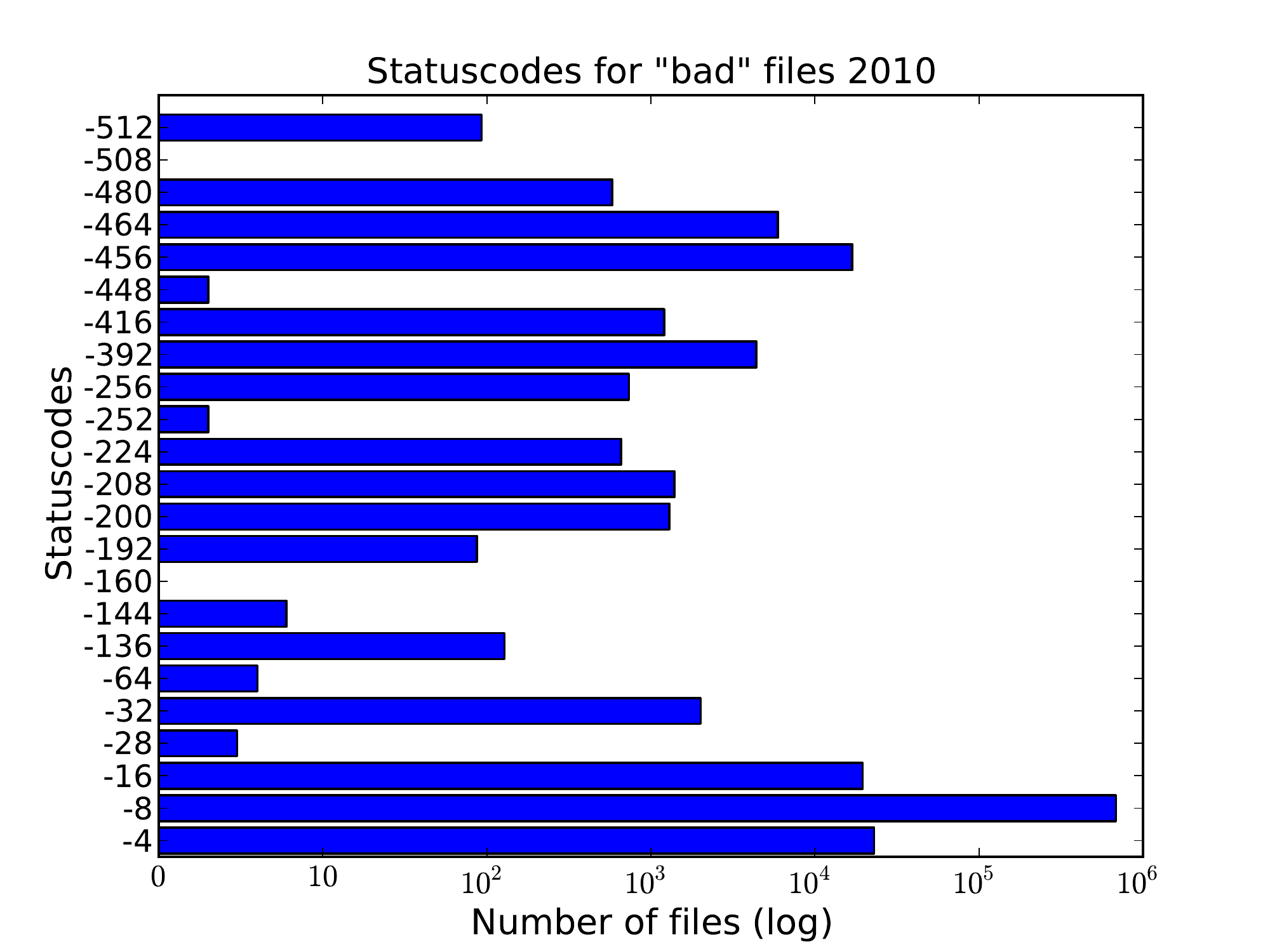}
\label{fig:bad_stat_log}}
\caption{(a) Number of frames for each error status code for data taken in 2009 (b) Number of frames for each error status code for data taken in 2010. Status codes that occur in 2010 but didn't in 2009 are combinations with status $64$ that didn't exist before due to the threshold change.}\label{fig:bad_status}
\end{figure}

\sec{Inspection statistics}
\label{sec:statistics}
\refpicl{fig:warn_status} shows how many image frames that were assigned with each status code. \refpicl{fig:bad_status} shows the same but for error status codes. The left histogram of each figure shows statistics from 2009 and the right histogram for 2010. The corresponding status codes can be found in \reftab{tab:crit} and a description of how to get the ''combined'' status codes are given in \refsec{sub:codes}.
\begin{table}[ht!]
  \begin{center}
    \begin{tabular}{lrr}\hline
      \textbf{Status}&\multicolumn{1}{c}{\textbf{2009}}&\multicolumn{1}{c}{\textbf{2010}}\\\hline
      Good&	853 171	(28 \%)&		292 581 (35 \%)\\
      Warning&	412 133	 (13 \%)&		161 437 (19 \%)\\
      Bad&	1 792 330 (59 \%)&		377 365 (45 \%)\\ \hline
      &	\multicolumn{1}{l}{\textbf{3 057 634}}&	\multicolumn{1}{l}{\textbf{831 383}}\\
    \end{tabular}
  \end{center}
  \caption{Number of files for each status \textit{good}, \textit{warning} and \textit{bad}.
}\label{tab:status}
\end{table}

A number of new (\textit{combined}) status codes were introduced in 2010 because of the threshold change in status code $64$. Some statuses were assigned in 2009 but haven't been assigned to any of the frames taken until the 13th of September 2010. These are mostly statuses which are combinations of statuses assigned when the intensity is low. These statuses become more common in the end of the observation season.

\newpage
As written in \refsec{sub:intensity} the intensity of \textit{bad} frames from 2009 was checked to see how close to the threshold they were and when they were taken. Almost all frames with lower subaperture intensity than 1600 were taken in October or later. Intensity values between that and the threshold (2000) were spread over the summer.

\newpage
Frames from 2009 that were deleted because of high RMS comes from two days, the 9th of July and the 24th of September. All other frames with high RMS are bad because of other things. Frames from the 9th of July have RMS higher than 20 but frames from the 24 of July have RMS values from 10 to 12, which is quite close to the threshold. These can not be examined by eye since all bad frames are deleted.

\begin{figure}[ht!]
  \centering
    \subfloat[]{\includegraphics[width=0.4\textwidth]{\imgpath 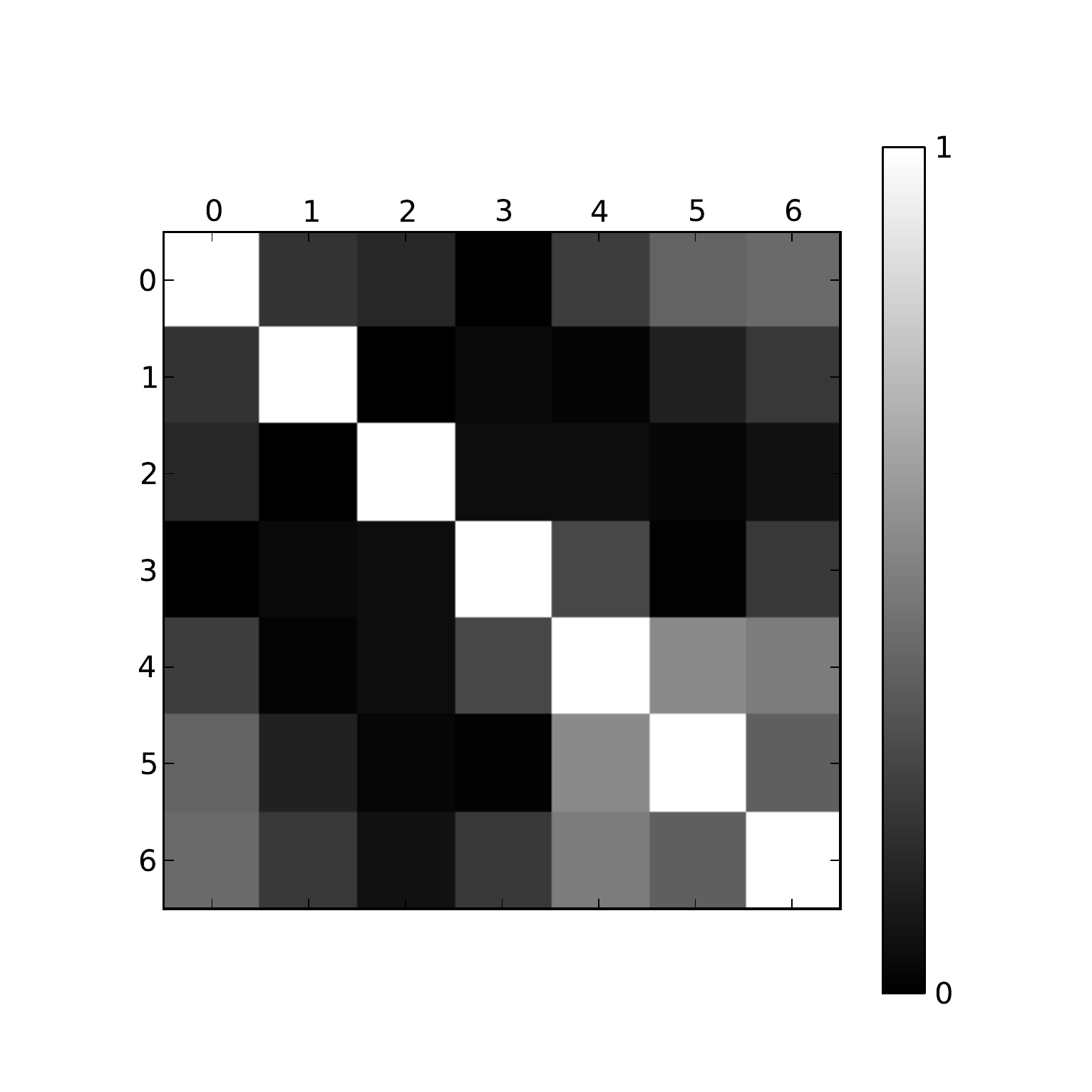}
    \label{fig:corr_wfwfs_lim1}}
  \centering
    \subfloat[]{\includegraphics[width=0.4\textwidth]{\imgpath 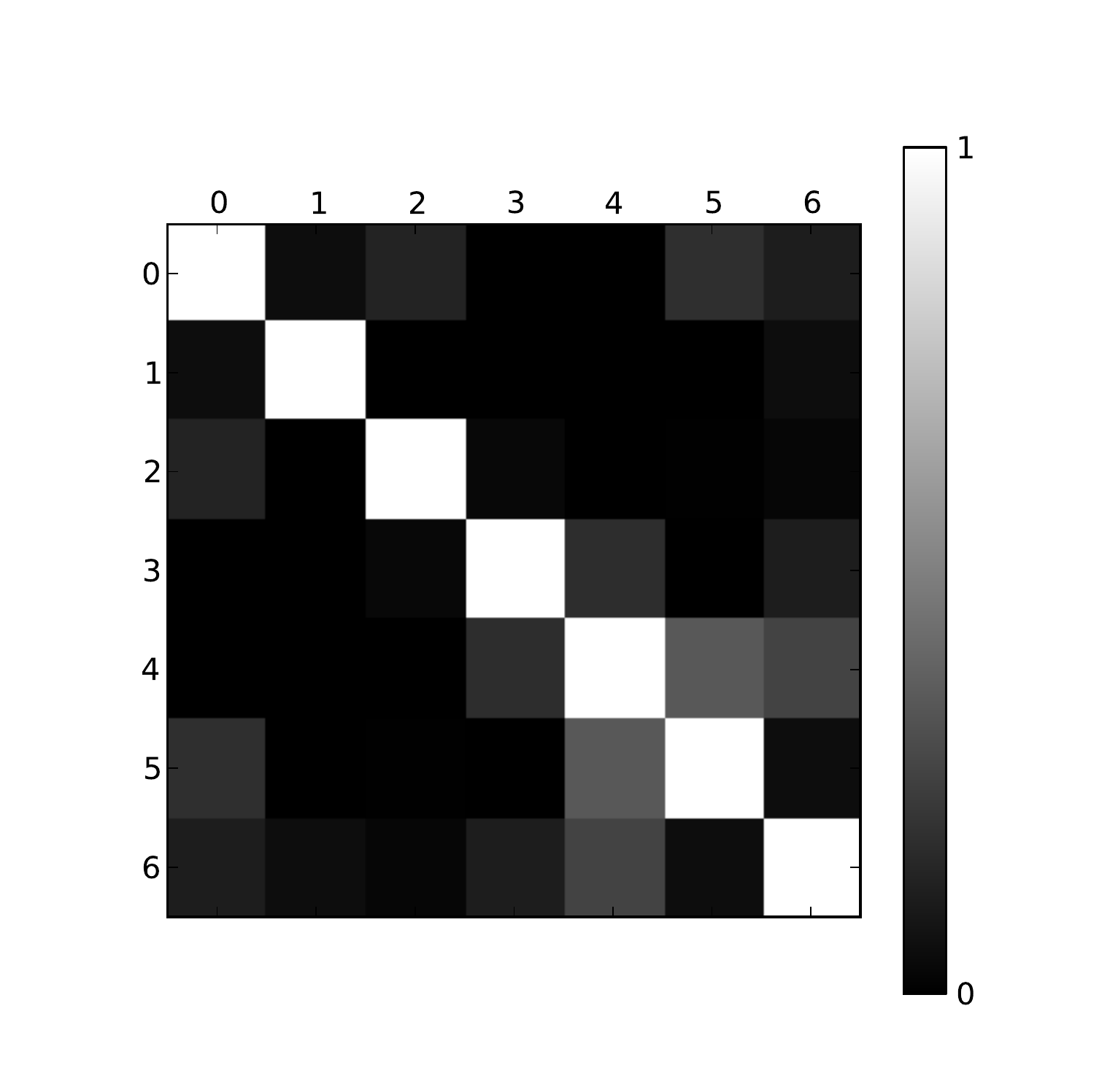}
    \label{fig:corr_wfwfs_lim75}}
  \caption{(a) Correlation matrix of $r_0$ for WFWFS layers. Black pixels indicate no correlation and white pixels show perfect correlation (when a layer is correlated with itself). The gray pixels show that there exist some correlation for layers at higher heights. (b) Correlation matrix of $r_0$ for WFWFS layers where the integrated $r_0<7.5$ cm have been excluded. The correlation between higher layers are much less when these values are excluded.}\label{fig:corr_wfwfs}
\end{figure}

\sec{$r_0$}
\label{sec:r0}
The correlation coefficients obtained when $r_0$ from different heights (measured the 9th of June) were correlated with each other, are presented in correlation matrices, \refpic{fig:corr_wfwfs}. The correlation coefficient is described in \refeqn{eq:corr_r} and \refsec{sec:shab_data}. In \refpicl{fig:corr_wfwfs_lim1} the correlation matrix, for all values of $r_0$, is shown. 

It can be seen that some correlation exists between layers at higher heights as well as between the lowest layer and these at higher heights. That the smallest values for $r_0$ at the highest heights were associated with poor ground layer seeing was found by Scharmer and van Werkhoven \citep{2010A&A...513A..25S}. 

The correlation matrix changes when $r_0<7.5$ cm are excluded. The new correlation matrix is shown in \refpic{fig:corr_wfwfs_lim75}. The correlation between heights (especially higher) are now less than in the previous case. The difference was first noticed when the limit was set to 7.5 cm. This limit was therefore used in the rest of this thesis.

\begin{figure}[ht!]
  \centering
    \includegraphics[width=0.65\textwidth]{\plotpath 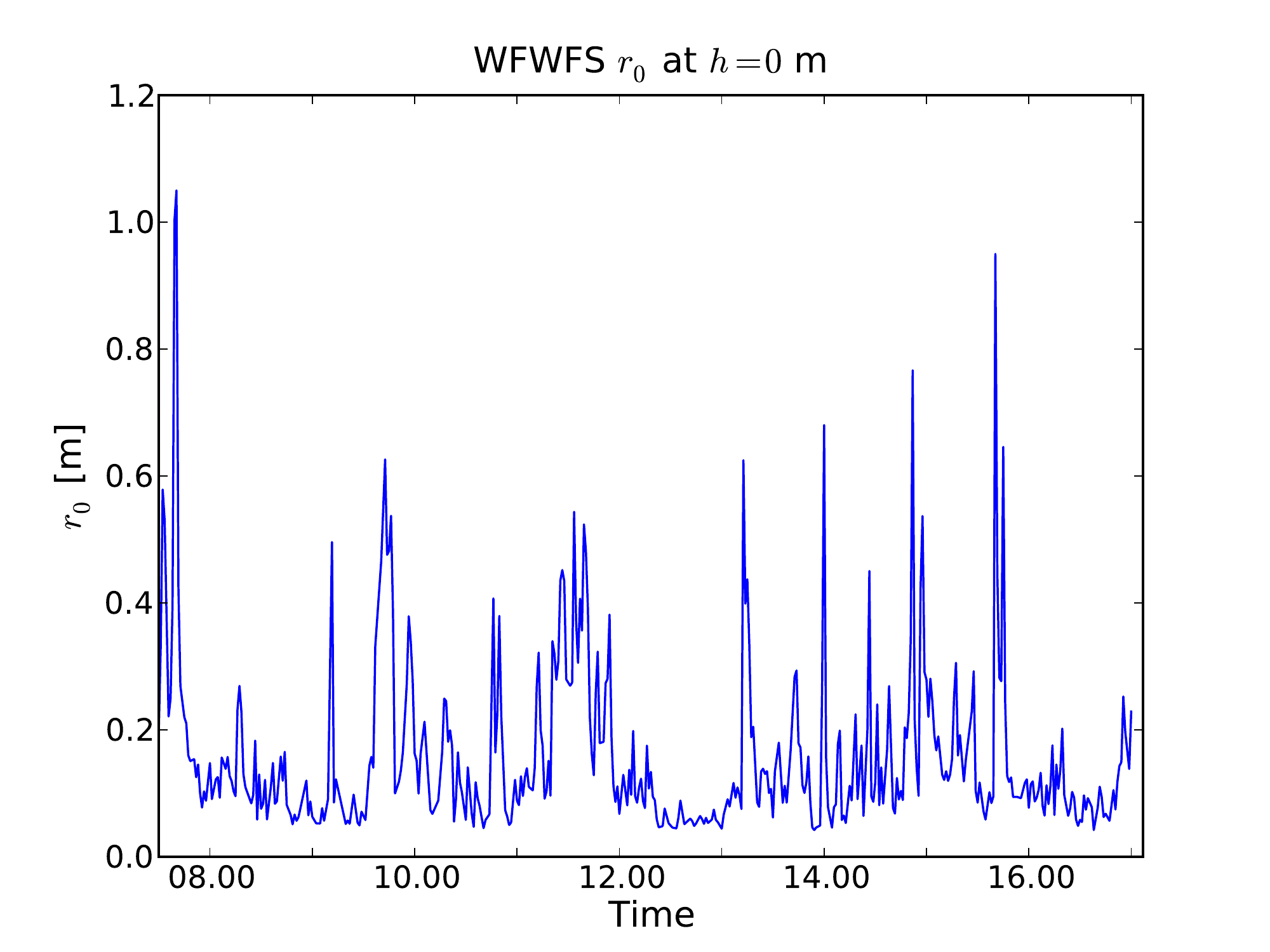}
  \caption{$r_0$ calculated from WFWFS measurements at $h=0$ m.}
  \label{fig:r0_wfwfs_0}
\end{figure}

The calculated $r_0$ values, from the 9th of June, for the 7 different WFWFS heights are shown in \refpics{fig:r0_wfwfs_0}{fig:r0_wfwfs2}. 

\begin{figure}[ht!]
\hspace*{-0.7cm}
  \centering
    \subfloat[500 m]{\includegraphics[width=0.55\textwidth]{\plotpath 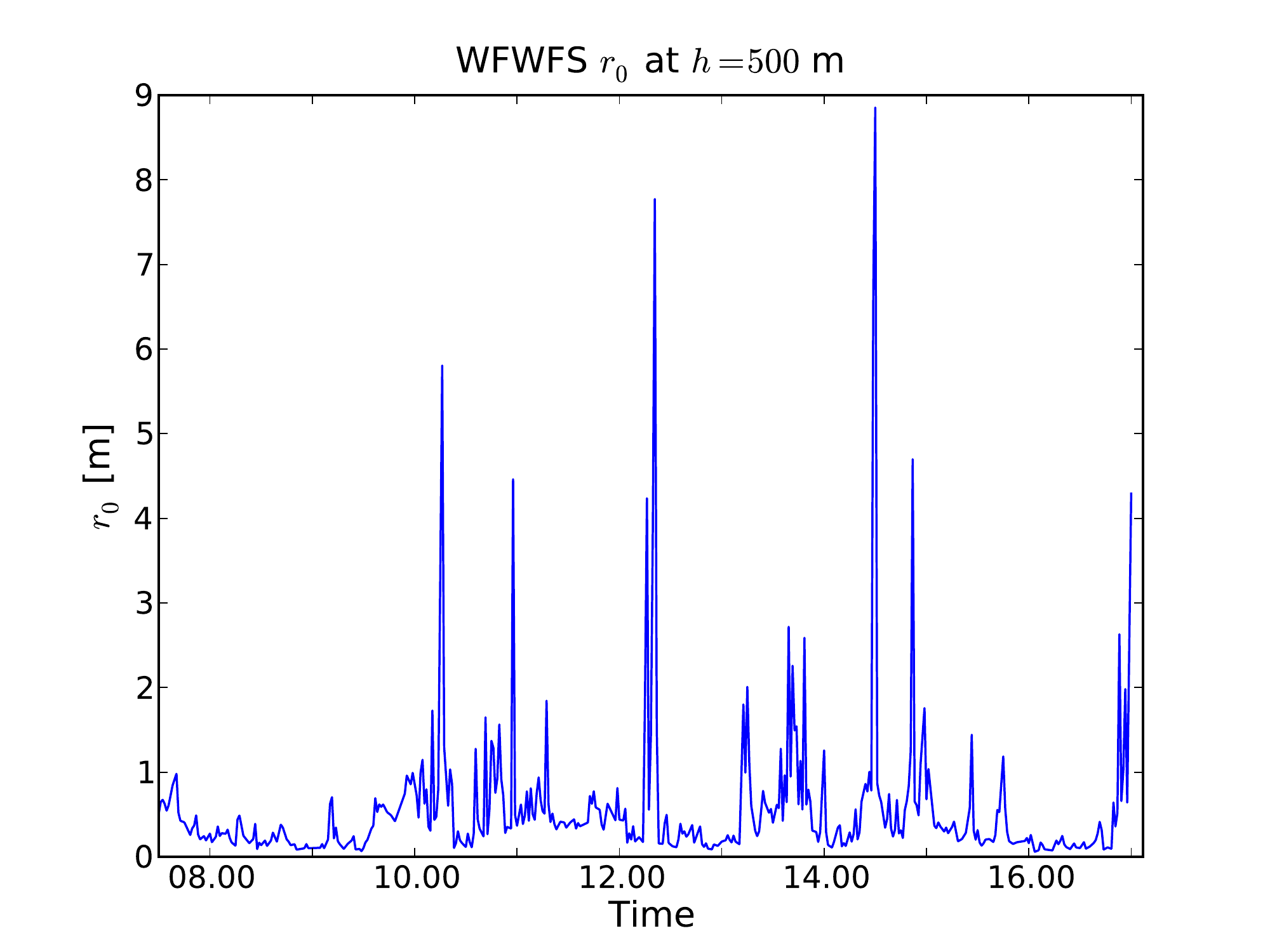}
  \label{fig:r0_wfwfs_500}}
  \centering
    \subfloat[1300 m]{\includegraphics[width=0.55\textwidth]{\plotpath 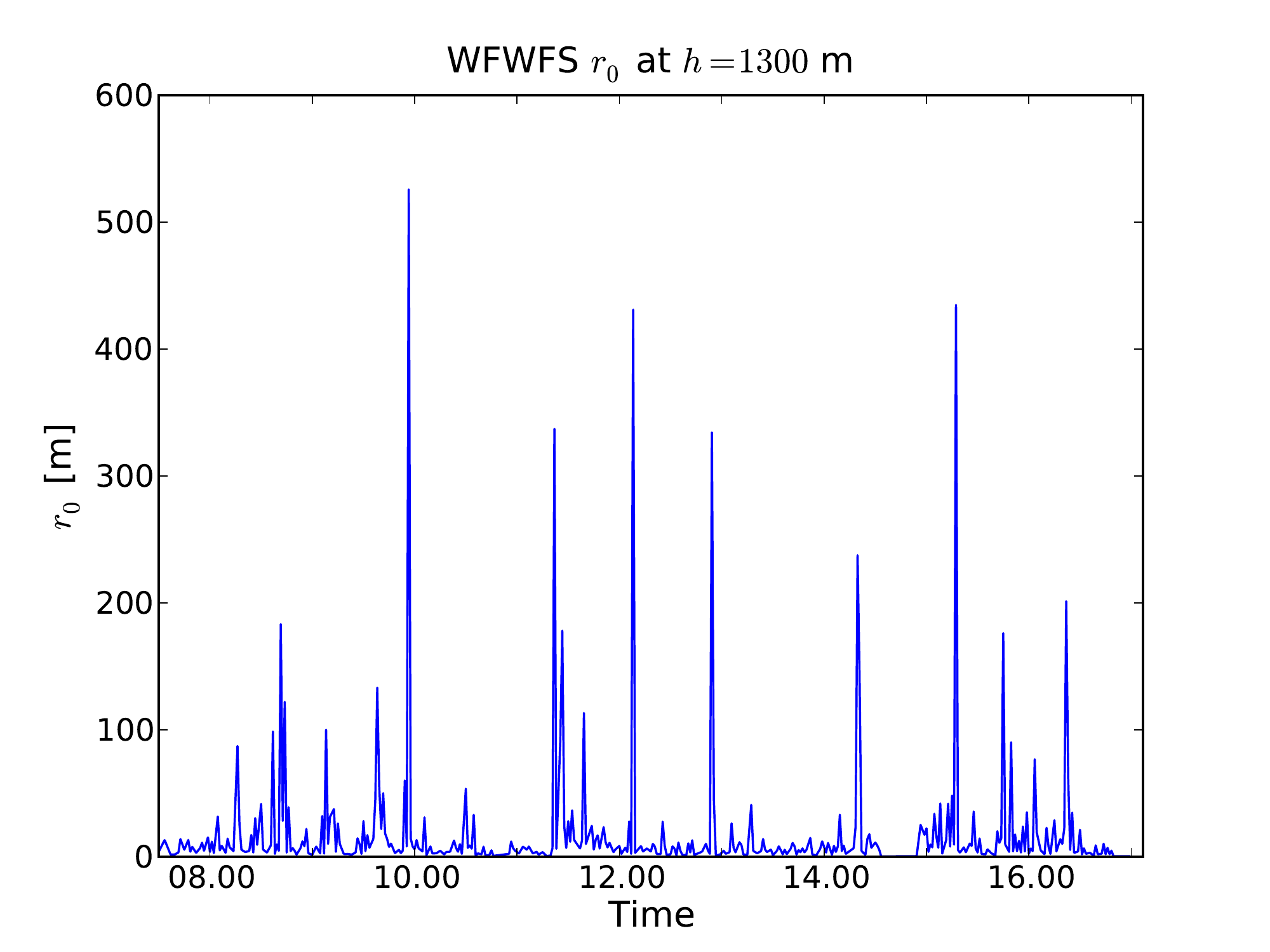}
  \label{fig:r0_wfwfs_1300}}\\
\hspace*{-0.7cm}
  \centering
    \subfloat[3200 m]{\includegraphics[width=0.55\textwidth]{\plotpath 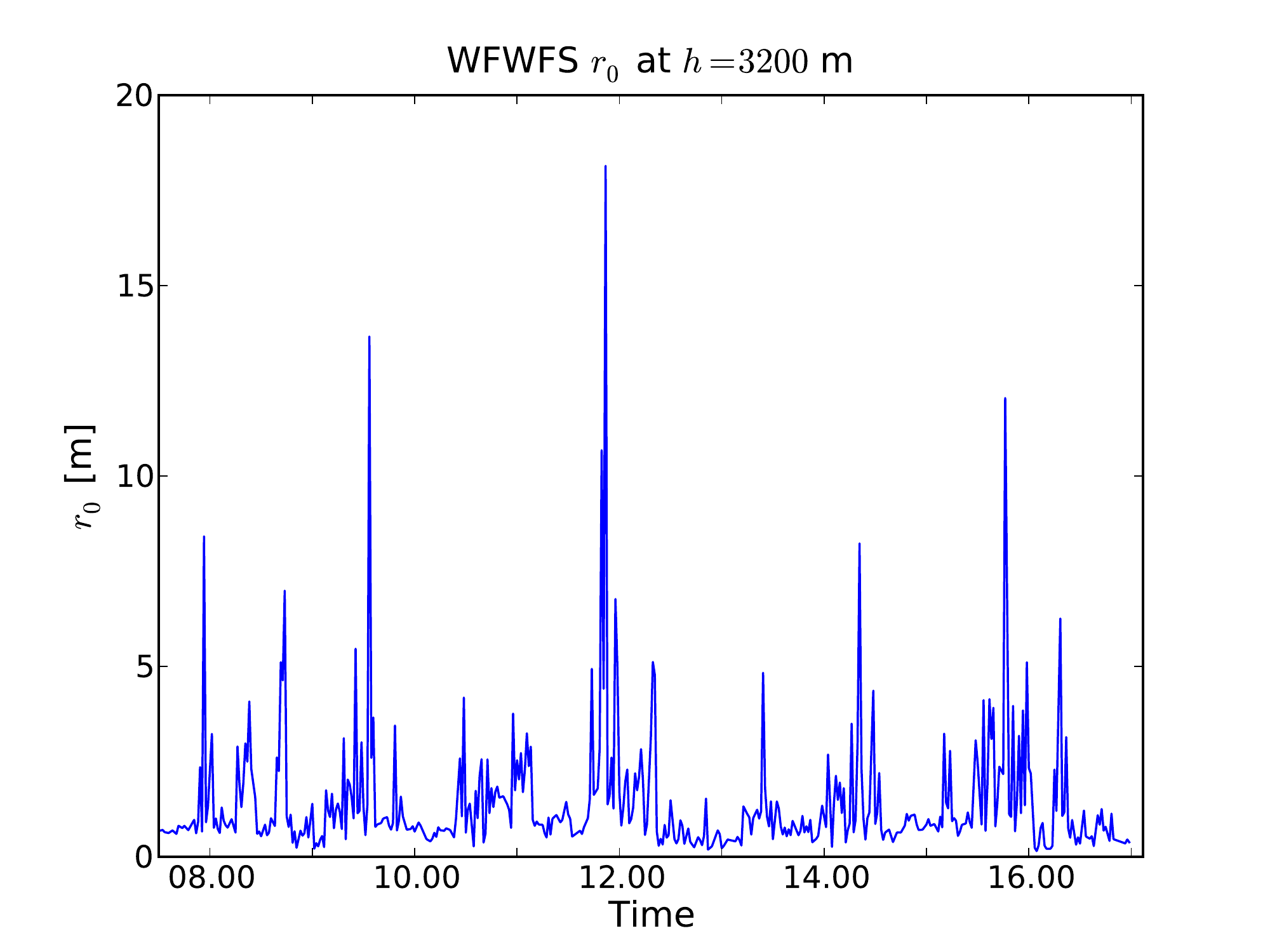}
  \label{fig:r0_wfwfs_3200}}
  \centering
    \subfloat[5800 m]{\includegraphics[width=0.55\textwidth]{\plotpath 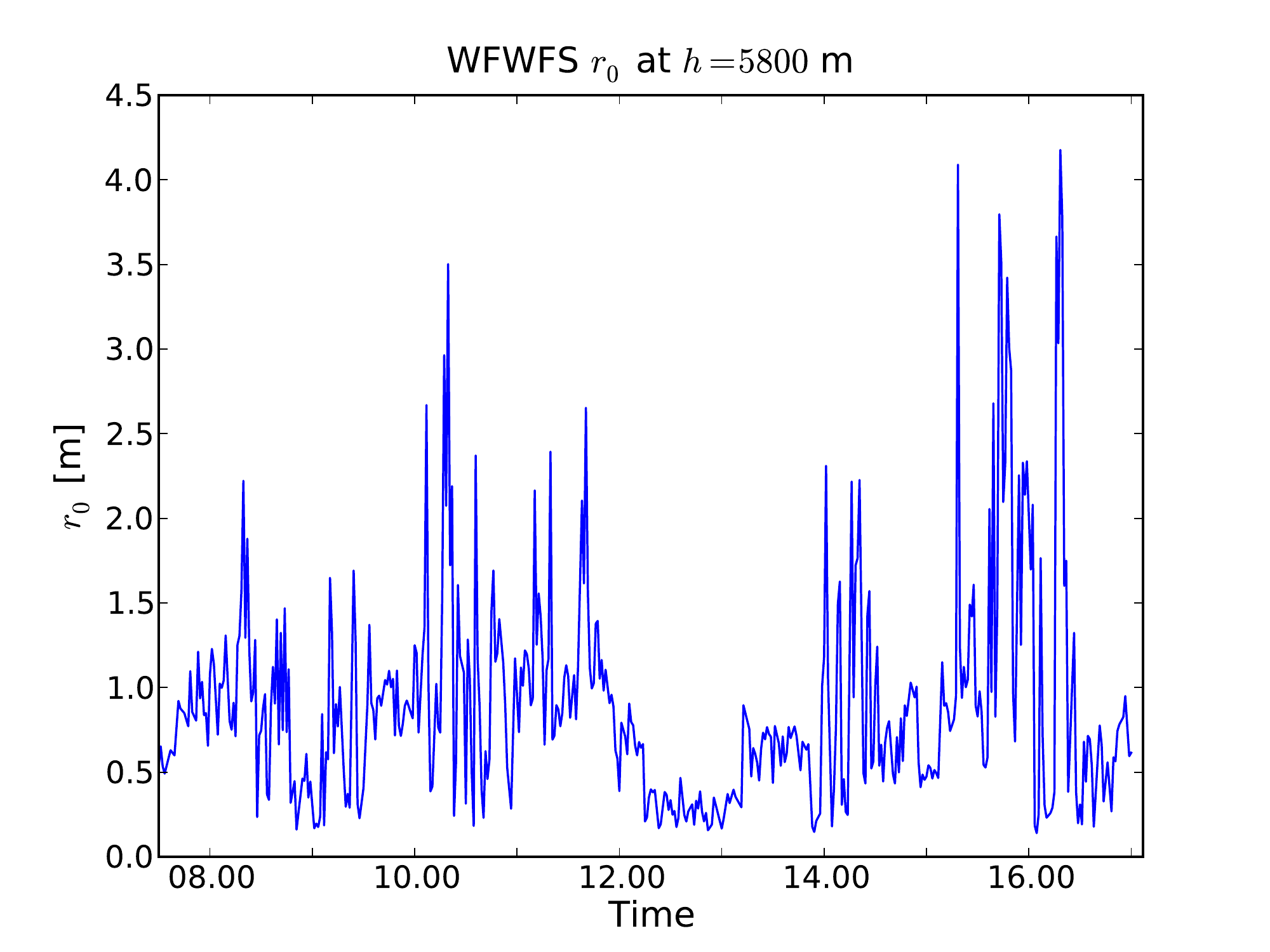}
  \label{fig:r0_wfwfs_5800}}
\caption{$r_0$ calculated from WFWFS measurements at different heights \\ $h=[500,1300,3200,5800]$ m.}\label{fig:r0_wfwfs}
\end{figure}

\begin{figure}[ht!]
\centering
\subfloat[]{\includegraphics[width=0.4\textwidth]{\imgpath 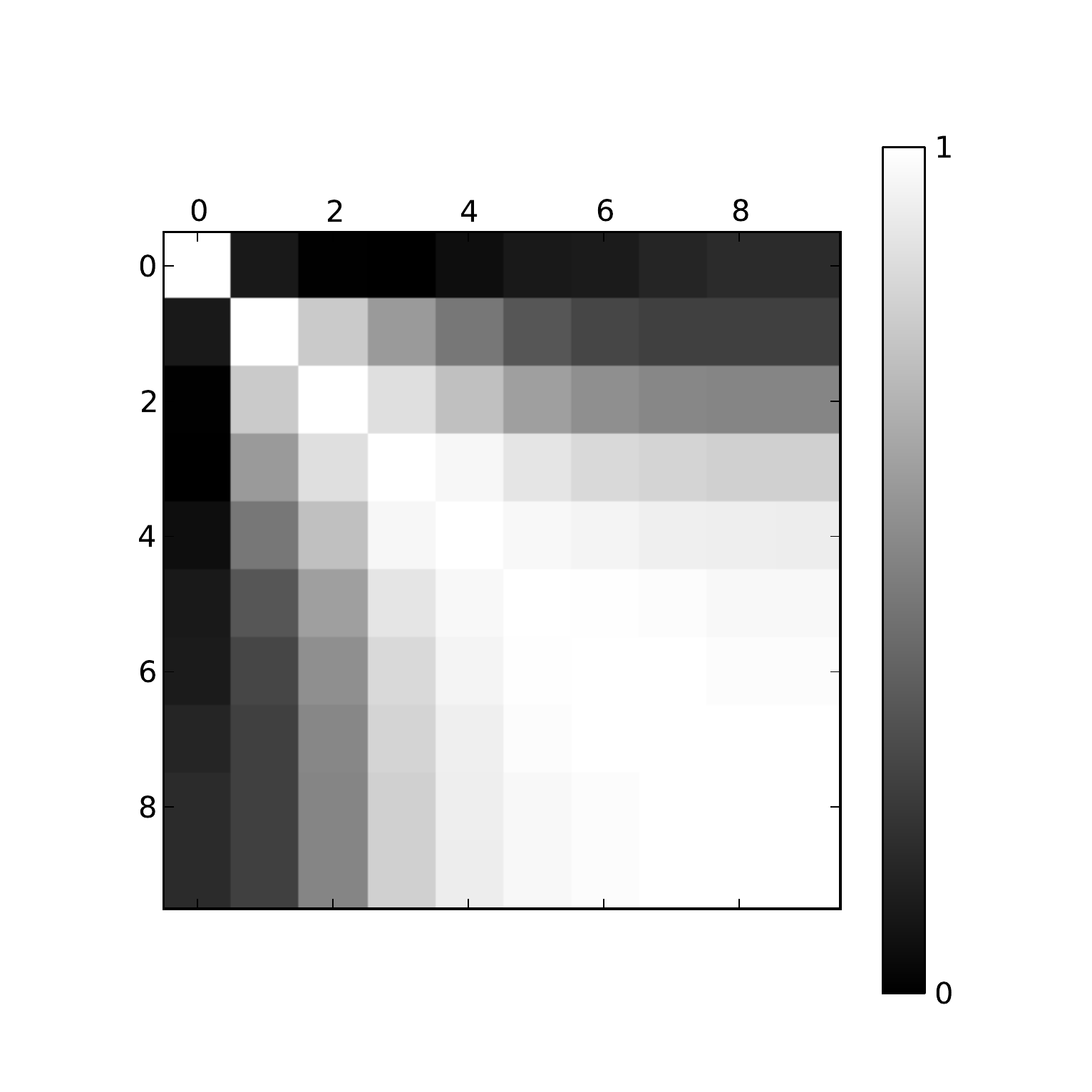}
\label{fig:corr_shabar_10}}
\centering
\subfloat[]{\includegraphics[width=0.4\textwidth]{\imgpath 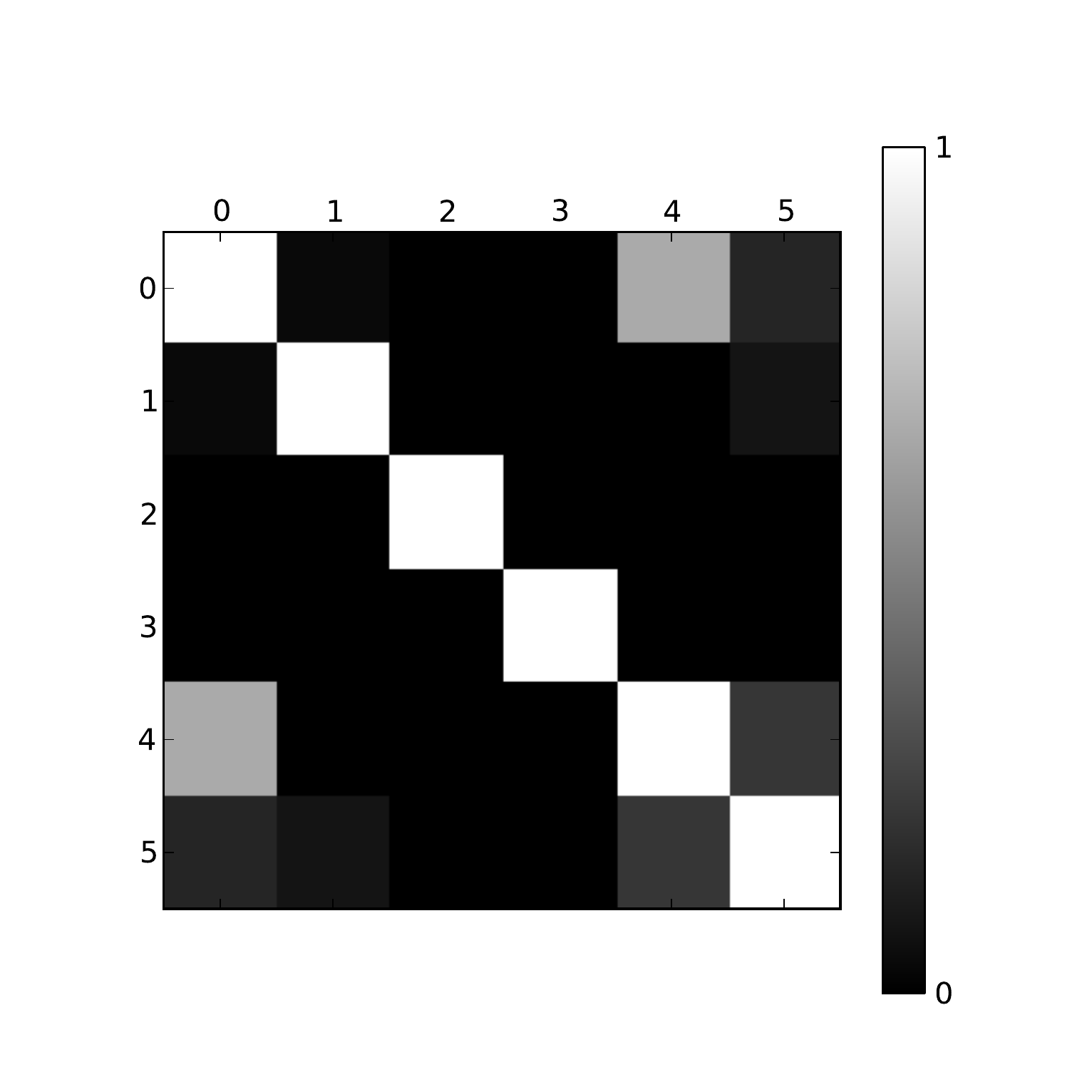}
\label{fig:corr_shabar_lim75_6}}
\caption{(a) Correlation matrix of $r_0$ for the 10 SHABAR layers first used. Black pixels indicate no correlation and white pixels show perfect correlation (when a layer is correlated with itself). The different white-gray pixels show how much correlation that exist between layers. (b) Correlation matrix for the 6 SHABAR layers chosen to minimize the correlation between layers.}\label{fig:corr_shabar}
\end{figure}

The high peaks for $r_0$ shown in \refpic{fig:r0_wfwfs_1300} are probably due to no turbulence region in this WFWFS layer. 
\newpage
\begin{figure}[ht!]
\hspace*{-0.7cm}
  \centering
    \subfloat[10800 m]{\includegraphics[width=0.55\textwidth]{\plotpath 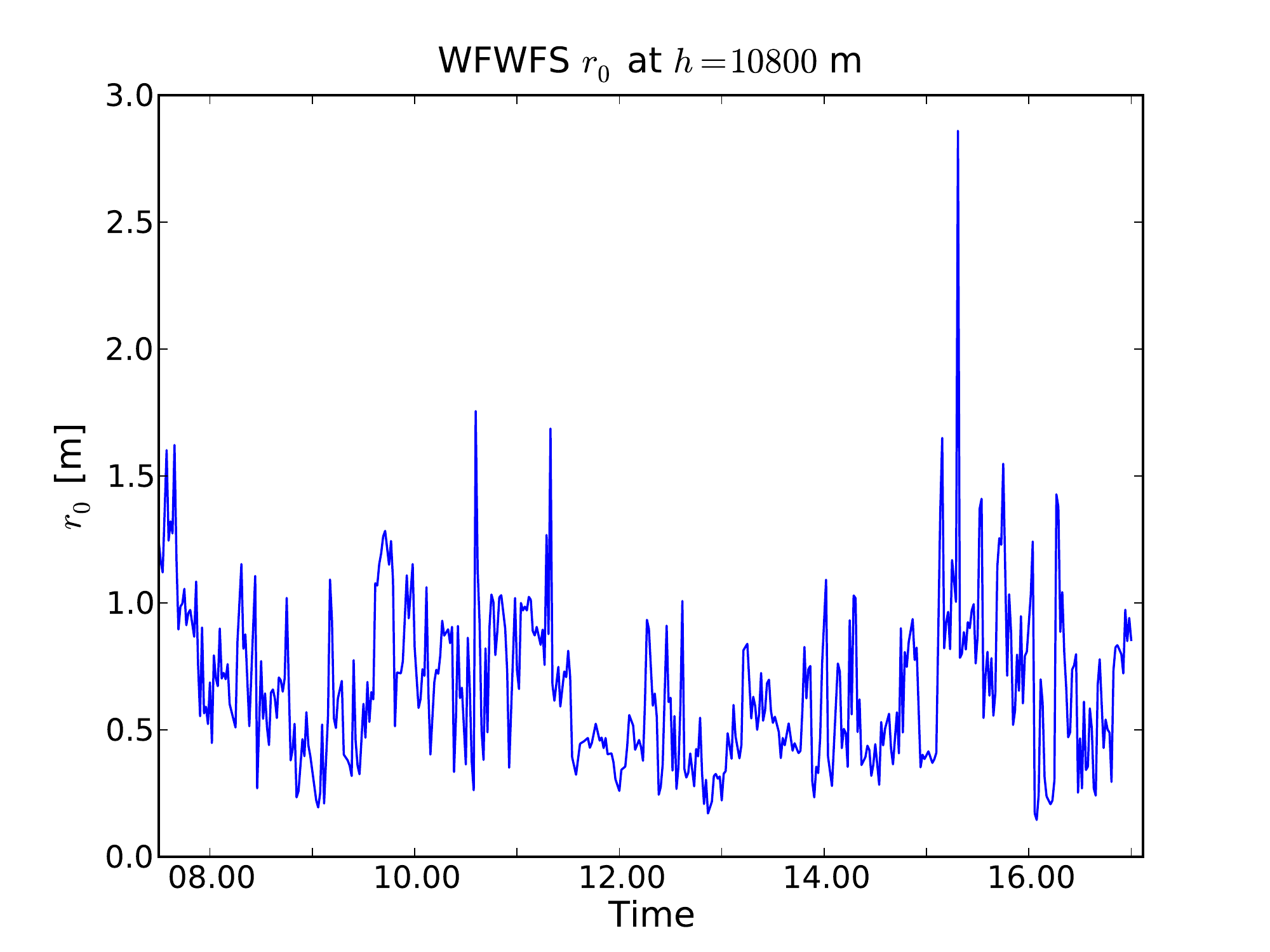}
  \label{fig:r0_wfwfs_10800}}
  \centering
    \subfloat[20000 m]{\includegraphics[width=0.55\textwidth]{\plotpath 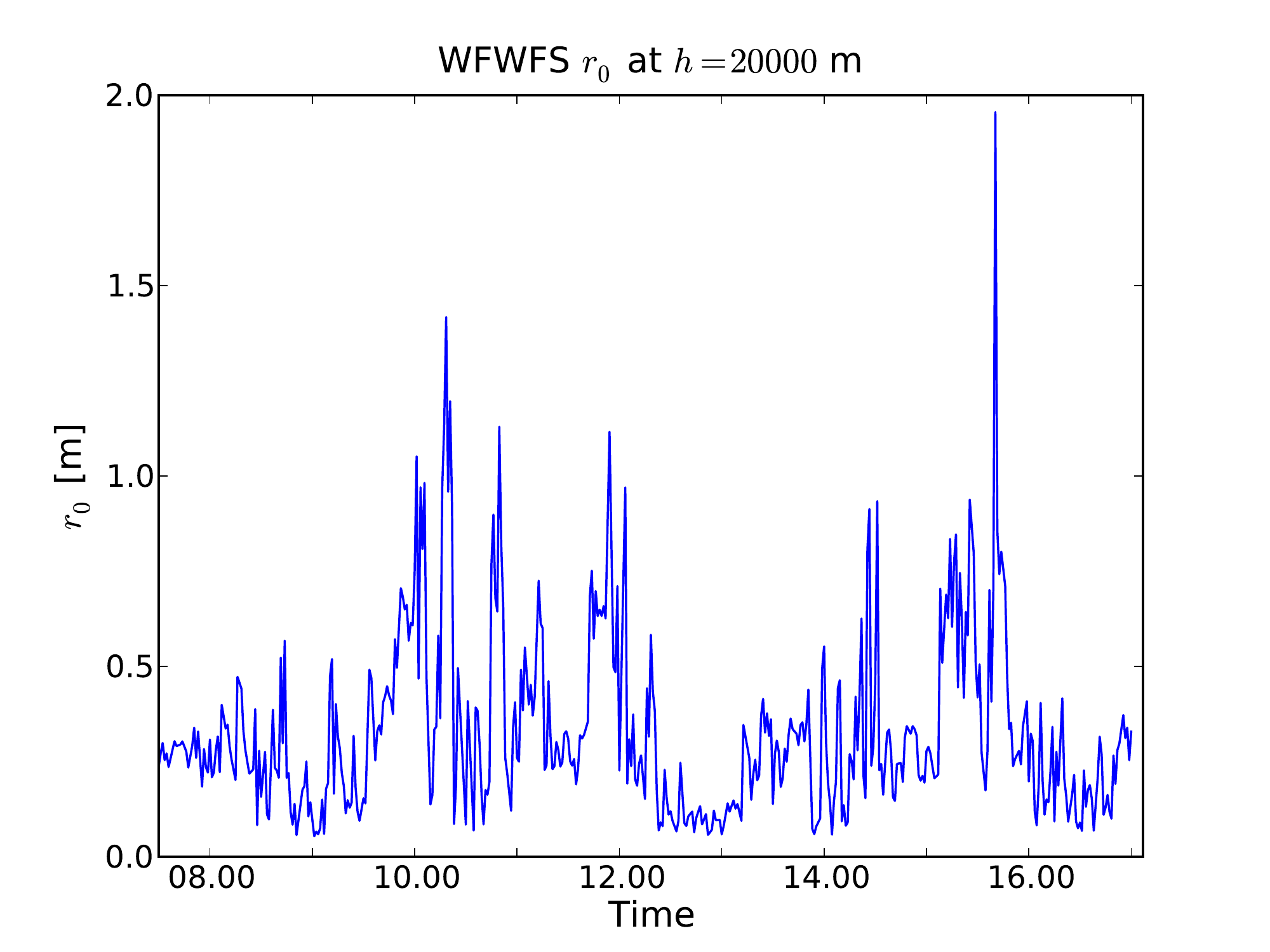}
  \label{fig:r0_wfwfs_20000}}
\caption{$r_0$ calculated from WFWFS measurements at different heights \\ $h=[10800,20000]$ m.}\label{fig:r0_wfwfs2}
\end{figure}
\newpage
\sec{WFWFS $r_0$ compared with SHABAR $r_0$}
\label{sec:shab_comp}
The correlation matrix of $r_0$ for the 10 SHABAR layers first used (described in \refsec{sec:shab_data}), \refpic{fig:corr_shabar_10}, shows that there were too many SHABAR layers. The measured $r_0$ values in the different layers correlate strongly with each other, which indicates that the measurements are not independent. The layers were therefore changed so that the correlation between them were as low as possible. The new boundaries between layers were chosen to be located at 1 m, 10 m, 30 m, 200 m, 800 m,  3 km and 20 km. These 7 new heights define boundaries between 6 new SHABAR layers which have average heights $h_{shabar}=[5, 20, 115, 500, 1900, 11500]$. The corresponding correlation matrix is shown in \refpic{fig:corr_shabar_lim75_6}. 

\begin{figure}[ht!]
  \centering
    \includegraphics[width=0.8\textwidth]{\plotpath 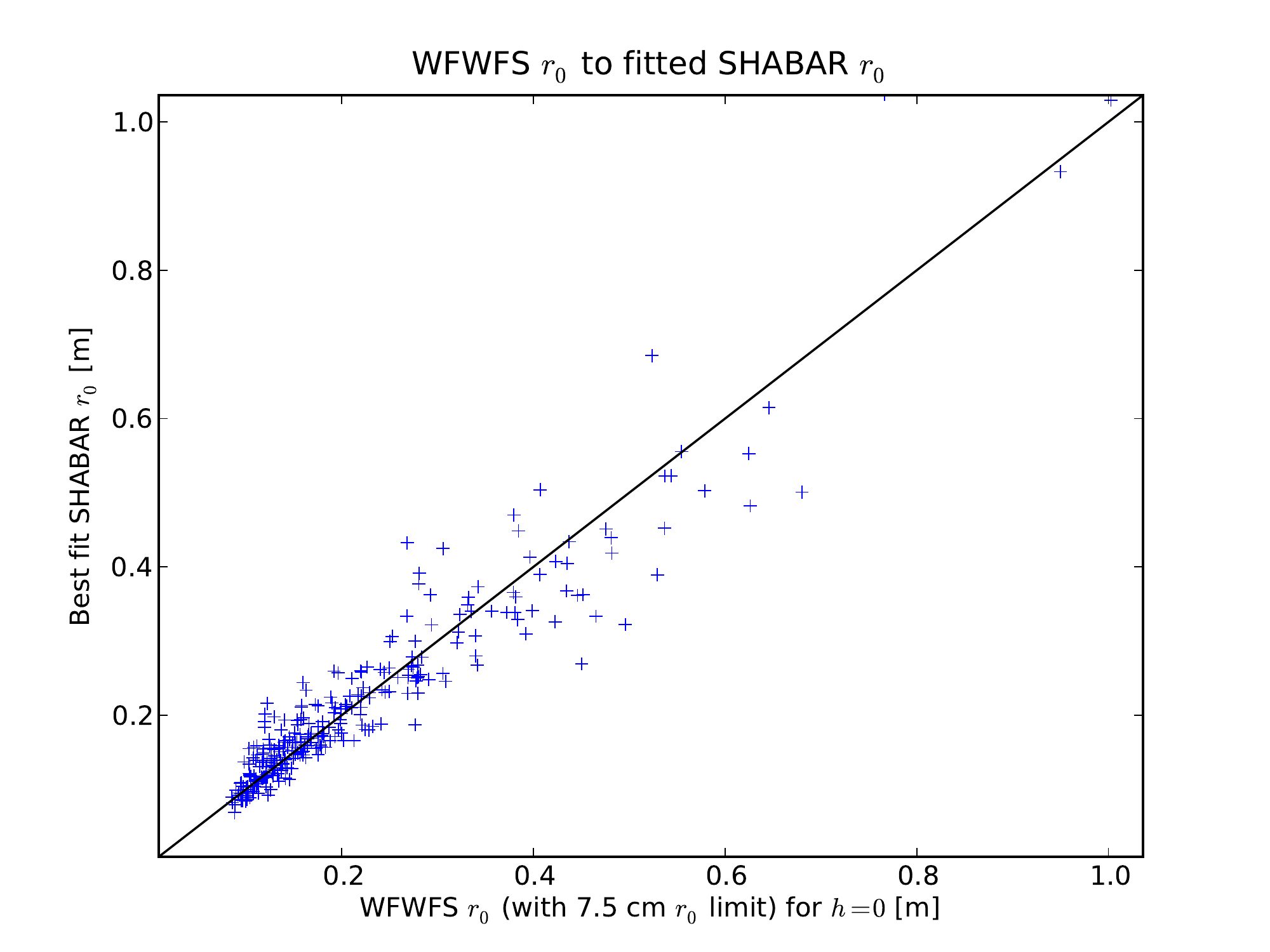}
  \caption{Correlation between best fit $r_0$ from SHABAR data and $r_0$ calculated from WFWFS measurements at $h=0$ m. Correlation coefficient 0.95.}
  \label{fig:fit_0}
\end{figure}

\begin{figure}[ht!]
  \centering
    \includegraphics[width=0.8\textwidth]{\plotpath 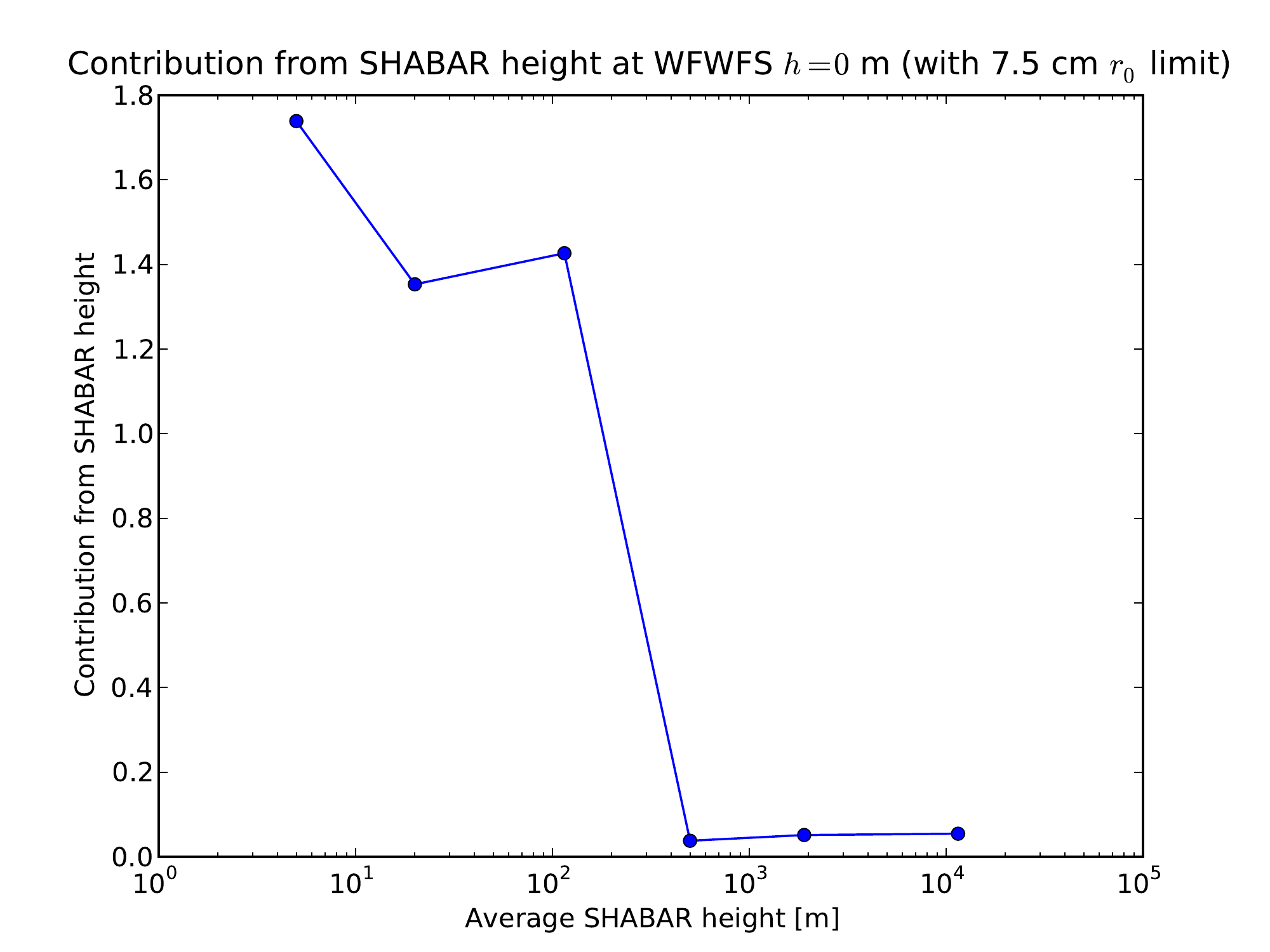}
  \caption{Contribution to best fit $r_0$ for WFWFS height $h=0$.}
  \label{fig:contr_0}
\end{figure}

The 6 new SHABAR layers were compared with the 7 WFWFS layers. The correlation between $r_0$ calculated from WFWFS measurements at $h=0$ m, plotted against the best fitted $r_0$ from SHABAR data is shown in \refpic{fig:fit_0}. The correlation coefficient is 0.95. How much each of the SHABAR layers contributed to the best fitted $r_0$ for that WFWFS height ($h=0$ m) is shown in \refpic{fig:contr_0}. This shows that the contribution from SHABAR data to the best fitted $r_0$ comes from layers below 200 meter. 

\begin{figure}[ht!]
  \centering
    \includegraphics[width=0.8\textwidth]{\plotpath 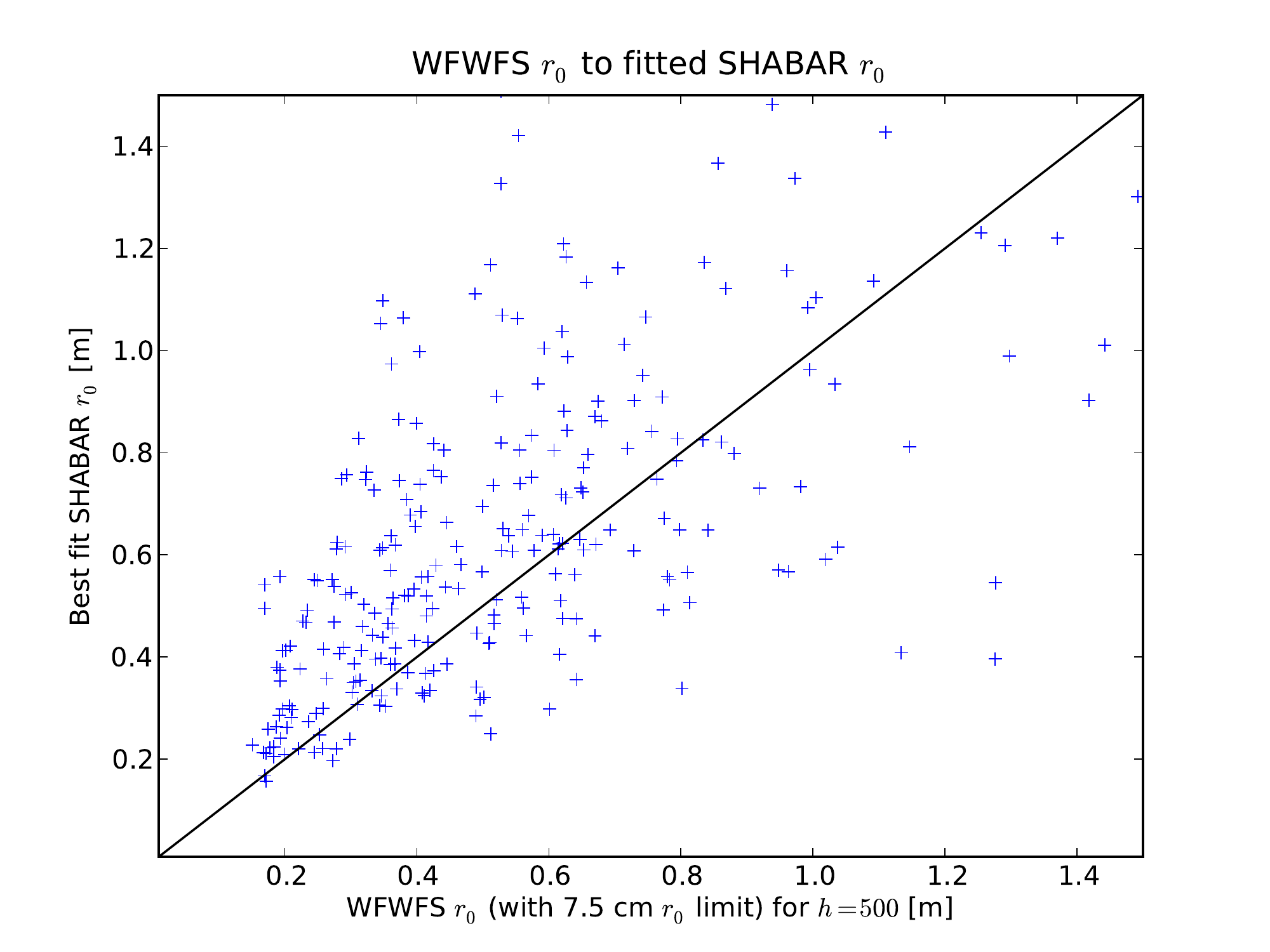}
  \caption{Correlation between best fit $r_0$ from SHABAR data and $r_0$ calculated from WFWFS measurements at $h=500$ m. Correlation coefficient 0.27.}\label{fig:fit_500}
\end{figure}

\begin{figure}[ht!]
  \centering
    \includegraphics[width=0.8\textwidth]{\plotpath 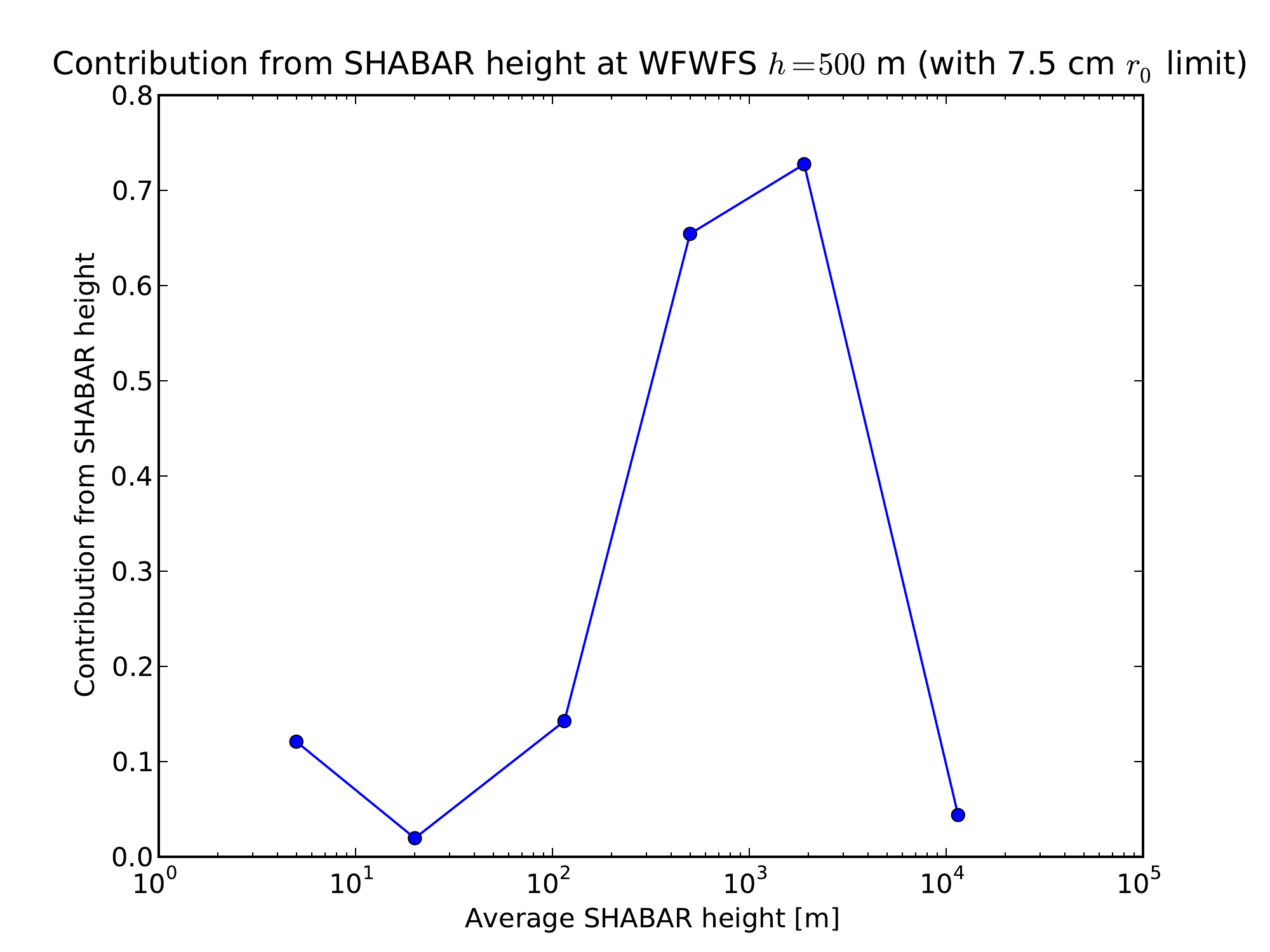}
  \caption{Contribution to best fit $r_0$ for WFWFS height $h=500$.}
  \label{fig:contr_500}
\end{figure}

\refpicl{fig:fit_500} shows the correlation between WFWFS measurements and best fitted SHABAR $r_0$ for WFWFS-height $h=500$ m. The correlation coefficient is 0.27 which means that the correlation is not as strong as for the first layer but still there. The corresponding contribution from the different SHABAR layers is shown in \refpic{fig:contr_500}, which shows that the highest contributions come from the two layers located at $h=500$ m and $h=1900$ m.

\newpage
The kernels for the SHABAR inversion converge at heights larger than 1 km \citep{ATST-RPT-0014,2010SPIE.7733E.144S} which should set the limit for the SHABAR range. This means that the SHABAR most likely cannot distinguish layers at higher heights. 

No comparison between WFWFS height 1300 m and SHABAR data is shown in the figures below since these $r_0$ values were too high. No turbulence regions seem to be located at this height. 

It is hard to see any correlation between the WFWFS and the SHABAR for $h=3200$ m, where the correlation coefficient is 0.07. The contribution to the fitted $r_0$ still comes mostly from expected layers ($h=500, 1900$), even though the correlation is low. 

The correlation coefficient is even lower, -0.01, for $h=5800$ m. The contribution from the lowest layer gets stronger for this height but the peak is still at higher heights ($h_{shabar}=500, 2000$). Almost no correlation is found for WFWFS height $h=10800$ m either, where the correlation coefficient is 0.16. The contribution to the best fit comes mostly from the first layer. 

Some correlation seems to be present again at higher WFWFS heights. Especially for $h=20000$ m, where the correlation coefficients is 0.22. This effect is however not real. The seeing is worse here than for the closest previous layers and the small values of $r_0$ come from the poor ground layer seeing. The correlation found is because of the correlation at lower heights.

\newpage

\pagebreak
\pagebreak
\begin{figure}[ht!]
  \centering
    \includegraphics[width=0.8\textwidth]{\plotpath 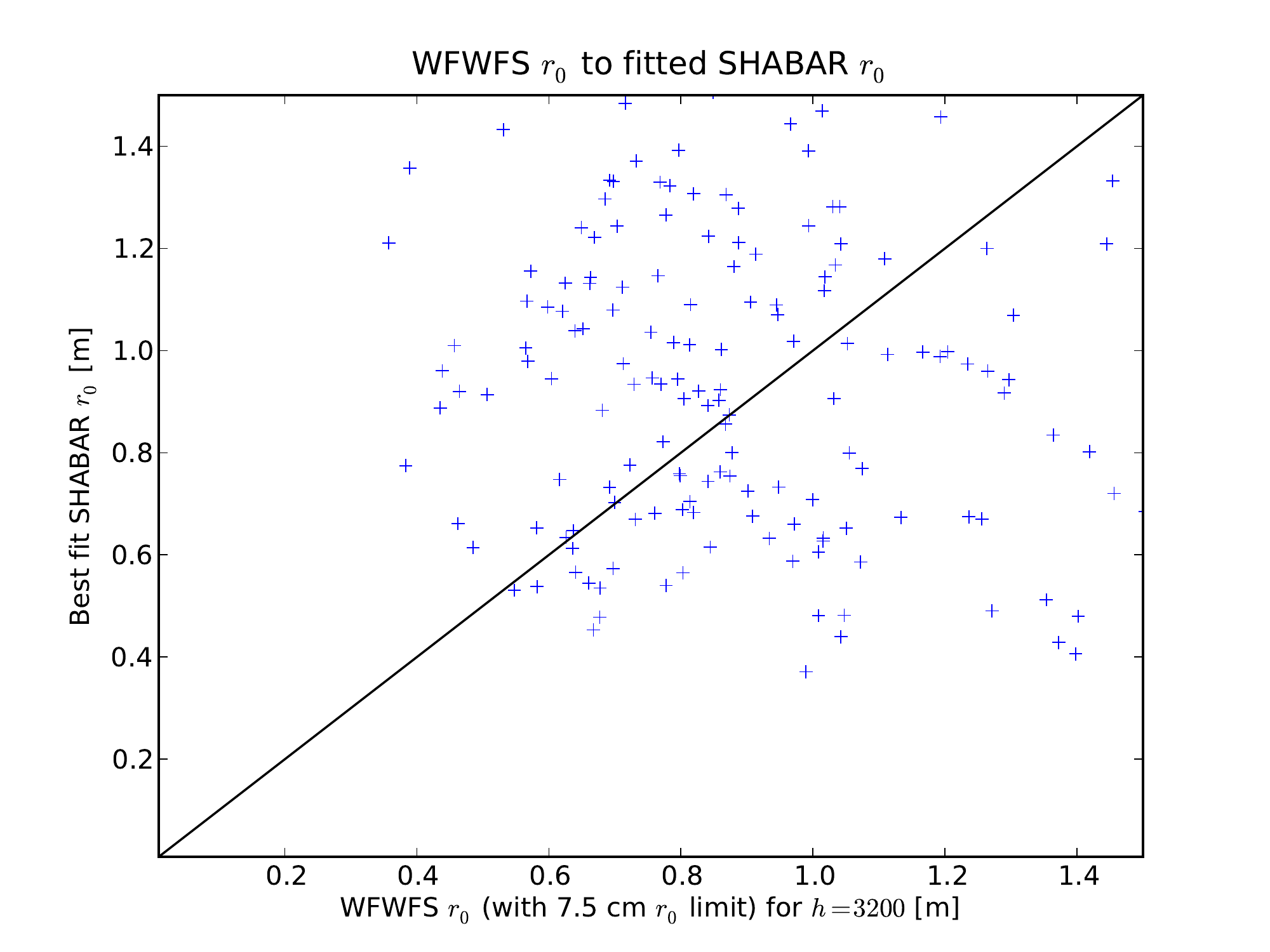}
  \caption{Correlation between best fit $r_0$ from SHABAR data and $r_0$ calculated from WFWFS measurements at $h=3200$ m. Correlation coefficient 0.07.}\label{fig:fit_3200}
\end{figure}
\begin{figure}[ht!]
  \centering
    \includegraphics[width=0.8\textwidth]{\plotpath 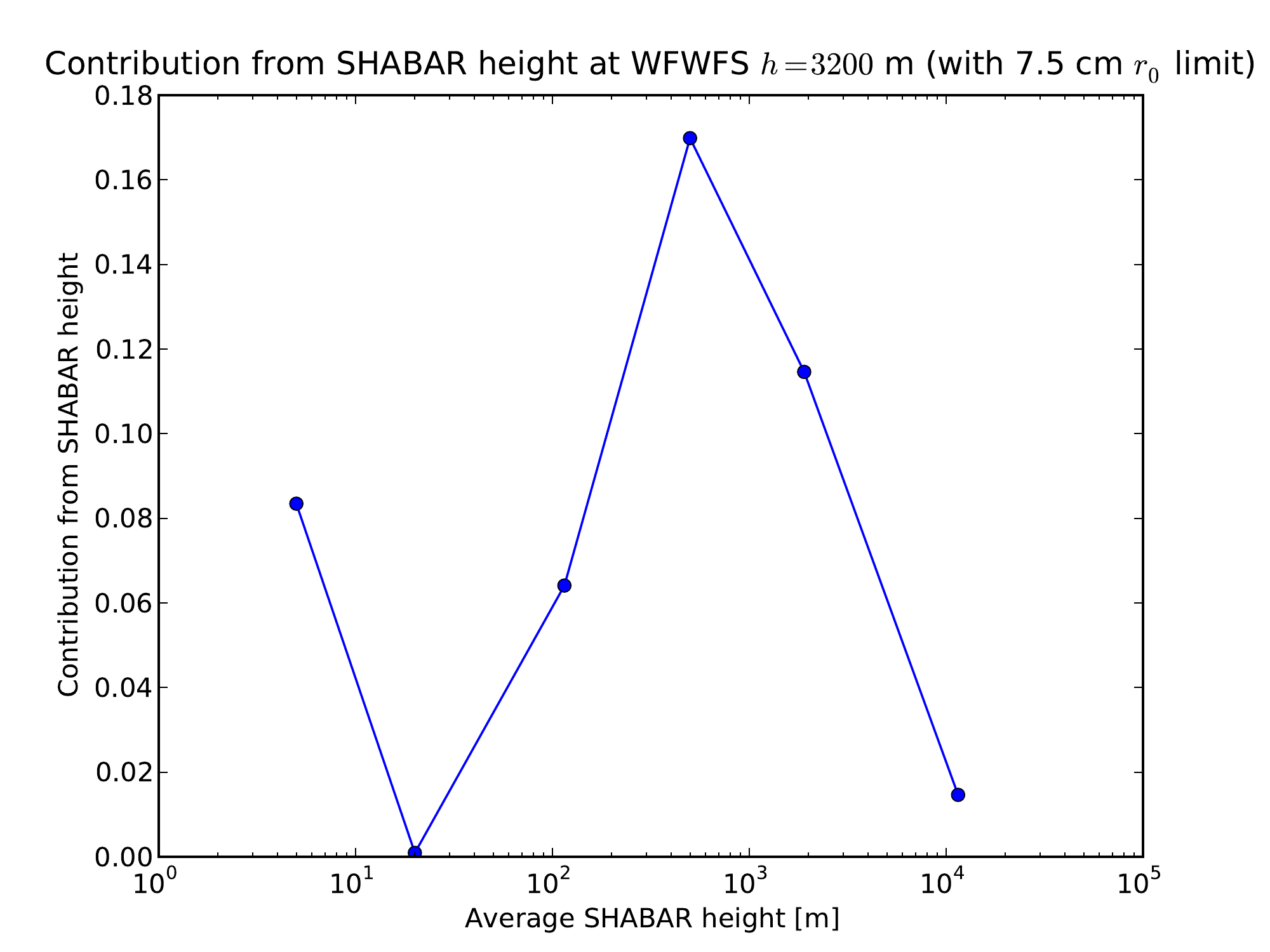}
  \caption{Contribution to best fit $r_0$ for WFWFS height $h=3200$.}
  \label{fig:contr_3200}
\end{figure}
\newpage
\begin{figure}[ht!]
  \centering
    \includegraphics[width=0.8\textwidth]{\plotpath 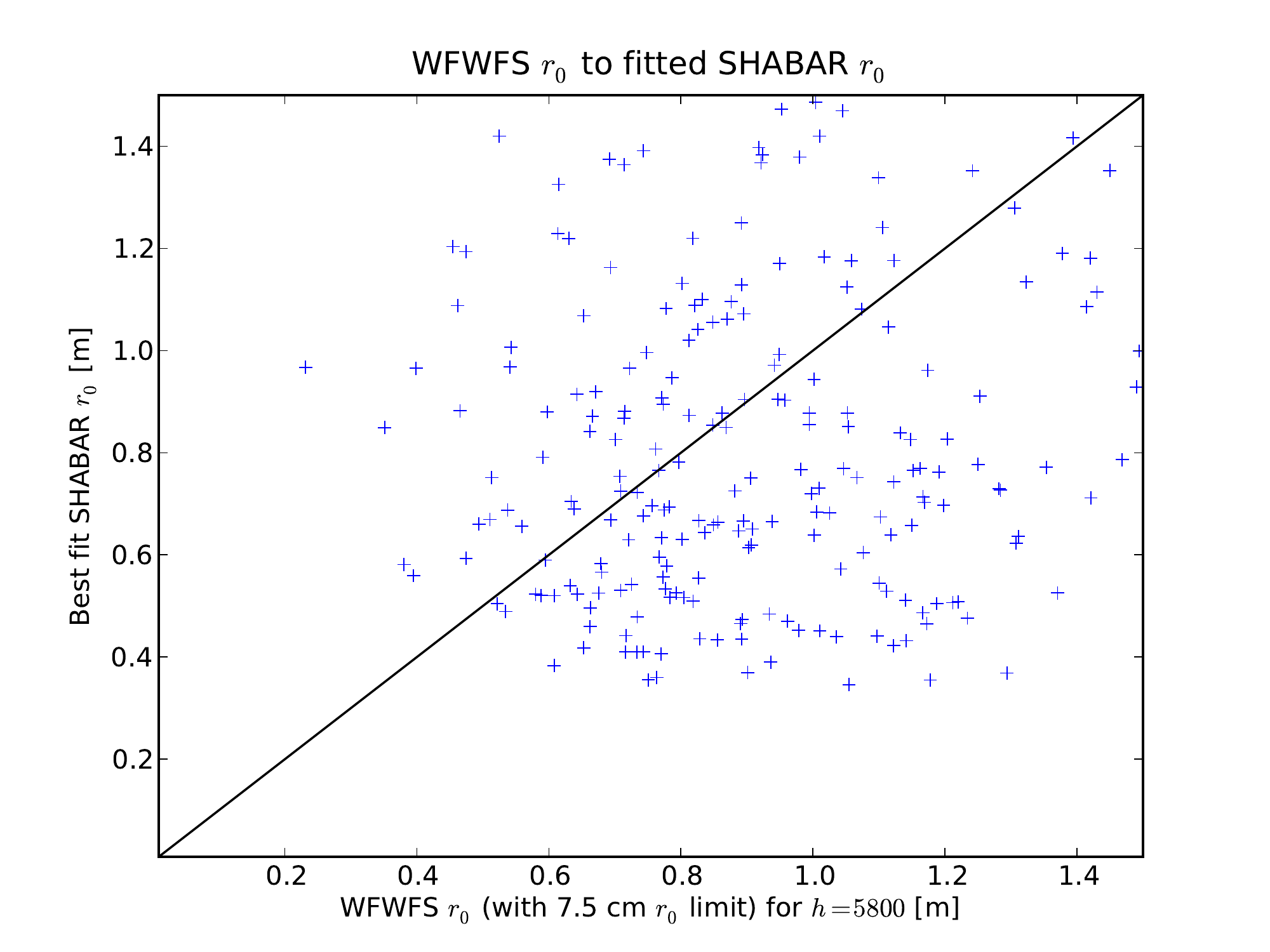}
  \caption{Correlation between best fit $r_0$ from SHABAR data and $r_0$ calculated from WFWFS measurements at $h=5800$ m. Correlation coefficient -0.01.}\label{fig:fit_5800}
\end{figure}
\begin{figure}[ht!]
  \centering
    \includegraphics[width=0.8\textwidth]{\plotpath 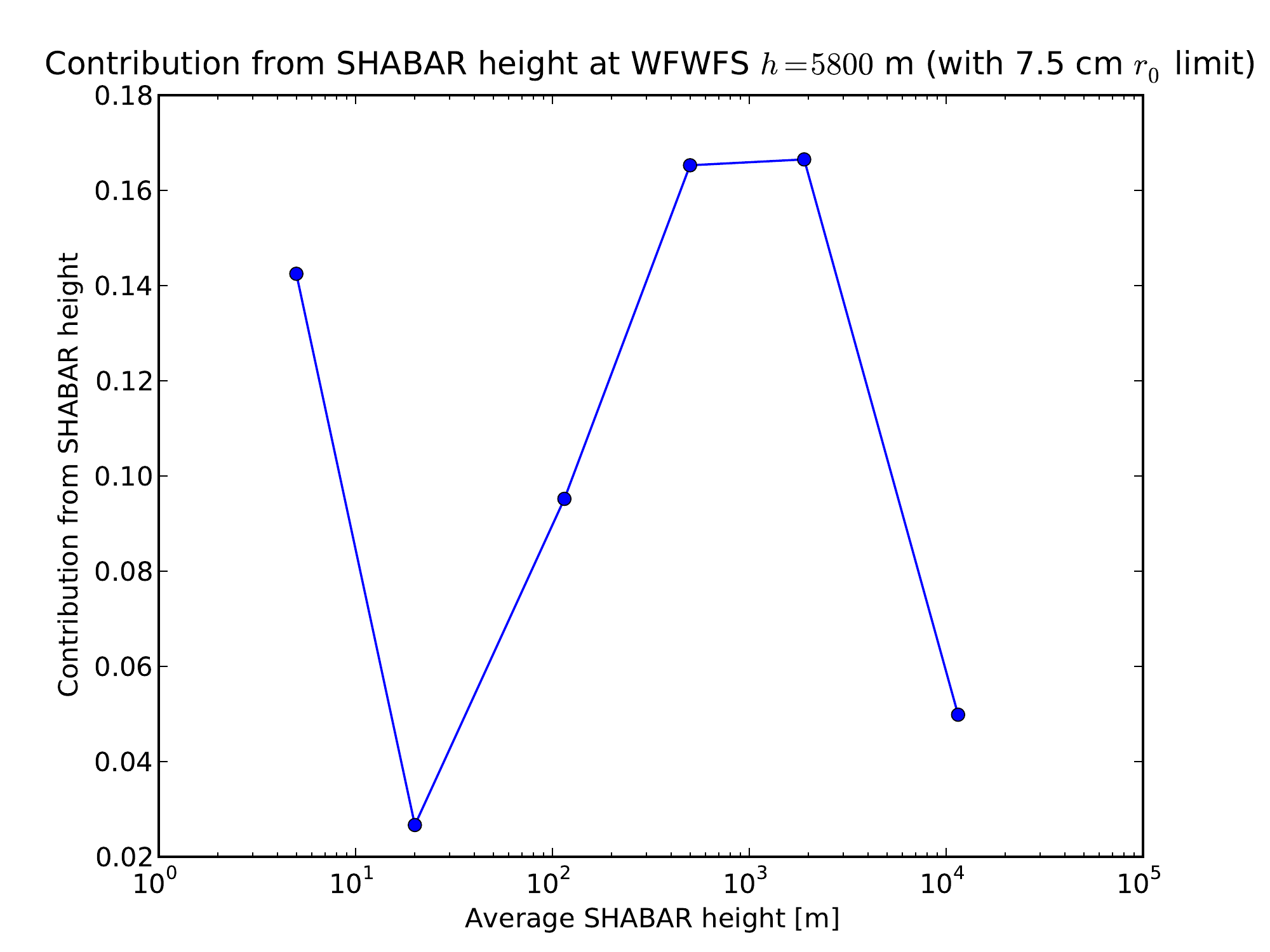}
  \caption{Contribution to best fit $r_0$ for WFWFS height $h=5800$.}
  \label{fig:contr_5800}
\end{figure}
\newpage
\begin{figure}[ht!]
  \centering
    \includegraphics[width=0.8\textwidth]{\plotpath 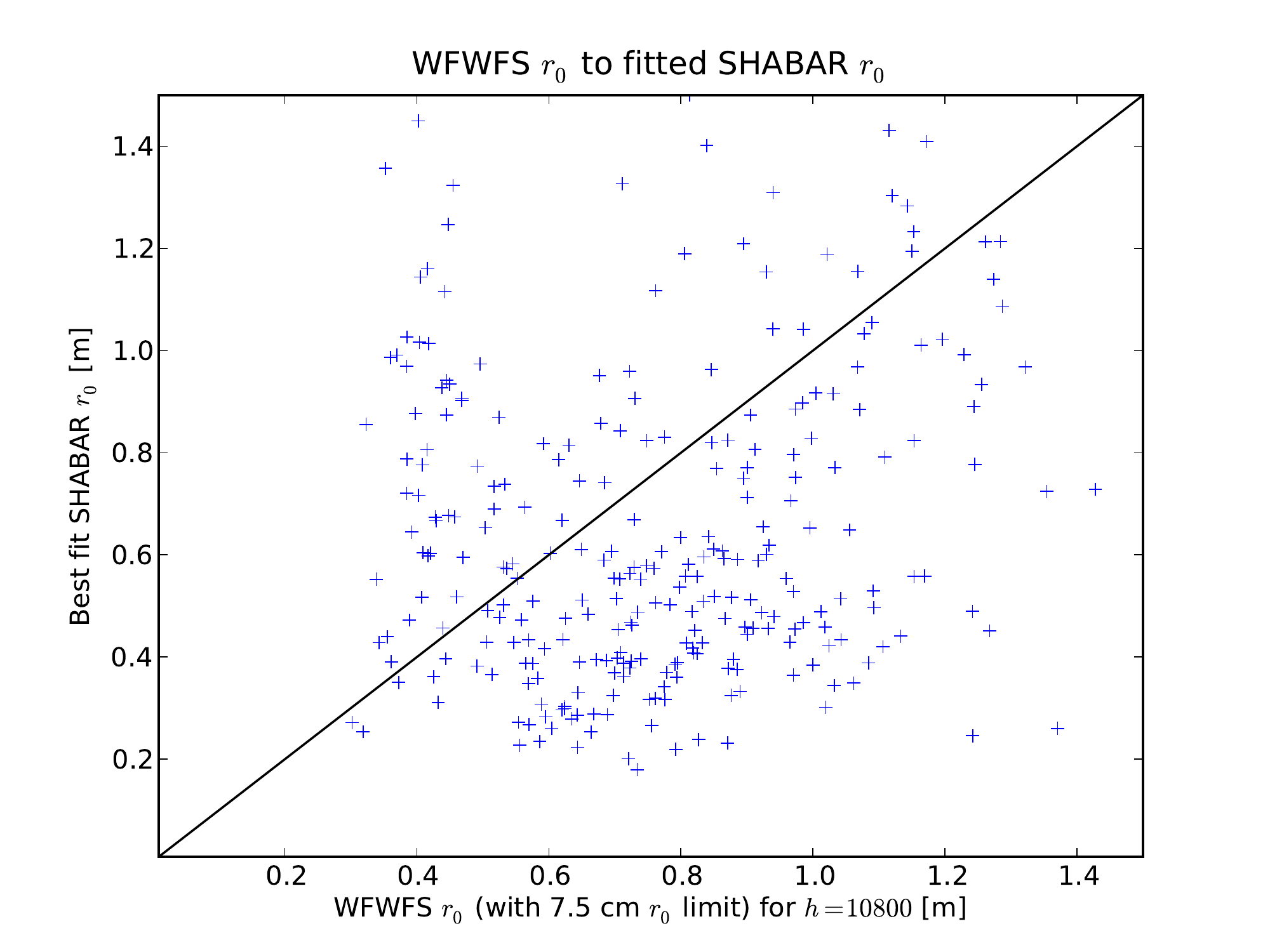}
  \caption{Correlation between best fit $r_0$ from SHABAR data and $r_0$ calculated from WFWFS measurements at $h=10800$ m. Correlation coefficient 0.16.}\label{fig:fit_10800}
\end{figure}
\begin{figure}[ht!]
  \centering
    \includegraphics[width=0.8\textwidth]{\plotpath 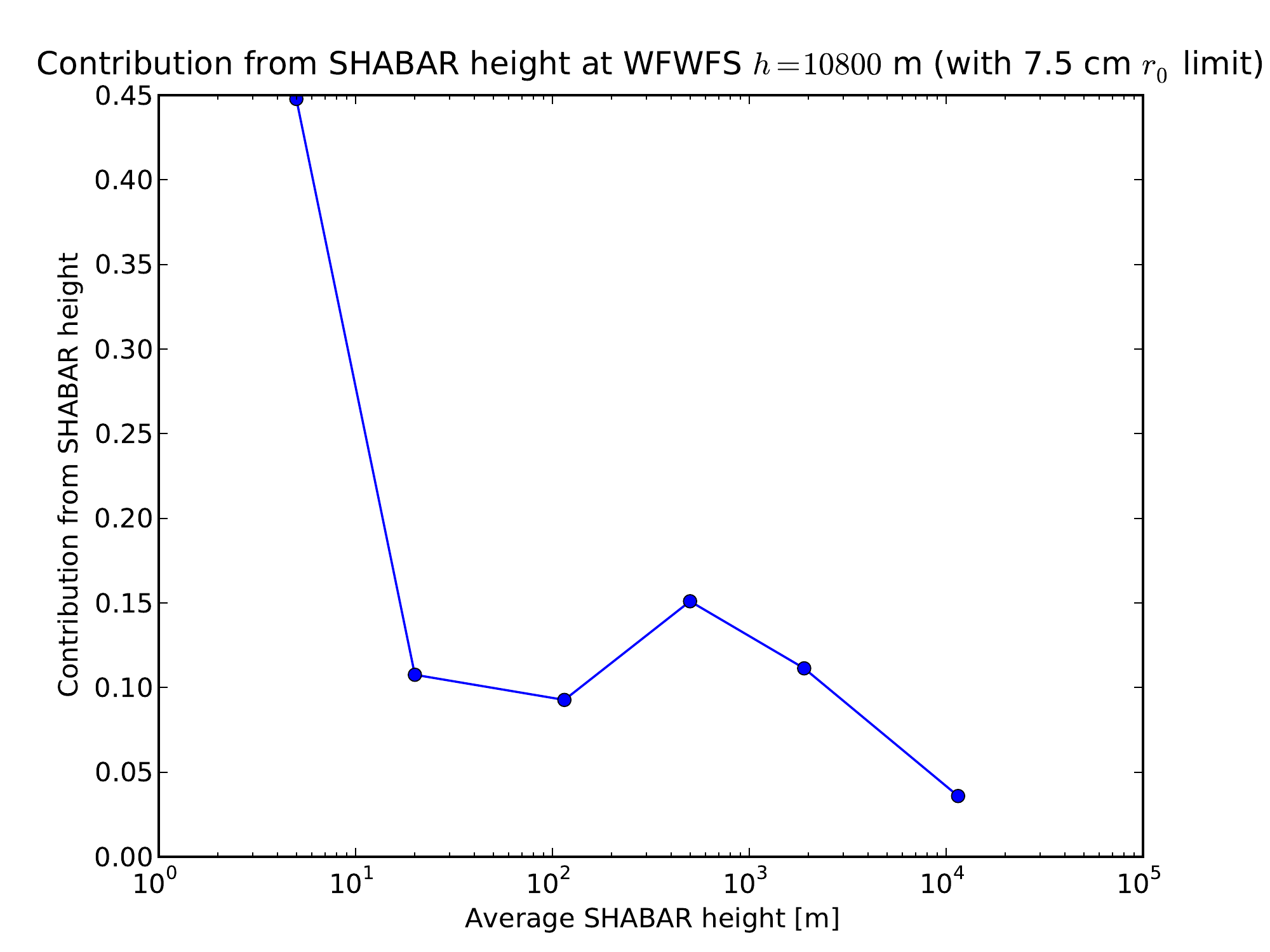}
  \caption{Contribution to best fit $r_0$ for WFWFS height $h=10800$.}
  \label{fig:contr_10800}
\end{figure}
\newpage
\begin{figure}[ht!]
  \centering
    \includegraphics[width=0.8\textwidth]{\plotpath 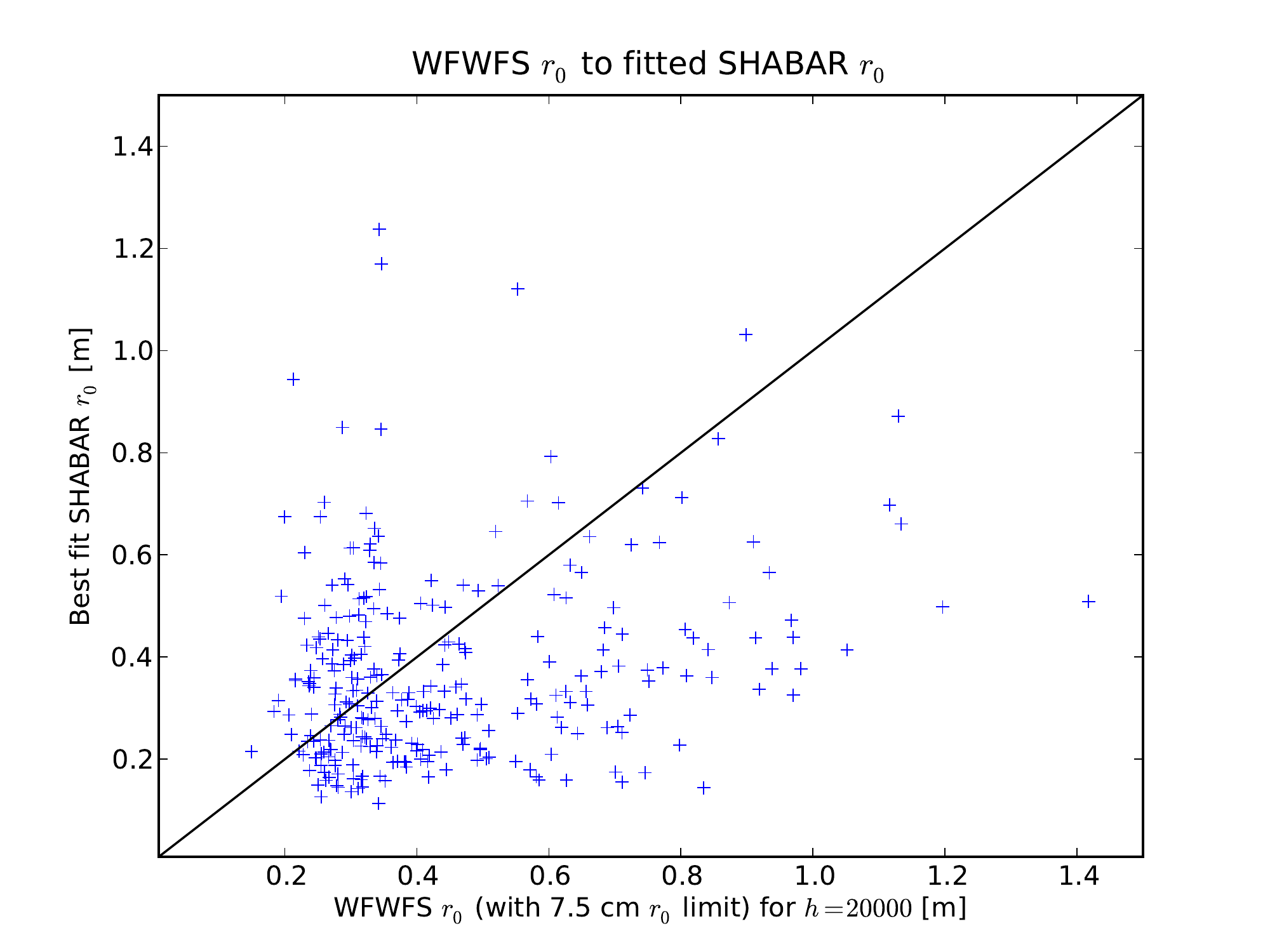}
  \caption{Correlation between best fit $r_0$ from SHABAR data and $r_0$ calculated from WFWFS measurements at $h=20000$ m. Correlation coefficient 0.22.}\label{fig:fit_20000}
\end{figure}
\begin{figure}[ht!]
  \centering
    \includegraphics[width=0.8\textwidth]{\plotpath 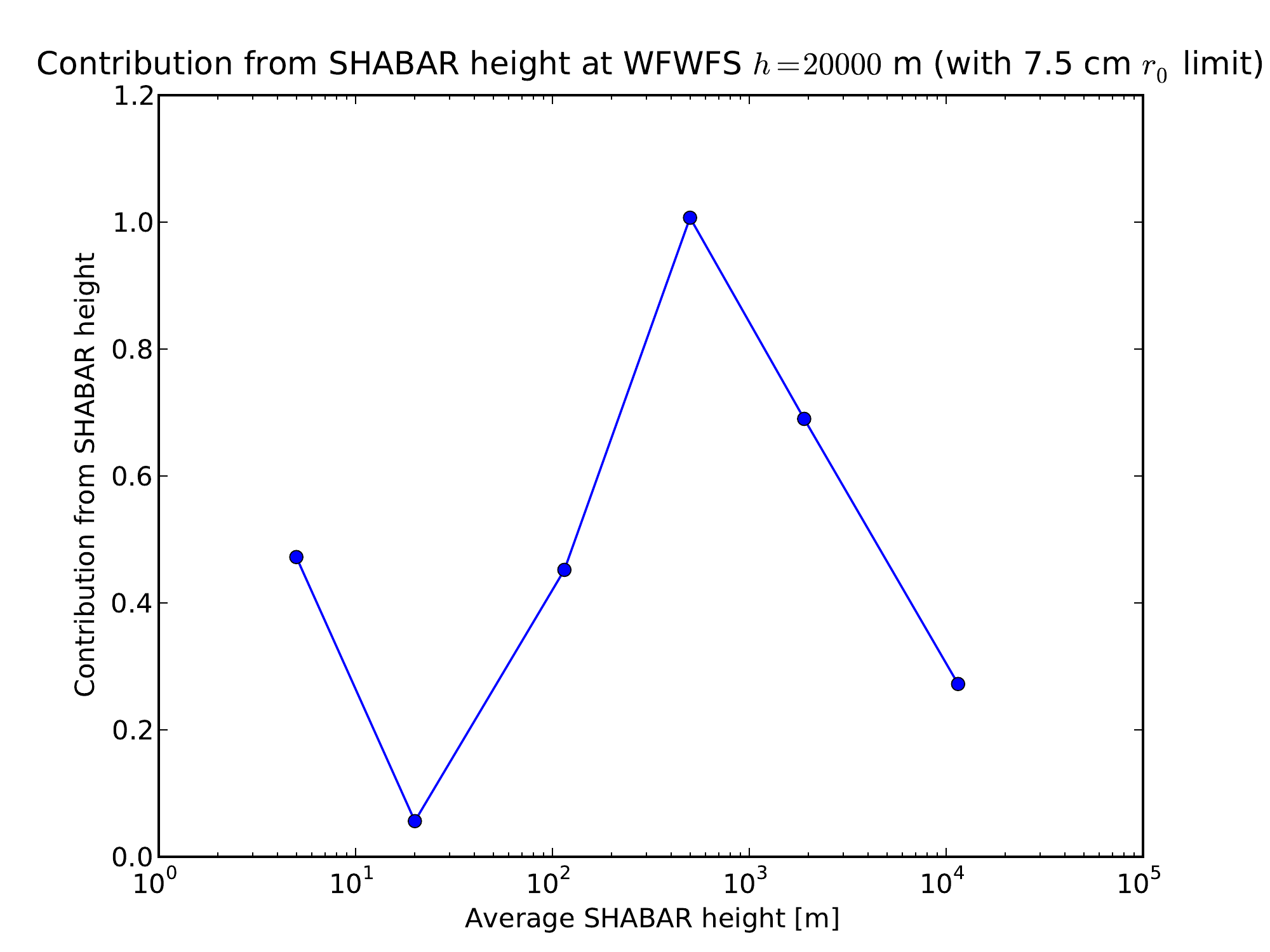}
  \caption{Contribution to best fit $r_0$ for WFWFS height $h=20000$.}
  \label{fig:contr_20000}
\end{figure}
\begin{figure}[ht!]
 \centering
  \includegraphics[width=0.71\textwidth]{\plotpath 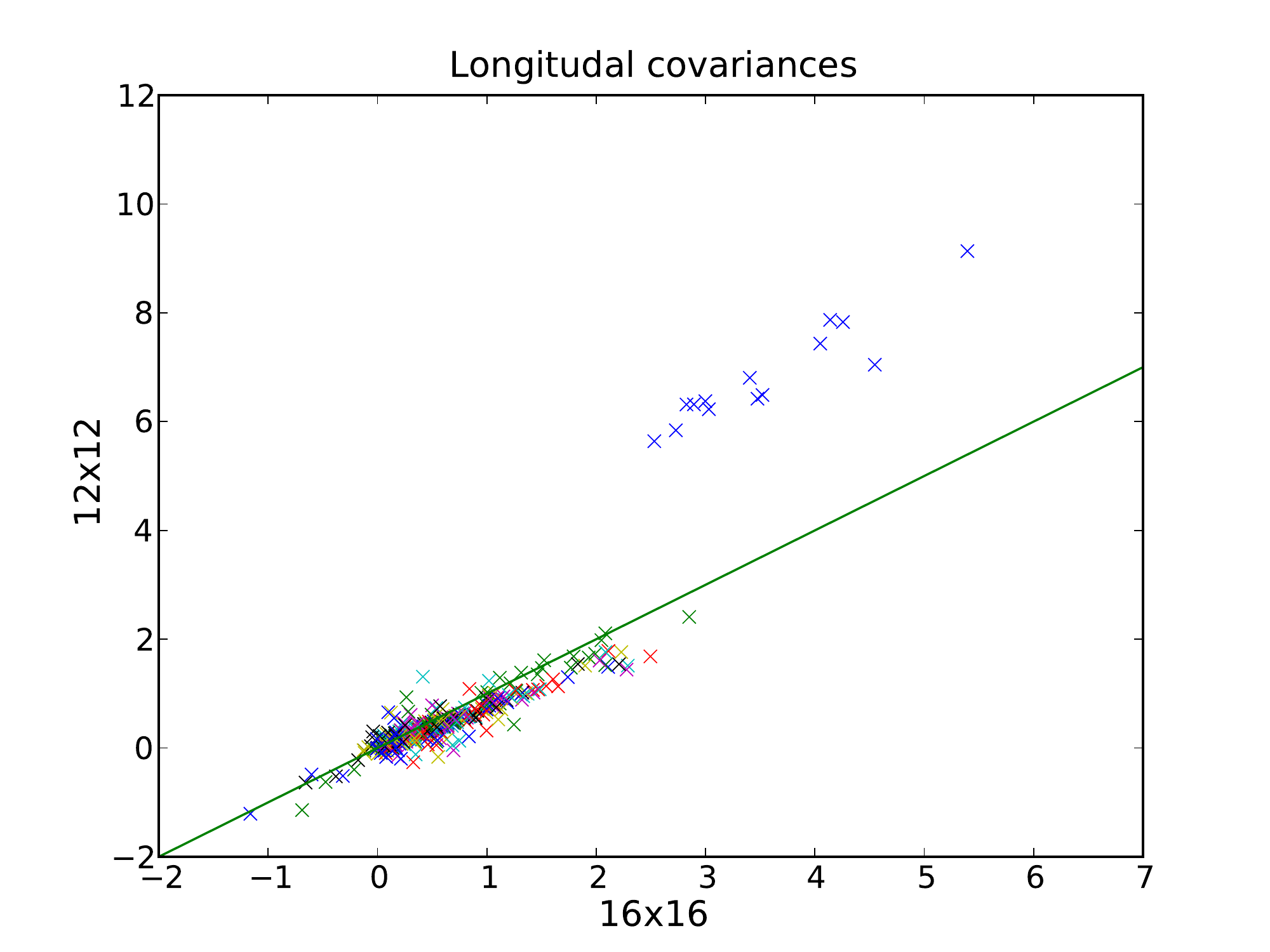}
  \label{fig:cov_16_12}\\
 \centering
  \subfloat[]{\includegraphics[width=0.71\textwidth]{\plotpath 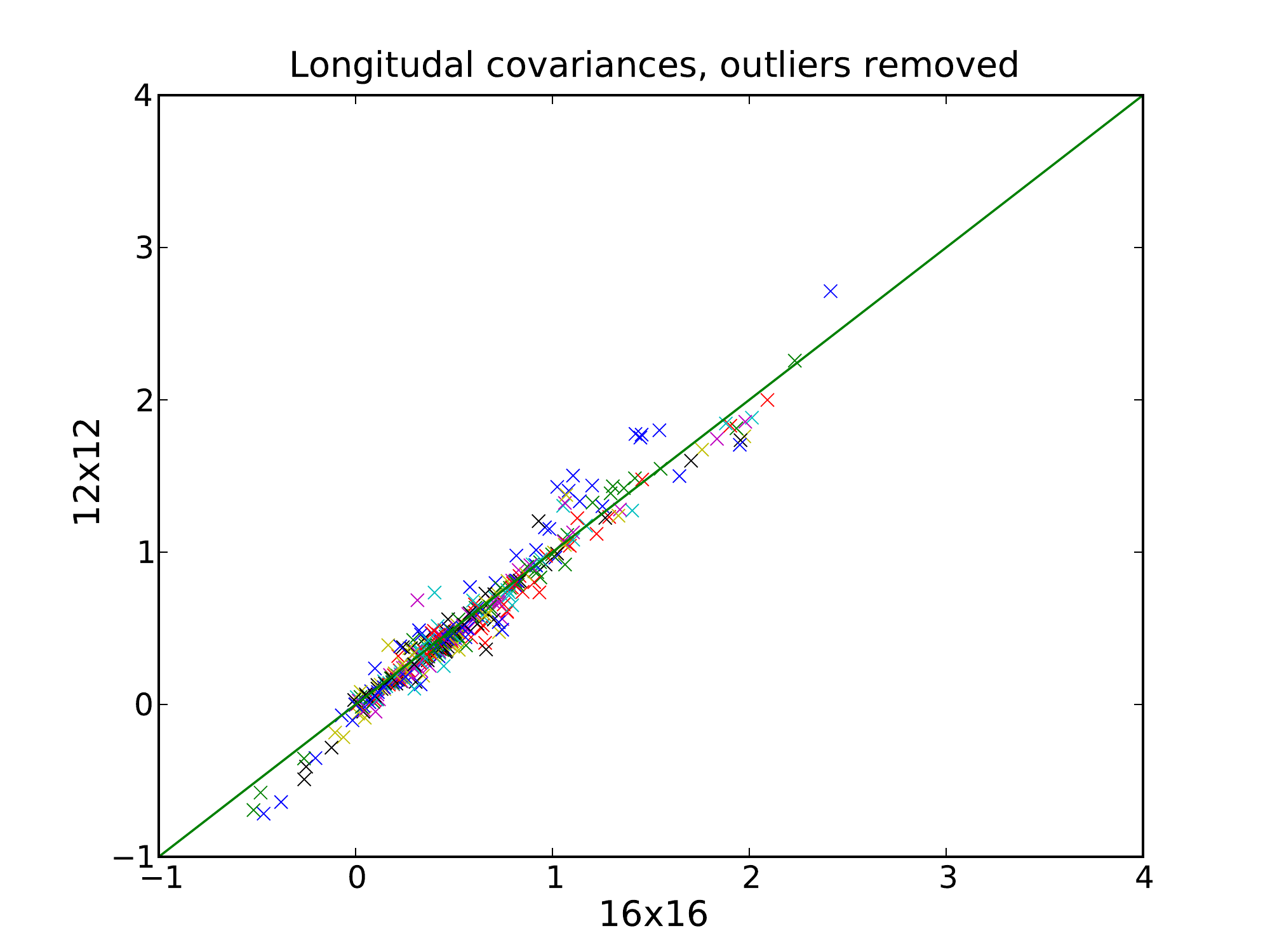}
  \label{fig:cov_16_12_out}}
 \caption{Resulting covariances obtained with a $16 \times 16$ circular subfield mask for image shift calculations compared with the resulting covariances when a $12 \times 12$ circular subfield mask was applied. The different colours show different angular separation. (a) Covariances calculated with all shift measurements. The blue points that have a much higher covariance when using the 12 pixel mask corresponds to no subaperture separation $s=0$. (b) Covariances calculated after shift measurements larger than $3\sigma$ was removed.}
 \label{fig:cov_16_12_2}
\end{figure}

\newpage
\sec{Smaller subfield size and removed outliers}
\label{sec:mask}
The resulting covariances achieved after changing the circular subfield mask (applied when calculating the image shift) are compared to the original covariances in \refpics{fig:cov_16_12_2}{fig:cov_16_8_2}. 

\begin{figure}[ht!]
  \centering
  \subfloat[]{\includegraphics[width=0.71\textwidth]{\plotpath 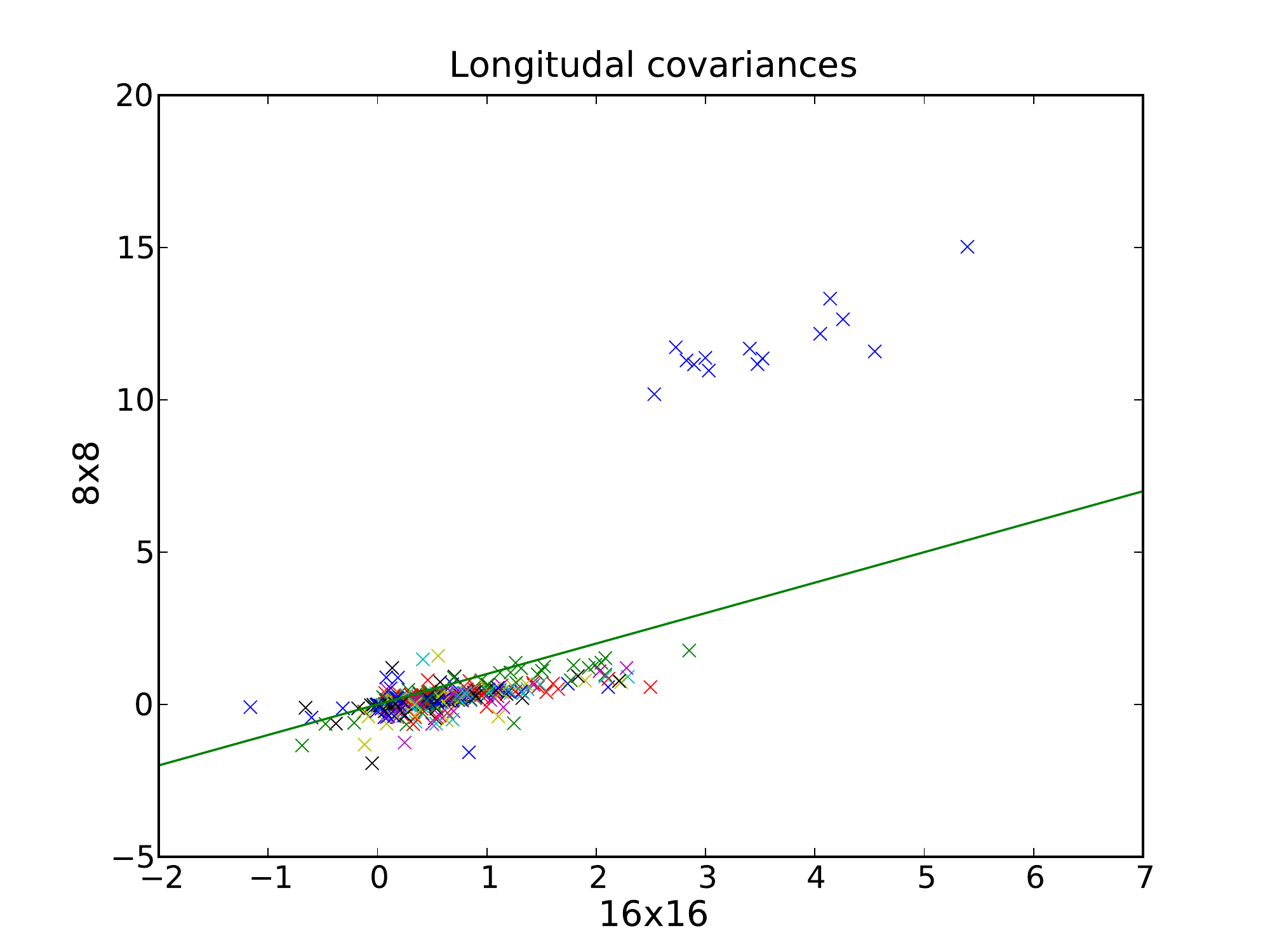}
  \label{fig:cov_16_8}} \\
 \centering
  \subfloat[]{\includegraphics[width=0.71\textwidth]{\plotpath 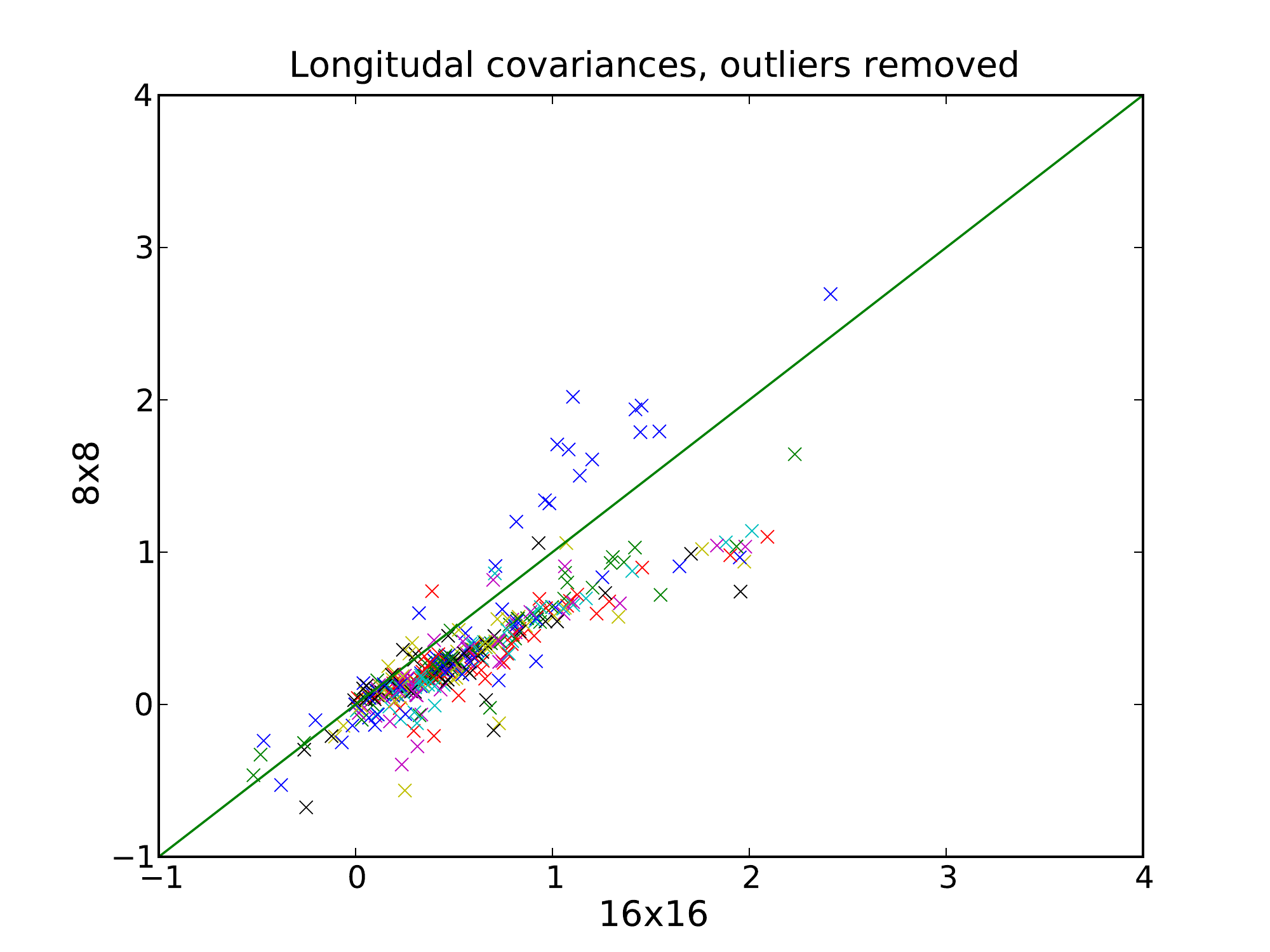}
  \label{fig:cov_16_8_out}}
 \caption{Resulting covariances obtained with a $16 \times 16$ circular subfield mask for image shift calculations compared with the resulting covariances when a $8 \times 8$ circular subfield mask was applied. The different colours show different angular separation. (a) Covariances calculated with all shift measurements. The blue points that have a much higher covariance when using the 12 pixel mask corresponds to no subaperture separation $s=0$. (a) Covariances calculated with all shift measurements. \\ (b) Covariances calculated after shift measurements larger than $3\sigma$ was removed.}
 \label{fig:cov_16_8_2}
\end{figure}

Changing the mask size from 16 to 12 pixels increased the number of failed shifts (larger than $3\sigma$) from 1.35 \% to 3.29 \%. The percentage of failed shifts for the 8 pixel mask was 7.43 \%.
\newpage
Both the 12 and 8 pixel mask showed lower covariances than the 16 pixel mask, before the outliers where removed. This was still the case for the 8 pixel mask after the outliers where removed, see \refpic{fig:cov_16_8_out}. As can be seen in \refpic{fig:cov_16_12_out}, the 12 pixel mask and the 16 pixel mask correlate well after removal of the failed shifts.



\chap{Conclusions}
\label{chap:conclusions}
Some conclusions drawn from the results in \refchap{chap:results} are presented in this chapter together with some thoughts about next step and  future improvements.

\sec{$r_0$ at different heights}
Scharmer and van Werkhoven \citep{2010A&A...513A..25S} found that the smallest values for $r_0$ at the highest height were associated with poor ground layer seeing. The images get noisy and the cross-correlations fail if the small values of $r_0$ is kept. They determined the value for when $r_0$ should be rejected to 7.5 cm, which is about 75 \% of the subaperture size. 

As discussed in \refsec{sec:r0} and shown in \refpic{fig:corr_wfwfs}, the same result was found during this project. It was therefore concluded that this limit shall be used.

The values of $r_0$ differs a lot from height to height, which is shown in \refpics{fig:r0_wfwfs_0}{fig:r0_wfwfs2}. $r_0$ is small for the first height, $h=0$ m, as well as for the highest height $h=20$ km. It can therefore be concluded that one turbulent layer is located at the ground level and one at higher altitude.

The intermediate layers have $r_0$ larger than 40 cm and $r_0$ is sometimes really large for $h=1300$ m. It can be concluded that the turbulent layer at this height is weak.

\sec{Correlation between WFWFS and SHABAR}
The correlation between calculated $r_0$ from the WFWFS and from the SHABAR is best for lower layers, $h=0$ and $h=500$ m. 
Next WFWFS node $h=1300$ showed very varying and high $r_0$ and no comparison with the SHABAR was therefore possible.

The small correlation found for higher layers still have most contribution from layer $h_{shabar}=500$ m. It therefore seems likely that the SHABAR cannot distinguish layers at these heights. This is consistent with the convergence of the SHABAR kernels \citep{ATST-RPT-0014,2010SPIE.7733E.144S}.

\sec{New subfield mask and outliers removed}
\label{sec:mask_out}
The comparison between the covariances achieved with different sizes of the circular subfield mask applied when calculating the image shifts, \refpics{fig:cov_16_12_2}{fig:cov_16_8_2}, showed that the 12 pixel mask is useful but the 8 pixel mask is not.

Removal of outliers is an improvement that will be used in the further analysis.

\sec{Future}
\label{sec:future}
Even though data from several days are reduced it is only one day (2010-06-09) that is compared in detail with the SHABAR and this comparison was done before the outliers were removed from the WFWFS data. A new comparison, without these outliers, would be interesting. Comparing more days both in the beginning and end of the observation season would also be interesting. The reduction script can, after being started, work autonomously with reduction of data from a whole year. The problem is where to store the reduced data. The resulting covariance map takes up little space but the steps in between produce quite large files. One days reduction takes up 20--170 GB depending on the number of good sets and the number of references used in the reduction.

A similar WFWFS will be placed at the Vacuum Tower Telescope (VTT) on Tenerife. After comparison of the data from the two WFWFS, the different sites can be evaluated in terms of strength, location and stratification of the seeing.

Comparison with a long baseline SHABAR and a WFWFS (at both sites) will also be done. This SHABAR will have a baseline of 3.5 m.

After all these further reductions and comparisons, it might be possible to decide where to build the European Solar Telescope and how to make a good design for its Multi Conjugate Adaptive Optic system.



\chap{Author's contributions}
\label{chap:contributions}
This chapter is a short summary of what I have done during my master project.

\subsub*{Inspection}
I changed some thresholds for the inspection to easier distinguish between \textit{bad} and \textit{good} frames. I also wrote several inspection scripts that makes it easier to compare the results from the automated inspection. 

All data from 2009 and most of 2010 were inspected using these scripts and the result can now be used to make a decision of which files to save.

\subsub*{Reduction in one go}
I wrote a reduction script so that all the reduction steps could be evaluated after each other. It is now possible to reduce data from a whole year in one go.

\subsub*{Change of subfield mask}
I changed the circular subfield mask used when calculating the image shifts to test if better height resolution could be achieved. 
Two new sizes where tested.

\subsub*{Removing outliers}
The S-DIMM+ analysis was changed so that outliers (failed shift measurements) are removed before the covariances are calculated.

\subsub*{Comparison between the WFWFS and the SHABAR}
I compared the results from the Wide Field Wavefront Sensor with the Shadow Band Ranger mounted on the SST. 

\subsub*{Documentation}
I have documented the changes and updated the existing documentation for \Astooki \, written by T.I.M. van Werkhoven \citep{Astooki}.

\subsub*{Updating software}
I have had discussions with T.I.M. van Werkhoven about implementing my changes into his software permanently.

\appendix
\renewcommand{\chapord}[1]{}



\chap{Inspection scripts}
\label{app:inspection}
All inspection scripts store the output to a text file after every step. Not needed if you're only interested in the output of the last step but it turned out to be quite handy to be able to look at all of them at different times.

Shown below is a small bash script that inspects which status codes all bad frames had. The path to all log files from 2010 are stored in \verb!logfiles.txt!. 

\begin{lstlisting}[style=codesnippet, caption={Inspection of bad frames (\textbf{inspection.sh})},label=code:bad]
# ---------------------------------------------------------
# Locate logfiles and sort out files with only bad data
# ---------------------------------------------------------
# Store in a badlog
logfiles=`cat logfiles.txt`
for logfile in $logfiles
do
  bad=`cat $logfile | grep -c BAD`
  if [ $bad != 0 ]
  then
    echo $logfile:$bad >> badlog.txt
  fi
done
# ---------------------------------------------------------
# Read path to logfile from badlog.txt
FILES=`awk -F ':' '{print $1}' badlog.txt`

# ---------------------------------------------------------
# Find status codes and store in a file 
# ---------------------------------------------------------
for FILE in $FILES
do
  # Display status for files with error status
  grep BAD $FILE | cut -d',' -f 4 | sort -u | cut -d'-' -f 2
  
done>statuses.txt

# Make a sorted list of available status codes
sort -nu statuses.txt >statuscodes_bad.txt
rm statuses.txt

# ---------------------------------------------------------
# List all directories (data sets) and number of bad files
# ---------------------------------------------------------

# Loading the statuses into an "array"
STATUSCODES=`awk '{print $1}' statuscodes_bad.txt`

for FILE in $FILES
do
  # Display which directory your looking in
  id=`echo $FILE | cut -f 4 -d/`
  echo "Processing directory: $id"
  
  # Display status for files with bad status
  grep BAD $FILE | cut -f 4 -d',' | cut -d'-' -f 2 >filestatuses.txt
    
  for code in $STATUSCODES
  do
    count=`grep -wc "$code" filestatuses.txt`
    echo "Number of files with status $code: $count"
  done
done>bad.txt
rm filestatuses.txt

# ---------------------------------------------------------
# Read the bad files and store them in an array
# ---------------------------------------------------------
# Number of statuscodes
columns=$(( `wc -l statuscodes_bad.txt | cut -f 1 -d' ' ` + 1 ))
rows=$(( `wc -l badlog.txt | cut -f 1 -d' '` + 1 ))

row=1
column=1
while [ $row -le $rows ]
do
  while [ $column -le $(( $row * $columns )) ]
  do
    text=`head -n $column bad.txt | tail -n 1 | cut -f 2 -d':' | cut -f 2 -d' '`
    printf "%s \t" $text
    let column++
  done
  echo ""
  let row++
done>badarray.txt
\end{lstlisting}



\chap{Reduction script}
\label{app:reduction}
The code snippet in \ref{code:reduce} is a part of the bash script \textbf{reduce.sh}, that calls and sends input variables to \textbf{pyatk.py}. \textbf{pyatk.py} is the base class for all other ''tools'' provided by \textit{Astooki}. 

The different steps in the reduction described in \refsec{sec:astooki} are shown below. The beginning of the script, where flat fields and dark frames are chosen, are not shown below. Neither are the loops that make the reduction run autonomously for all specified days.

\begin{lstlisting}[style=codesnippet, caption={Reduction script (\textbf{reduce.sh})},label=code:reduce]
#-----------------------------------------------------------
# Subimage mask
# Equal for all data sets (runs)
pyatk.py samask -vv -d /scratch/wfwfs/proc/simask \
  --file simask.csv \
  --rad 1024 \
  --shape circular \
  --sasize 175,154 \
  --pitch 194,166 \
  --xoff 0,0.5 \
  --disp 1017,1000 \
  --scale=1 \
  --plot 

#-------------------------------------------------------
# Optimizing the subimage mask for that set
pyatk.py saopt -vv -d /scratch/wfwfs/proc/$day/$imset-simask-fopt \
  --mf /scratch/wfwfs/proc/simask/simask-origin.csv \
  --ff /scratch/wfwfs/$ffday/$ffday-$ffset/$ffbest --fm 300 \
  --file simask-orig-fopt.csv \
  --saifac 0.8 \
  --rad 1024 \
  --plot

#-------------------------------------------------------
# Get sasize values (from prev. step)
sasize=`grep subimage /scratch/wfwfs/proc/$day/$imset-simask-fopt/astooki-log | cut -f 3 -d '(' | cut -f 1 -d ')'| sort -u`

# Subfield mask
pyatk.py sfmask -vv -d /scratch/wfwfs/proc/$day/$imset-sfmask \
  --file sfmask-big.csv \
  --sfsize=113,89 \
  --sasize=$sasize \
  --overlap=0,0 \
  --border=30,30 \
  --plot

#-------------------------------------------------------
# Static offset
pyatk.py shifts -vv -d /scratch/wfwfs/proc/$day/$imset-statoff \
  --ff /scratch/wfwfs/$ffday/$ffday-$ffset/$ffbest --fm 300 \
  --df /scratch/wfwfs/$ddday/$ddday-$ddset/$ddbest --dm 300 \
  --safile /scratch/wfwfs/proc/$day/$imset-simask-fopt/simask-orig-fopt.csv \
  --sffile /scratch/wfwfs/proc/$day/$imset-sfmask/sfmask-big.csv \
  --range 7 --nref 1 --mask none --plot \
  /scratch/wfwfs/$day/$day-$imset/wfwfs_survey_im* 
  
#-------------------------------------------------------
# Updated subimage mask (with offset)
pyatk.py saupd -vv -d /scratch/wfwfs/proc/$day/$imset-statoff \
  --mf /scratch/wfwfs/proc/$day/$imset-simask-fopt/simask-orig-fopt.csv \
  --offset /scratch/wfwfs/proc/$day/$imset-statoff/static-offsets.csv \
  --plot  

#-------------------------------------------------------
# Subfield mask with smaller subfields
# Pick the smallest sasize (from prev. step)
xsasize=`cat /scratch/wfwfs/proc/$day/$imset-statoff/simask-orig-fopt-updated.csv | head -n 3 | tail -n 1 | cut -f 1 -d ,`
ysasize=`cat /scratch/wfwfs/proc/$day/$imset-statoff/simask-orig-fopt-updated.csv | head -n 3 | tail -n 1 | cut -f 2 -d ,`

pyatk.py sfmask -vv -d /scratch/wfwfs/proc/$day/$imset-sfmask \
  --file sfmask-16x16.csv \
  --sfsize=16,16 \
  --sasize=$xsasize,$ysasize \
  --overlap=0.7,0.7 \
  --border=7,7 
  
#-------------------------------------------------------
# Measure the subfield shifts
pyatk.py shifts -vv -d /scratch/wfwfs/proc/$day/$imset-subshift-16x16 \
  --ff /scratch/wfwfs/$ffday/$ffday-$ffset/$ffbest --fm 300 \
  --df /scratch/wfwfs/$ddday/$ddday-$ddset/$ddbest --dm 300 \
  --safile /scratch/wfwfs/proc/$day/$imset-statoff/simask-orig-fopt-updated.csv \
  --sffile /scratch/wfwfs/proc/$day/$imset-sfmask/sfmask-16x16.csv \
  --range 7 --nref 2 --mask circular \
  /scratch/wfwfs/$day/$day-$imset/wfwfs_survey_im*

#---------------------------------------------------------
# Subaperture mask for the lenslets
# Equal for all runs
pyatk.py samask -vv -d /scratch/wfwfs/proc/samask \
  --file samask-ll.csv \
  --rad 0.52 \
  --shape circular \
  --sasize 0.098,0.098 \
  --pitch 0.098,0.0849 \
  --xoff 0,0.5 \
  --disp 0,0 \
  --scale=1 --plot
    
#-------------------------------------------------------
# S-DIMM+ analysis
pyatk.py sdimm -vv -d /scratch/wfwfs/proc/$day/$imset-sdimm-16x16 \
  --skipsa -1 \
  --shifts /scratch/wfwfs/proc/$day/$imset-subshift-16x16/image-shifts.npy \
  --safile /scratch/wfwfs/proc/samask/samask-ll-centroid.csv \
  --sffile /scratch/wfwfs/proc/sfmask/sfmask-16x16.csv

\end{lstlisting}



\small
\bibliographystyle{natbib}
\bibliography{references}

\end{document}